\documentclass[final,1p,times]{elsarticle}

\usepackage{graphicx}
\usepackage[usenames,dvipsnames]{color}

\usepackage{url}
\usepackage{dcolumn}
\usepackage{bm}
\usepackage{color}
\usepackage[normalem]{ulem}
\usepackage{empheq}

  \usepackage{amsmath} 
  \usepackage{amssymb} 
  \usepackage{amsthm}
  
\usepackage{mathtools}
\usepackage{color}

\newcommand{\be}{\begin{equation}}
\newcommand{\ee}{\end{equation}}
\newcommand{\bea}{\begin{eqnarray}}
\newcommand{\eea}{\end{eqnarray}}

\newcommand{\ddt}{\frac{\partial}{\partial t}}
\newcommand{\ihddt}{i\hbar\frac{\partial}{\partial t}}

\newcommand{\op}[1]{\hat{#1}}
\newcommand{\ave}[1]{\langle #1 \rangle}
\newcommand{\comm}[2]{\left[ #1,\,#2 \right]_{-}}
\newcommand{\trace}[1]{{\rm Tr}\left[ #1 \right]}
\newcommand{\traceN}[1]{{\rm Tr}_{\rm N}\left[ #1 \right]}

\newcommand{\absN}[1]{{| #1 \rangle}}
\newcommand{\absNS}[1]{{\langle #1 |}}

\newcommand{\ident}[0]{\mathbb{I}}

\newcommand{\group}[1]{{\{#1\}}}
\newcommand{\absNN}[1]{{| #1 \rangle}_{\rm N}}
\newcommand{\absNC}[1]{{| #1 \rangle}_{\rm C}}
\newcommand{\absCC}[2]{{|#1 \rangle_{\rm C\,C}\langle #2 |}}
\newcommand{\ex}[1]{{#1}_{\rm ex}}

\newcommand{\exN}[1]{{#1}_{\rm ex,N}}
\newcommand{\opNC}{{\op{N}_{\rm C}}}
\newcommand{\opNN}{{\op{N}_{\rm N}}}

\newcommand{\dopNC}{{\delta\op{N}_{\rm C}}}

\newcommand{\toX}{\xrightarrow{\rm ex}}
\newcommand{\NC}{N_{\rm C}}
\newcommand{\NN}{N_{\rm N}}
\newcommand{\nuN}{F_{\rm N}}

\newcommand{\rhoC}{\rho_{\rm BEC}}

\newcommand{\avex}[1]{{\langle #1 \rangle}_{\rm ex}}

\journal{Annals of Physics}

\begin{document}

\begin{frontmatter}



\title{}

\title{Excitation picture of an interacting Bose gas}
\author{M.~Kira}
\address{Department of Physics, Philipps-University Marburg, Renthof 5, D-35032
Marburg, Germany}

\begin{abstract}
Atomic Bose-Einstein condensates (BECs) can be viewed as macroscopic objects where atoms form correlated atom clusters to all orders. Therefore, the presence of a BEC  makes the direct use of the cluster-expansion approach --- lucrative e.g.~in semiconductor quantum optics --- inefficient when solving the many-body kinetics of a strongly interacting Bose. An excitation picture is introduced with a nonunitary transformation that exclusively describes the system in terms of atom clusters within the normal component alone. The nontrivial properties of this transformation are systematically studied, which yields a cluster-expansion friendly formalism for a strongly interacting Bose gas. Its connections and corrections to the standard Hartree-Fock Bogoliubov approach are discussed and the role of the order parameter and the Bogoliubov excitations are identified. The resulting interaction effects are shown to visibly modify number fluctuations of the BEC. Even when the BEC has a nearly perfect second-order 
coherence, the BEC number fluctuations can still resolve interaction-generated non-Poissonian fluctuations.
\end{abstract}


\end{frontmatter}

\section{Introduction}
\label{sec:intro}

The atomic Bose- and Fermi-gas investigations have become increasingly more ingenious ever since the discovery of atomic Bose-Einstein condensates (BECs) \cite{Anderson:1995,Bradley:1995,Davis:1995} in the mid 1990s. Nowadays, one can routinely confine multiple atomic clouds in free space\cite{Andrews:1997,Orzel:2001} or on a lattice\cite{Jaksch:1998,Greiner:2002,Bakr:2009}, and even make BECs interact with each other\cite{Orzel:2001,Bloch:2008}, or prepare a Fermi gas to exhibit quantum degeneracy\cite{DeMarco:1999,Schreck:2001,Truscott:2001}, just to mention few highlights. At the same time, the development to control atom--atom interactions through a Fesbach 
resonance\cite{Kohler:2006,Giorgini:2008,Chin:2010} has opened the possibility to systematically study\cite{Inouye:1998,Courteille:1998,Roberts:1998,Cornish:2000,Takemura:2013,
Partridge:2005} the many-body quantum kinetics of strongly interacting Bose/Fermi gas. Conceptually, these atomic investigations start to approach many-body problems that have been studied, e.g., in nonlinear semiconductor 
optics\cite{Schmitt-Rink:1989,Kira:1999,Shah:1999,Khitrova:1999,Chemla:2001,Rossi:2002,Kira:2006,Meier:2007,Haug:2008,Haug:2009,Book:2011,Jahnke:2012,Malic:2013} for decades. Therefore, it clearly is interesting to explore which complementary insights many-body techniques -- refined for the semiconductor studies -- could provide for the strongly interacting Bose gas. In this paper, I develop a theoretic framework to connect these seemingly different many-body investigations, with the aim to identify the complementary aspects between typical semiconductor and BEC approaches.

Close to the equilibrium, an interacting Bose\cite{Dalfovo:1999,Leggett:2001,Andersen:2004,Blakie:2008,Proukakis:2008} or Fermi\cite{Chen:2005,Gurarie:2007,Giorgini:2008} gas can be accurately described with many sophisticated methods, which has provided detailed understanding of, e.g., many-body ground-state properties\cite{Ensher:1996,Giorgini:1999,Lee:2006}, BEC coherences\cite{Wright:1996,Javanainen:1996,Andrews:1997,Castin:1997,Ruostekoski:1998,Bloch:2000}, 
BEC dynamics\cite{Castin:1996,Milburn:1997,Javanainen:1999,Ohberg:1999,Santos:2000,Davis:2001,Davis:2001b}
superfluidity\cite{Carlson:2003,Zwierlein:2005,Lobo:2006,Leyronas:2007,Gurarie:2007}, vortices\cite{Marzlin:1997,Matthews:1999,Madison:2000,AboShaeer:2001,Ruostekoski:2001,Weiler:2008}, 
spectroscopic properties\cite{Kinnunen:2004,Yu:2006,Stewart:2008,Papp:2008,Braaten:2010},
so-called Tan relations\cite{Tan:2008a,Tan:2008b,Tan:2008c} and their consequences \cite{Blume:2009,Combescot:2009,Stewart:2010,Drut:2011,Barth:2011,Werner:2012,Wild:2012},
so-called BCS-BEC crossover\cite{Astrakharchik:2004,Burovski:2008,Bulgac:2008,Zhang:2009},
strong atom--atom interactions\cite{Cornish:2000,Greiner:2002,Viverit:2004,Carlson:2005,Bulgac:2006,Kinnunen:2006,Braaten:2008,Navon:2011,Smith:2012},
and Efimov physics\cite{Efimov:1973,Kraemer:2006,Stecher:2009,Williams:2009,Pollack:2009,Castin:2010} in a strongly interacting atom gas.
It also is interesting to study  situations where the BEC is somehow excited far from the equilibrium, such as in the Bosenova experiments\cite{Donley:2001,Altin:2011} where the BEC collapses due to a change in the atom--atom interactions. To explain the many-body quantum kinetics of the BEC, various perturbative approaches have been successfully used for weak interactions. One possibility is to apply the 
Hartree-Fock-Bogoliubov (HFB)\cite{Baranger:1961,Goodman:1974,Mang:1975,Zaremba:1999,Bender:2003, Milstein:2003, Wuster:2005} equations that couple the generalized Gross-Pitaevskii equation with the mean-field many-body dynamics of normal-component density and anomalous density. This approach qualitatively explains the spatial changes in the atom cloud during, e.g., the Bosenova implosion and eventual collapse of the atom cloud.

However, the HFB analysis cannot explain quantitatively the properties of BEC too far from equilibrium because it is based on the perturbation theory. For example, the HFB produces a collapse time that is up to 100\% longer\cite{Wuster:2005} than in the Bosenova experiment\cite{Donley:2001}. This analysis was extended in Ref.~\cite{Wuster:2007} to compare the HFB approach with the truncated Wigner approximation (TWA)\cite{Werner:1997,Steel:1998,Sinatra:2002,Norrie:2006} which produced essentially the same results; the outlook of this work concludes that one must extend both the TWA and the HFB approach to systematically include higher-order many-body correlations in order to quantitatively explain the nonperturbative phenomena such as the Bosenova. 

The concepts of semiconductor quantum optics \cite{Kira:2006b,Book:2011,Jahnke:2012} could provide a complementary description for such correlations because they already provide an extremely accurate and nonperturbative nonequilibrium treatment\cite{Kira:1999,Kira:2006,Kira:2008,Kira:2011,Book:2011} of the many-body and quantum-optical interaction effects\cite{Kira:2006b,Koch:2006,Khitrova:2006,Smith:2010,Hunter:2014} among fermionic electrons and bosonic photons\cite{Yoshie:2004,Reithmaier:2004,Ulrich:2007,Gies:2009} and phonons\cite{Forstner:2003,Nielsen:2004,Carmele:2010} far from equilibrium. When extending this approach for the strongly interacting Bose gas, one must first understand what happens when the atom--atom interactions become so strong that they can eject a large fraction of atoms from the BEC to the normal component. This process appears even at 0\,K because the {\it interactions} among normal-component atoms may result to a lower energy than atoms have inside the BEC. This phenomenon is 
often referred to as {\it quantum 
depletion}\cite{Xu:2006,Wuster:2007,Cui:2013} in 
contrast to thermal 
depletion of the BEC. The simplest description of such a process follows from Bogoluibov excitations, as experimentally demonstrated in Refs.~\cite{StamperKurn:1999,Jin:1996,Mewes:1996,Utsunomiya:2008} for a relatively weakly interacting Bose gas. As the interactions become stronger, significant modifications are expected based on the HFB insights discussed above.

My conceptualization of semiconductor quantum optics is founded on the general properties of the quantum statistics which is any representation defining uniquely {\it all} quantum properties of the many-body system, as formulated in Ref.~\cite{Kira:2011}. For example, a density matrix or a Wigner function are possible choices for the quantum statistics. Alternatively, one may apply the cluster-expansion approach\cite{Wyld:1963,Fricke:1996,Kira:2006,Kira:2008,Book:2011} to determine quantum statistics in terms of correlated particle clusters within the many-body system. Physically, clusters with $N$ particles correspond, e.g., to molecular states as well as correlated transition amplitudes. Therefore, the cluster expansion provides a natural way to identify stable cluster configurations within many-body systems. Even more so, it can be systematically applied to include the dynamics among {\it all} particle clusters up to a user-defined particle number. As a principal feature, pair-wise many-body interactions 
create 
higher-order 
clusters 
only sequentially in time\cite{Book:2011,Mootz:2012}, which makes the cluster-expansion 
extremely efficient approach when solving the quantum-kinetic evolution from low-order clusters to more complex clusters. Such a quantum-kinetic method is nonperturbative\cite{Kira:2006,Book:2011} and extremely successful in explaining quantitative properties of a great variety of systems; in nuclear physics and quantum chemistry, the coupled-cluster approach\cite{Coester:1958,Coester:1960,Cizek:1966,Bartlett:2007} has become one of the most accurate many-body methods. Likewise, the cluster-expansion approach describes the nonequilibrium quantum kinetics of many-body systems with utmost accuracy and predictability, as demonstrated in both semiconductor optics\cite{Haug:2008} and  semiconductor quantum optics\cite{Book:2011,Jahnke:2012}.

However, the interacting Bose gas poses a major challenge for an efficient description of the cluster generation because atoms inside the BEC are already correlated to all orders, in the atom-cluster sense. In other words, already the initial state of the quantum depletion is extremely highly correlated such that it is not clear how interaction-induced evolution from low- to high-rank atom clusters can be efficiently isolated in a strongly interacting Bose gas. This means that the BEC clusters ``overshadow'' the ones being generated by the quantum depletion, which makes the direct application of the standard cluster-expansion approach inefficient. One of the main goals of this paper is to find a way to focus the investigation on the generated clusters instead of the ones already present in the BEC.

In this paper, I convert the interacting Bose gas problem into a format where application of the powerful cluster-expansion techniques becomes directly possible. 
Section \ref{sec:Hamiltonian} presents the standard many-body Hamiltonian and cluster-expansion properties of the BEC.
I will then introduce the excitation picture in Sec.~\ref{sec:Excitation-pic}; much of this work involves finding a proper transformation that focuses the analysis onto the clusters that are generated by the quantum depletion. This essentially ``shifts'' the representation of quantum statistics to a ``frame'' where the BEC appears as a particle vacuum such that all atom clusters describe correlated normal-component atoms excited by the quantum depletion, hence the name ``excitation picture''. The found transformation is nonunitary such that its special properties must be carefully analyzed as is done in Secs.~\ref{sec:Excitation-trafo}--\ref{sec:EP-relations}. After that, the quantum statistical aspects of quantum depletion on BEC are studied in 
Sec.~\ref{sec:QS-basics}. These formulations yield the exact excitation-picture system Hamiltonian $\ex{\op{H}}$ that is presented in Sec.~\ref{sec:H-excitation}; the resulting $\ex{\op{H}}$ serves as a general starting point for generic cluster-expansion studies that can be carried out in full analogy to the semiconductor investigations. Sections~\ref{sec:Bogloliubov}--\ref{app:Coherence} present how the semiconductor-based approach complements and extends the standard BEC concepts\cite{StamperKurn:1999} such as the Bogoliubov excitations, coherence of the BEC, and the HFB approach. 

Altogether, this paper sets up a cluster-expansion friendly framework for a strongly interacting Bose gas. The explicit quantum kinetics of atom clusters is derived and studied in Ref.~\cite{Kira:HBE}, on this solid basis, which completes the work started in this paper. While working through these details, I will systematically refer to ``we'' because the derivations will require active participation of the reader.

\section{System Hamiltonian for interacting Bose gas}
\label{sec:Hamiltonian}

Following the consensus of BEC studies\cite{Dalfovo:1999,Leggett:2001,Andersen:2004,Blakie:2008,Proukakis:2008}, also we start the investigations from  a many-body Hamiltonian
\begin{align}
  \op{H} = 
    \int d^3r\;\op{\Psi}^\dagger({\bf r}) H_0({\bf r} ) \op{\Psi}({\bf r})
    + 
    \frac{1}{2} \int d^3r\,d^3r'\;
    \op{\Psi}^\dagger({\bf r}) \op{\Psi}^\dagger({\bf r}') 
    V({\bf r}'-{\bf r}) \op{\Psi}({\bf r}')\op{\Psi}({\bf r})
\label{eq:Hamiltonian}\;,
\end{align}
containing bosonic field operators $\op{\Psi}^\dagger({\bf r})$ and $\op{\Psi}({\bf r})$ for the atoms involved. The atom trapping is described by a single-particle contribution $H_0({\bf r}) \equiv -\frac{\hbar^2 \nabla^2}{2m}+U({\bf r})$ where atoms with mass $m$ are also subjects to a potential $U({\bf r})$ created e.g.~by an atom trap. The many-body aspects arise from the pair-wise interaction  potential $V({\bf r})$ between the atoms. Despite its name, it actually generates multi-atom interaction processes at all levels due to the inevitable BBGKY-hierarchy problem\cite{Yvon:1935,Bogoliubov:1946,Born:1946,Kirkwood:1946}. 
The systematic description of the BBGKY hierarchy constitutes the most challenging part of many-body problems, and it still remains unsolvable,  being approachable only through suitable approximations. We intend to convert the BEC problem 
into a form where the cluster-expansion approach can be systematically, accurately, and efficiently applied to approximate the BBGKY hierarchy, as is done in the semiconductor quantum optics\cite{Kira:2006,Kira:2008,Kira:2011,Book:2011}.

Since many-body physics can be solved only approximatively, it is very important to limit the investigations to the most relevant aspects of the problem at hand. As the first decision, we must choose which aspects of the atoms must be included to the $\op{\Psi}({\bf r})$ in order to describe the physics of the strongly interacting Bose gas. Fundamentally, atoms are constituent particles of electrons, protons, and neutrons. Consequently, the atom--atom interactions may involve all length scales from the long-range Coulomb coupling to extremely short length scales of the strong force inside the nuclei. However, we are studying here ultracold atoms, which makes the atom energies so low that internal atomic configurations can hardly be changed, let alone ionized into electron--ion plasma, when the ultracold atoms collide with each other. Therefore, the many-body aspects of BECs can indeed be described with an atomic $\op{\Psi}({\bf r})$ that ignores the internal atomic substructure by treating atoms as 
``elementary'' particles whose total spin is integer valued for the interacting Bose gas; we follow this common choice\cite{Dalfovo:1999,Leggett:2001,Andersen:2004,Blakie:2008,Proukakis:2008}. A very different situation emerges in semiconductors when one considers excitons that are bound, composite, pairs of electrons and holes (electronic vacancies in the valence band). While the internal structure of atoms can be largely ignored, the fermionic electron--hole substructure of excitons often dominates the properties of semiconductors; for a textbook discussion cf.~Ref.~\cite{Book:2011}.

Furthermore, we choose to focus on studying strongly interacting Bose gas at low-temperatures where atoms remain essentially bound to their $s$-shell even when they become strongly interacting. Therefore, the $s$-shell atoms interact with the so-called $s$-wave scattering that yields  radially symmetric pair-wise interaction $V({\bf r})=V(|{\bf r}|)$ where $|{\bf r}|$ denotes the atom--atom separation.\cite{Dalfovo:1999} At large atom distances, $V({\bf r})$ stems from the dipole--dipole attraction defined by the van der Waals force. At small distances, atom--atom interactions become repulsive due to ion--ion and electron--electron repulsion. Many of these aspects can be successfully described by replacing  $V(|{\bf r}|)$ by a contact-potential\cite{Leggett:2001,Giorgini:2008} that is a nonanalytic function, but produces possibilities to solve several nontrivial aspects of the many-body problem analytically.\cite{Tan:2008a,Tan:2008b} We do not specify $V({\bf r})$ explicitly in this paper because we want to 
develop a flexible framework that has analogy to the semiconductor studies.

Typically, the many-body interactions involve much shorter length scales (nanometer scale) than the size of the atom cloud (micrometer scale). Therefore, the essential many-body effects are generated within regions where the atom cloud appears to be locally homogeneous. In the spirit of local-density approximation (LDA), we choose to investigate a homogeneous many-body system in order to find the relevant structure to tackle the BBGKY hierarchy problem with the cluster expansion. After a cluster-expansion friendly formulation is found, the theory for the inhomogeneous systems can be developed straightforwardly. The development toward that direction is discussed in Sec.~\ref{sec:HFB}.

For homogeneous systems, it is convenient to set $U({\bf r})$ to zero because the potential cannot change across regions where the LDA is valid. We then express the field operators using a plane-wave expansion:
\begin{align}
  \op{\Psi}({\bf r}) = \frac{1}{{\cal L}^{3/2}} \sum_{\bf k} e^{i{\bf k} \cdot {\bf r}} \, B_{\bf k}\,,
  \qquad
  \op{\Psi}^\dagger({\bf r}) = \frac{1}{{\cal L}^{3/2}} \sum_{\bf k} e^{-i{\bf k} \cdot {\bf r}} \, B^\dagger_{\bf k}
\label{eq:Field-ops}\,,
\end{align}
where $B_{\bf k}$ and $B^\dagger_{\bf k}$ are boson operators of an atom having momentum $\hbar{\bf k}$, expressed here with the help of the wave vector ${\bf k}$. The quantization lenght is given by ${\cal L}$. For later use, we summarize the standard boson commutation relations
\begin{align}
  \comm{B_{\bf k}}{B^\dagger_{{\bf k}'}} = \delta_{{\bf k},{\bf k}'}\,,
  \qquad
  \comm{B_{\bf k}}{B_{{\bf k}'}} = 0 = \comm{B^\dagger_{\bf k}}{B^\dagger_{{\bf k}'}}
\label{eq:boson-comm}\,.
\end{align}
To simplify the bookkeeping, we have normalized the plane waves inside a quantization box that has a volume ${\cal L}^3$.  For noninteracting systems, atom ${\bf k}$ has the energy $E_{\bf k} = \frac{\hbar^2{\bf k}^2}{2m}$ defined by its kinetic energy because the trapping potential is neglected. 

In general, $U({\bf r})$ does not directly influence the many-body effects even though it affects how the atom cloud spreads. The spreading dynamics itself can be described with the Gross-Pitaevskii equation\cite{Dalfovo:1999,Leggett:2001,Andersen:2004,Blakie:2008,Proukakis:2008} or its generalizations, discussed further in Sec.~\ref{sec:HFB}. At the same time, the different hyperfine levels of atoms can have a qualitatively different  $U({\bf r})$ resulting to the so-called open (closed) channel when the atom scattering has only unbound (bound molecular) solutions within the relevant energy range. The hyperfine levels can be coupled by a magnetic field to produce a Feshbach resonance\cite{Feshbach:1958} which dispersively modifies the strength of atom--atom interactions. By tuning the system through a Feshbach resonance, one can control both the sign and magnitude of the atom--atom interactions with an external magnetic field.\cite{Stwalley:1976,Tiesinga:1993,Kohler:2006,Bloch:2008,Chin:2010} 
The physics of Feshbach resonance can be included by reducing a multi-level model\cite{Feshbach:1958,Kokkelmans:2002} to an effective single-channel analysis\cite{Leggett:2001,Wuster:2007,Giorgini:2008,Chin:2010} with a freely tunable interaction strength $V({\bf r})$. We use explicitly an effective $V({\bf r})$ whereas the multi-level extension is briefly outlined after the full Hamiltonian is worked out, in the end of Sec.~\ref{sec:Separation}.
 
\subsection{Separation of BEC and normal component}
\label{sec:Separation}

The basis choice \eqref{eq:Field-ops} also yields a simple classification of BEC vs.~normal-component atoms: the BEC atoms have a macroscopic occupation only at the single-particle ground state ${\bf k}=0$ while the normal-component atoms are found only at states with a nonzero momentum, i.e.~${\bf k} \neq 0$. We utilize this separation when we insert the field operators \eqref{eq:Field-ops} into Hamiltonian \eqref{eq:Hamiltonian}, producing
\begin{eqnarray}
  \op{H} 
  &=& 
    \sum_{\bf k}{}' E_{\bf k} \, B^\dagger_{\bf k} B_{\bf k} 
    + 
    \frac{V_0}{2} 
    \left( 
      B^\dagger_0 B^\dagger_0 B_0\, B_0
      + 2 \sum_{\bf k}{}' B^\dagger_0 B^\dagger_{\bf k} B_{\bf k} \, B_0
      + \sum_{{\bf k},{\bf k}'}{}' B^\dagger_{\bf k} B^\dagger_{{\bf k}'} B_{{\bf k}'} \, B_{\bf k}
    \right)
\nonumber\\
  &+&\sum_{\bf k}{}' V_{\bf k} B^\dagger_0 B^\dagger_{\bf k} B_{\bf k} \, B_0
  +\frac{1}{2}\sum_{\bf k}{}' V_{\bf k} 
  \left[ 
    B^\dagger_0 B^\dagger_0 B_{\bf k} \, B_{-{\bf k}} 
    +
    \, B^\dagger_{-{\bf k}} B^\dagger_{\bf k} B_0 B_0 
  \right]
\nonumber\\
  &+&
  \sum_{{\bf k},{\bf k}'}{}' V_{\bf k} 
  \left[ 
    B^\dagger_0 B^\dagger_{{\bf k} + {\bf k}'} B_{{\bf k}'}\,  B_{\bf k} 
    +
    B^\dagger_{\bf k} \, B^\dagger_{{\bf k}'}  B_{{\bf k} + {\bf k}'} B_0
  \right]
  +
  \frac{1}{2} \sum_{{\bf k}\neq {\bf k}'}{}' \sum_{{\bf q}\neq({\bf k},{\bf k}')} \, V_{{\bf k}-{\bf k}'} 
  B^\dagger_{{\bf k}} B^\dagger_{{\bf q}-{\bf k}} B_{{\bf q}-{\bf k}'}\,  B_{{\bf k}'}
\label{eq:H-in-B}\,,\quad
\end{eqnarray}
after having introduced the Fourier transform of the pair-wise interaction
\begin{eqnarray}
  V_{\bf q} \equiv \frac{1}{{\cal L}^3} \int d^3r\, V({\bf r}) \,e^{-i{\bf q} \cdot {\bf r}}
\label{eq:V-Fourier}\;,
\end{eqnarray}
where ${\cal L}^3$ is the quantization volume.
To separate the BEC from the normal-component atoms, we have also introduced a set of normal-component sums: 
\begin{eqnarray}
  \sum_{{\bf k}}{}' \equiv \sum_{{\bf k} \neq 0}\,, \qquad\qquad
  \sum_{{\bf k},{\bf k}'}{}' \equiv \sum_{{\bf k} \neq 0}\sum_{{\bf k}'\neq 0}\,,\qquad\qquad
  \sum_{{\bf k}\neq {\bf k}'}{}' \equiv \sum_{{\bf k} \neq 0}\sum_{{\bf k}'\neq \group{0,\,{\bf k}}}  
\label{eq:Special-sums}
\end{eqnarray}
that exclude the zero-momentum element corresponding to the BEC. Hamiltonian \eqref{eq:H-in-B} does not introduce approximations to the interactions because the basis choice \eqref{eq:Field-ops} is valid for any bosonic many-body system and the separation in the BEC and normal components is exact.

At this point, we may generalize the treatment to include several hyperfine levels of the ultracold atoms following the phenomenological approach of Refs.~\cite{Stwalley:1976,Tiesinga:1993,Kohler:2006,Bloch:2008,Chin:2010,Kokkelmans:2002}. One simply adds the hyperfine states $\absN{\lambda}$ to the basis states. For example, the plane-wave state $e^{i{\bf k} \cdot {\bf r}}$ becomes $e^{i{\bf k} \cdot {\bf r}}\, \absN{\lambda}$ such that the corresponding boson operator is attached with an additional quantum number, i.e.~$B_{\bf k} \rightarrow B_{\lambda,{\bf k}}$. By inserting the field operator $\op{\Psi}({\bf r}) = \frac{1}{{\cal L}^{3/2}} \sum_{\lambda,{\bf k}} e^{i{\bf k} \cdot {\bf r}}\, \absN{\lambda}\, B_{\lambda,{\bf k}}$ into Eq.~\eqref{eq:Hamiltonian}, we obtain
\begin{eqnarray}
  \op{H} 
  &=& 
    \sum_{\lambda,{\bf k}} E^\lambda_{\bf k} \, B^\dagger_{\lambda,{\bf k}} B_{\lambda,{\bf k}} 
    + 
  \frac{1}{2} \sum_{\lambda,\nu,\nu',\lambda'}\sum_{{\bf k},{\bf k}',{\bf q}} \, V^{\lambda,\nu,\nu',\lambda'}_{{\bf k}-{\bf k}'} 
  B^\dagger_{\lambda,{\bf k}} B^\dagger_{\nu,{\bf q}-{\bf k}} B_{\nu',{\bf q}-{\bf k}'}\,  B_{\lambda',{\bf k}'}
\label{eq:H-in-B-hyper}\,,
\end{eqnarray}
where both single-particle energies $E^\lambda_{\bf k}$ and interactions $V^{\lambda,\nu,\nu',\lambda'}_{{\bf k}-{\bf k}'}$ depend on the hyperfine levels involved. In case we included only one $\absN{\lambda}$ level, Eq.~\eqref{eq:H-in-B-hyper} reduces to Hamiltonian \eqref{eq:H-in-B} after the BEC and normal-component are separated from one another. Hamiltonian \eqref{eq:H-in-B-hyper} has an identical structure compared to semiconductor nanostructures, with the exception that $\lambda$ refers to different electronic states and the operators are fermionic. However, this resemblance is not enough to automatically guarantee an efficient cluster-expansion approach for the BEC, as discussed in Sec.~\ref{sec:PLE_correlations}.

The influence of a Feshbach resonance on the BEC can be often described by including one open channel ($\absN{\lambda=o}$) and one closed channel ($\absN{\lambda=c}$) that contains strongly bound atom-molecules, the dimers. In this situation, the atom--atom interactions produce 16 different $(\lambda,\nu,\nu',\lambda')$ combinations. Quite often, only a subset of these are needed to model the properties of the Feshbach resonances, cf.~Ref.~\cite{Roberts:1998,Kokkelmans:2002,Gurarie:2005,Chin:2010} for more details. Once the atom--atom interactions are tuned to a specific value, we do not need to follow the open--close channel coupling, but the interaction effects that are created by the modified atom--atom interactions. This can be performed with a single effective $\absN{\lambda}$ state and $V({\bf r})$, which is used as a common starting point of many-body investigations\cite{Dalfovo:1999,Leggett:2001,Andersen:2004,Blakie:2008,Proukakis:2008}. Therefore, we explicitly apply the single-boson level 
formulation throughout this paper.

In the corresponding the Hamiltonian \eqref{eq:H-in-B}, we also have made a division into two subsets -- the BEC (${\bf k}=0$) and normal-component (${\bf k} \neq 0$) atoms. This division produces only 12 (9 topologically) different combinations, which is four (seven) fewer that open--closed channel separation. This reduction originates from the momentum conservation because Hamiltonian \eqref{eq:H-in-B} contains only such combinations of boson operators where the sum of creation-operator momenta is equal to the sum of annihilation-operator momenta. With this constraint, pairwise interactions cannot induce processes involving three BEC operators and one normal-component operator because the momentum sum of the creation and annihilation parts cannot then be matched. For example, two atoms within a BEC cannot scatter into a normal and a BEC state. 

\begin{figure}[t]
\includegraphics*[scale=0.35]{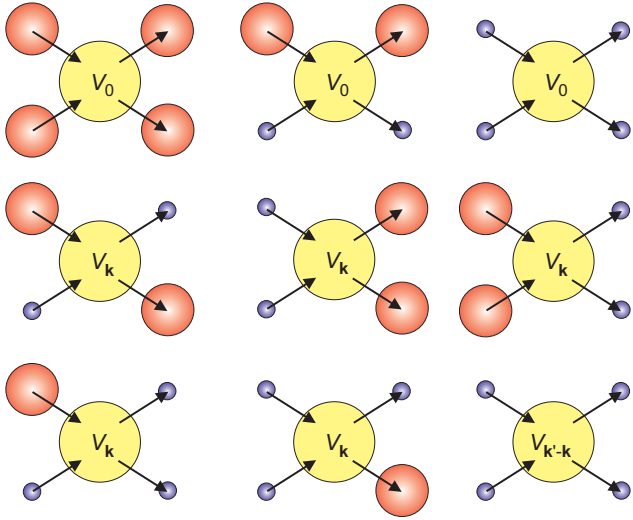}
\caption{(Color online) Diagrammatic representation of atom--atom interactions \eqref{eq:H-in-B}.
A large shpere identifies a condesate atom while a small sphere denotes a normal-component atom. The arrows entering and exciting the interaction vertex (yellow-shaded circle) signify atom annihilation and creation, respectively.}\label{Vdiagams}
\end{figure}

Figure \ref{Vdiagams} shows diagrammatically the interaction possibilities among the BEC and normal-component atoms allowed by the Hamiltonian \eqref{eq:H-in-B}; The annihilation operators are depicted as arrows entering the interaction vertex (large circle with the matrix element $V_{\bf q}$) while the exiting arrows denote the 
creation operators. Each large sphere identifies a BEC state whereas the small spheres refer to a normal-component atom. We observe that the contributions with three BEC operators are indeed completely missing from the diagrams, which introduces a reduction in the interaction possibilities. 

\subsection{Cluster-expansion representation}
\label{sec:CE_representation}

General quantum properties of interacting Bose gas can be represented, e.g., with the normally-ordered characteristics\cite{Walls:2008,Kira:2008} function,
\begin{eqnarray}
  \chi(\left\{ \alpha \right\}) \equiv 
  \ave{
    e^{\alpha_0 \, B^\dagger_0 + \sum_{\bf k}{}' \alpha_{\bf k} \, B^\dagger_{\bf k}}
    \;\;
    e^{-\alpha^\star_0 \, B_0 - \sum_{\bf k}{}' \alpha^\star_{\bf k} \, B_{\bf k}}
   }  
\label{eq:chi_N}\;,
\end{eqnarray}
that constitutes a quantum statistics as shown in Ref.~\cite{Kira:2008}. In this context,  $\left\{ \alpha \right\}$ refers to a group of all complex-valued $\alpha_{\bf k}$ and $\alpha^\star_{\bf k}$ arguments connected with the $B^\dagger_{\bf k}$ and $B_{\bf k}$ operators, respectively. By defining differentials,
\begin{eqnarray}
  \partial_{\bf k} \equiv \frac{\partial}{\partial \alpha_{\bf k}}\,,\qquad
  \partial^\star_{\bf k} \equiv \frac{\partial}{\partial \alpha^\star_{\bf k}}
\label{eq:differentials}\;,
\end{eqnarray}
we can connect $\chi(\left\{ \alpha \right\})$ to yet another quantum statistics, the expectation-value representation,
\begin{eqnarray}
  \ave{
    B^\dagger_{{\bf k}_1} \cdots B^\dagger_{{\bf k}_J}
    B_{{\bf k}'_L} \cdots B^\dagger_{{\bf k}'_1}
  }
  =
  (-1)^L
  \partial_{{\bf k}_1} \cdots \partial_{{\bf k}_J} \partial^\star_{{\bf k}'_L} \cdots \partial^\star_{{\bf k}'_1}  
  \left.
  \chi(\left\{ \alpha \right\}) \right|_{\{ \alpha =0 \}}
\label{eq:I-rep}\;,
\end{eqnarray}
where $|_{\{ \alpha =0 \}}$ denotes that all $\alpha_{\bf k}$ indices are set to zero after the differentiation. Any expectation value can be generally classified based on how many boson operators they contain. When it contains $J$ creation and $L$ annihilation operators, it is a $(J+L)$-particle operator. 

The cluster-expansion approach\cite{Kira:2006,Kira:2008,Book:2011} systematically identifies the correlations among particles within the many-body system. Hence, it is useful to introduce the correlation-generating function 
\begin{eqnarray}
  \xi(\left\{ \alpha \right\})
  \equiv 
  {\rm ln} 
    \left[ 
      \chi(\left\{ \alpha \right\})
    \right] 
    \qquad \Leftrightarrow \qquad
     \chi(\left\{ \alpha \right\}) = e^{\xi(\left\{ \alpha \right\})}
\label{eq:xi_N}\,
\end{eqnarray}
which provide a unique connection between $\chi$ and $\xi$ such that also $\xi$ is one possible quantum-statistical representation. More importantly, it uniquely identifies a specific particle cluster, i.e.~the many-body correlation
\begin{eqnarray}
  \Delta\ave{
    B^\dagger_{{\bf k}_1} \cdots B^\dagger_{{\bf k}_J}
    B_{{\bf k}'_L} \cdots B^\dagger_{{\bf k}'_1}
  }
  =
  (-1)^L
  \partial_{{\bf k}_1} \cdots \partial_{{\bf k}_J} \partial^\star_{{\bf k}'_L} \cdots \partial^\star_{{\bf k}'_1}  
  \left.
  \xi(\left\{ \alpha \right\}) \right|_{\{ \alpha =0 \}}
\label{eq:DI-rep}
\end{eqnarray}
that also constitutes quantum statistics as shown in Ref.~\cite{Kira:2008}. Physically, $\Delta \ave{\cdots}$ containing $J$ creation and $L$ annihilation operators is a $(J+L)$-particle correlation that exists only if $(J+L)$ particles are clustered together. This is the basis of the cluster-expansion representation of boson fields. Formally, expectation value \eqref{eq:I-rep} can also be factored in terms of clusters using the Wick's theorem.\cite{Wick:1950} We call single-, two-, three-, and four-atom clusters singlets, doublets, triplets, and quadruplets, respectively.

\subsection{Particle correlations of a noninteracting BEC}
\label{sec:PLE_correlations}

Below the critical temperature, the BEC emerges to the lowest-energy state when there is no continuous Bose-Einstein distribution that can accommodate all atoms.\cite{Pethick:2002} In other words, particle-number conservation is of central importance in realizing the BEC. Therefore, we follow the tradition of number-conserving theory\cite{Girardeau:1956,Gardiner:1997,Girardeau:1998,Castin:1998,Morgan:2003,Gardiner:2007,Gardiner:2012,Billam:2012,Mora:2003} 
to describe the BEC. Alternatively, the BECs have been successfully described by introducing coherence created by Beliaev broken symmetry\cite{Beliaev:1958}; one simply assumes that the BEC somehow becomes a coherent state, which violates the particle-number conservation, not possible for nonrelativistic particles\cite{Leggett:2001}. As a major benefit of this approach, it straightforwardly yields the Gross-Pitaveskii equation which provides the correct description of many central properties of the BEC, such as superfluidity\cite{Griffin:2012,Wright:2012}. 
As discussed in Sec.~\ref{app:Coherence} and Refs.~\cite{Leggett:2001,Wright:2012}, inclusion of further many-body effects to the Beliaev approach becomes difficult for strongly interacting Bose gas. Therefore, the number-conserving approach is more appropriate for the theory development of this paper; the connection of the developed number-conserving approach with the coherence is discussed further in Sec.~\ref{app:Coherence}.

To assess the amount of relevant clusters within an atom BEC, we evaluate the clusters in a noninteracting BEC at 0\,K. When the atom trap is well-enough isolated from the environment, the atom system becomes closed such that both the total energy and total particle number are constant, establishing a microcanocical ensemble. We denote the total number of atoms by ${\cal N}$. At 0\,K, each of the ${\cal N}$ noninteracting atoms must occupy the single-particle ground state, i.e.~the zero-momentum state associated with $B_0$. Therefore, the resulting 0\,K many-body wave function must necessarily be a Fock state $\absN{{\cal N}}$ of the ground state because any other state cannot contain exactly ${\cal N}$ atoms.

Since we consider here only the quantum statistics of the BEC, we set all other $\alpha_{\bf k}$ arguments of Eq.~\eqref{eq:chi_N} to zero, except $\alpha_0$ defining the BEC. This procedure introduces the characteristic function of the BEC:
\begin{eqnarray}
  \chi_{\rm BEC}(\alpha) \equiv 
  \ave{
    e^{\alpha \, B^\dagger_0 }
    \;\;
    e^{-\alpha^\star\, B_0}
   }  
   = \sum_{J,L=0}^\infty \frac{\alpha^J(-\alpha^\star)^L}{J!\,L!} \, \ave{[B^\dagger_0]^J [B_0]^L}
\label{eq:chi_cond}\;,
\end{eqnarray}
after Taylor expanding the exponential functions and omitting the explicit ``0'' index from $\alpha$ to shorten the notation. In the same way, the correlations follow from
\begin{eqnarray}
  \xi_{\rm BEC}(\alpha) \equiv 
  {\rm ln}
  \left[\chi_{\rm BEC}(\alpha)  \right]
  \equiv \sum_{J,L=0}^\infty \frac{\alpha^J(-\alpha^\star)^L}{J!\,L!} \Delta \ave{[B^\dagger_0]^J [B_0]^L}
\label{eq:xi_cond}\,,
\end{eqnarray}
based on definitions \eqref{eq:xi_N}--\eqref{eq:DI-rep}.

For a Fock state $\absN{{\cal N}}$, the $(J+L)$-particle expectation value becomes
\begin{eqnarray}
   \ave{[B^\dagger_0]^J [B_0]^L}
   = 
   \delta_{J,L} \frac{{\cal N}!}{({\cal N}-J)!}
\label{eq:I-Fock}\;,
\end{eqnarray}
which can be determined using the Basic properties of the Fock state given by Eq.~\eqref{eq:app-B_rules}. The factorials within this expression should be understood in a general sense, expressed through the gamma function $n!=\Gamma(n+1)$. Since $\Gamma(x)$ diverges for zero or negative-valued integer arguments, expectation value \eqref{eq:I-Fock} automatically vanishes for $J=L$ greater than the number of atoms ${\cal N}$. With this information, the characteristic function \eqref{eq:chi_cond} becomes
\begin{eqnarray}
  \chi_\absN{\cal N}(\alpha) 
 =\sum_{J=0}^{\cal N} \frac{{\cal N}!\, (-|\alpha|^2)^L}{J!\,J!\,({\cal N}-J)!} 
\label{eq:chi_Fock}
\end{eqnarray}
for the Fock state $\absN{\cal N}$. Based on definition \eqref{eq:xi_cond}, we also find
\begin{eqnarray}
  \xi_\absN{\cal N}(\alpha) \equiv 
  {\rm ln}
  \left[\sum_{J=0}^{\cal N} \frac{{\cal N}!\, (-|\alpha|^2)^L}{J!\,J!\,({\cal N}-J)!}  \right]
\label{eq:xi_Fock}\;.
\end{eqnarray}
Since $\xi_\absN{\cal N}(\alpha)$ does not depend on the phase of $\alpha$, all correlations $\Delta\ave{[B^\dagger_0]^J [B_0]^L}$ with an unequal number of creation and annihilation operators must vanish, in analogy to expectation-value expression \eqref{eq:I-Fock}, when they are computed with the help of Eq.~\eqref{eq:DI-rep}. More explicitly, the conversion formula \eqref{eq:DI-rep} produces the relevant two-, four- and six-atom correlations
\begin{eqnarray}
  \Delta\ave{B^\dagger_0 B_0} = {\cal N}\,,
  \quad
  \Delta\ave{[B^\dagger_0]^2 [B_0]^2} = -{\cal N}\left({\cal N} +1 \right)\,,
  \quad
  \Delta\ave{[B^\dagger_0]^3 [B_0]^3} = 2{\cal N}\left({\cal N} +1 \right)\left(2 {\cal N} +1 \right)
\label{eq:dI_Fock}\,,
\end{eqnarray}
respectively.

\begin{figure}[t]
\includegraphics*[scale=0.55]{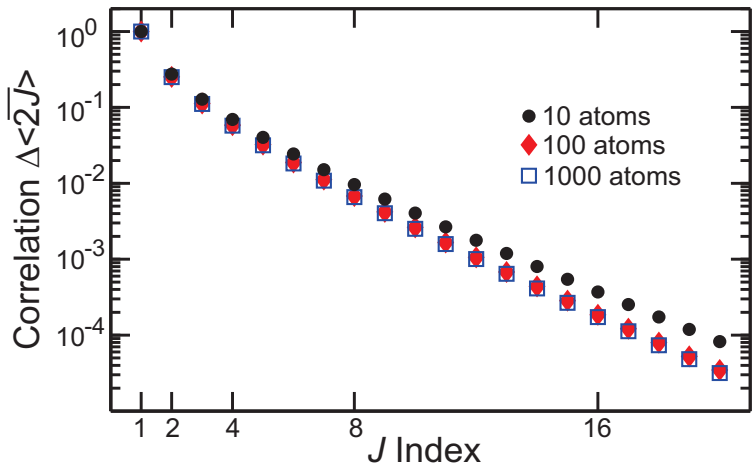}
\caption{(Color online) Atom-cluster correlations of a noninteracting BEC. Normalized $2J$-atom correlations $\Delta \bar{\ave{2J}}$ are plotted as function of the cluster number. The BEC contains ${\cal N}=10$ (black circles),  ${\cal N}=100$ (red diamonds),  ${\cal N}=1000$ (open squares) atoms. The BEC forms a closed system.
}\label{fig:FockCOR}
\end{figure}

We observe that the pure BEC results in particle correlations whose magnitude increases with the particle number. To analyze this even more transparently, we introduce a normalized $2J$-particle correlation
\begin{eqnarray}
  \Delta \bar{\ave{2J}}
  \equiv \frac{\Delta\ave{[B^\dagger_0]^J [B_0]^J}}{J!\,J!\, {\cal N}^J}
\label{eq:dI_Norm}\;,
\end{eqnarray}
which is the Taylor-expansion coefficient of $\xi_\absN{\cal N}(\alpha)$ normalized by the atom-number ${\cal N}$ to power $J$. Figure \ref{fig:FockCOR} presents $|\Delta \bar{\ave{2J}}|$ as function of cluster number $C=2J$ for a BEC having ${\cal N}=10$ (circle), ${\cal N}=100$ (diamond), and ${\cal N}=1000$ (open square) atoms. In all cases, the atom-correlations extend to a very high particle number while the normalized correlation approaches the same functional form for elevated ${\cal N}$. 

To see this very clearly, one can express the unnormalized correlations \eqref{eq:dI_Fock} to the leading order of ${\cal N}$; we find that $|\Delta\ave{[B^\dagger_0]^J [B_0]^J}|$ scales as ${\cal N}^J$ for large ${\cal N}$. We therefore conclude that the direct application of the cluster expansion to the interacting Bose gas must essentially include clusters to {\it all} orders. This is not entirely unexpected because the BEC atoms act collectively as a macroscopically correlated entity. In other words, the formation of the macroscopic entity induces atom--atom clusters that extend over all particles involved. Unfortunate for straightforward many-body investigations, this property seems to prevent an efficient use of the cluster-expansion approach whenever the many-body system contains a BEC. Next, we will develop a formalism to resolve this problematic issue.

\section{Interacting Bose gas in excitation picture}
\label{sec:Excitation-pic}

As motivated in Sec.~\ref{sec:PLE_correlations}, we describe the interacting Bose gas within the tradition of number-conserving approaches\cite{Girardeau:1956,Gardiner:1997,Girardeau:1998,Castin:1998,Morgan:2003,Gardiner:2007,Gardiner:2012,Billam:2012,Mora:2003}. 
These describe, e.g., situations where the atoms are removed slowly from the trap compared to the relevant many-body interaction time scales. When the trap is isolated enough, both the total atom number (${\cal N}$) and energy are fixed, yielding a microcanocical system. These ${\cal N}$ atoms can be distributed arbitrarily between the BEC and the normal component such that the particle number within each atom subsystem --- BEC or normal component --- is not fixed. Obviously, the subsystem energy is not fixed either, which makes the subsystems grand canonical ensembles. The inclusion of atom loss from the trap can be introduced as a simple loss after the relevant many-body interaction dynamics is solved within a microcanonical ensemble, which allows us to fully describe atom traps 
that are open systems. The generalization for all ensembles is outlined in Sec.~\ref{app:Coherence}.

The basis states of the normal-component atoms can conveniently  be identified using the number representation
\begin{eqnarray}
  \absNN{\group{n_{\bf k}}}
  \equiv
  \prod_{{\bf k}\neq 0} \absN{n_{\bf k}}_{\bf k}
\label{eq:Normal-state}\;,
\end{eqnarray}
where each $\absN{n_{\bf k}}_{\bf k}$ is a Fock state that contains {\it exactly} $n_{\bf k}$ atoms at the normal component ${\bf k}$. In total, $\absNN{\group{n_{\bf k}}}$ contains $\sum_{\bf k} n_{\bf k}$ normal-component atoms. Since the total system is microcanonical, this leaves exactly 
\begin{eqnarray}
  \NC^{\group{n_{\bf k}}} \equiv {\cal N} - \sum_{\bf k}{}' n_{\bf k}
\label{eq:NBEC}\;
\end{eqnarray}
atoms to the BEC. Therefore, the BEC is described by the Fock state $\absNC{\NC^{\group{n_{\bf k}}}}$ such that the generic  microcanonic wave function becomes
\begin{eqnarray}
  \absN{\Phi} = \sum_{\{n_{\bf k}\}} \phi_\group{n_{\bf k}} \, \absNC{N^C_{\{n_{\bf k} \}}} \otimes \absN{\{n_{\bf k} \}}
\label{eq:Phi-MC}\;,
\end{eqnarray}
where $\sum_{\group{n_{\bf k}}}$ is performed over all those normal-component occupations that leave the BEC occupation positive, i.e.~$N^C_\group{n_{\bf k}} \ge 0$. The coefficients $\phi_\group{n_{\bf k}}$ determine the amplitude of each occupation configuration within the microcanonical wave function $\absN{\Phi}$. The generalization of $\absN{\Phi}$ into a density matrix is straightforward when one uses the basis states $\absN{N^C_{\{n_{\bf k} \}}} \otimes \absN{\{n_{\bf k} \}}$ to present it, as shown in Sec.~\ref{sec:EP-relations}. Nevertheless, already the wave-function form \eqref{eq:Phi-MC} provides useful insights for good strategies when solving problems involving an interacting Bose gas. 

\subsection{Excitation-picture transformation}
\label{sec:Excitation-trafo}

At zero temperature, all atoms occupy the condensed state for a weakly interacting atom gas. The corresponding many-body wave function is then described by a single Fock state, $\absN{\Phi} =  \absNC{\cal N}$, which implies correlations among all of the ${\cal N}$ particles involved, as shown in Sec.~\ref{sec:PLE_correlations}. Due to these atom--atom correlations, one cannot {\it directly} describe the entire interacting Bose gas using only a few particle clusters, which can potentially make the cluster-expansion approach inefficient. At the same time, the normal component of the Bose gas does not contain quantum-degenerate states, which makes it much less correlated than the BEC part. Therefore, it is likely that the normal component of the Bose gas can be described with only a few clusters. We will next seek for a specific transformation that yields a cluster-expansion-friendly treatment for {\it both} normal and BEC components of the system.

For this purpose, we introduce BEC lowering and rising operators:
\begin{eqnarray}
  \op{L} = \sum_{n=0}^\infty \absCC{n}{n+1}\,,
  \qquad
  \op{L}^\dagger = \sum_{n=0}^\infty \absCC{n+1}{n}
\label{eq:L-operators}\;,
\end{eqnarray}
respectively, where $\absN{n}_{\rm C}$ is a Fock state of the BEC. It is straightforward to express $\op{L}$ and $\op{L}^\dagger$ also in terms of creation and annihilation operators,
\begin{eqnarray}
  \op{L} = \frac{1}{\sqrt{1+B^\dagger_0 B_0}}\,B_0,
  \qquad
  \op{L}^\dagger = B^\dagger_0 \,\frac{1}{\sqrt{1+B^\dagger_0 B_0}}
\label{eq:L-operators-with-B}\;,
\end{eqnarray}
by applying property \eqref{eq:app-B_rules}. Using the orthonormality of the Fock states, it is straightforward to derive the following properties:
\begin{eqnarray}
  \op{L} \, \op{L}^\dagger 
  &=& 
  \ident\,,\qquad  
  \op{L}^\dagger \, \op{L} = \ident - \absCC{0}{0}\,,\qquad
  \comm{\op{L}}{\op{L}^\dagger} =   \absCC{0}{0}  
\label{eq:L-property1}\;,
\\ 
  \op{L} \absNC{N+1} 
  &=& \absNC{N}\,, \qquad
  \op{L}^\dagger \absNC{N} = \absNC{N+1}\,, \qquad
  \left(\op{L}^\dagger\right)^{N} \absNC{0} = \absNC{N}
\label{eq:L-property2}\;,
\end{eqnarray}
for $N>0$. As usual, we classify operator sequences as normally (antinormally) ordered when all the creation and rising operators are ordered to the left (right). We observe that $\op{L}$ is almost unitary: the antinormally ordered product \eqref{eq:L-property1} yields identity while the normal-order product deviates from identity by the vacuum contribution. This seemingly harmless feature will introduce interesting properties for any $\op{L}$-based transformations, as shown below. We also introduce BEC-number operator
\begin{eqnarray}
  \opNC \equiv {\cal N} - \sum_{{\bf k}}{}' B^\dagger_{\bf k} B_{\bf k}
\label{eq:C-number-op}\;,
\end{eqnarray}
inspired by the microcanonical relation \eqref{eq:NBEC}; more specifically, $B^\dagger_{\bf k} B_{\bf k}$ is the number operator for the normal-component ${\bf k}$ and we have replaced $n_{\bf k}$ in Eq.~\eqref{eq:NBEC} by it.

By using property $\opNC \absNN{\group{n_{\bf k}}} = \NC^\group{n_{\bf k}} \absNN{\group{n_{\bf k}}}$ and relation \eqref{eq:L-property2}, we can rewrite the microcanonical wave function \eqref{eq:Phi-MC} in the form
\begin{eqnarray}
  \absN{\Phi} &=& \sum_{\{n_{\bf k}\}} \phi_\group{n_{\bf k}} \, \left( \op{L}^\dagger \right)^{\NC^{\group{n_{\bf k}}}} \absNC{0} \otimes \absN{\{n_{\bf k} \}}
    = \sum_{\{n_{\bf k}\}} \phi_\group{n_{\bf k}} \, \left( \op{L}^\dagger \right)^{\opNC} \absNC{0} \otimes \absN{\{n_{\bf k} \}}
\nonumber\\
    &=&
    \left( \op{L}^\dagger \right)^{\opNC} \absNC{0} \otimes  \sum_{\{n_{\bf k}\}} \phi_\group{n_{\bf k}} \absN{\{n_{\bf k} \}}
\label{eq:Phi-MC2}\;,
\end{eqnarray}
where the last step follows because $\left( \op{L}^\dagger \right)^{\opNC}$ and the vacuum state do not depend on the normal-component configuration. This expression allows us to directly identify a transfer operator 
\begin{eqnarray}
  \op{T}_{\rm ex}^\dagger \equiv  \left( \op{L}^\dagger \right)^{\opNC}
\label{eq:T-operator}
\end{eqnarray}
that produces the microcanonical wave function $\absN{\Phi}$ when it acts upon the product of BEC vacuum and the normal-component wave function,
\begin{eqnarray}
  \absNN{\Phi_T} \equiv \sum_{\{n_{\bf k}\}} \phi_\group{n_{\bf k}} \absN{\{n_{\bf k} \}}
\label{eq:Phi_T}\;.
\end{eqnarray}
The transfer operator \eqref{eq:T-operator} is only almost unitary because
\begin{eqnarray}
  \op{T}_{\rm ex} \op{T}_{\rm ex}^\dagger = \ident\,,
  \qquad
   \op{T}_{\rm ex}^\dagger \op{T}_{\rm ex}
   = 
   \left\{ 
      \begin{array}{cc}
	\ident\,, & {\rm if} \; {N}_C = 0 
	\\
	\ident - \sum_{j=0}^{N_C-1} \absCC{j}{j}\,, & {\rm otherwise}
      \end{array}
   \right.
\label{eq:T-unitarity}\;,
\end{eqnarray}
which follows directly from properties \eqref{eq:L-property1}. Here, $N_C$ should be understood as a number that is obtained when the operators eventually act upon a many-body state. Like in connection with Eq.~\eqref{eq:L-property1}, only the product of antinormally ordered operators yield identity whereas the normally ordered products yield additional contributions. Since $\op{T}_{\rm ex}$ is not unitary, we cannot benefit from many simple transform relations that are directly valid for unitary operators. Instead, we must carefully analyze the properties of each normally-ordered operator sequence, as is done in \ref{app:T-properties}. 

In several number-conserving approaches, different variants of $\op{L}$ and $\op{L}^\dagger$ have been successfully applied to either introduce {\it unitary} transformations\cite{Girardeau:1956,Girardeau:1998} to approximate the Hamiltonian via the Bogoliubov transformation or phonon/noise operators\cite{Gardiner:1997,Castin:1998,Mora:2003,Gardiner:2007} to include the interaction effects between the BEC and normal component perturbatively. 
To the best of my knowledge, $\op{L}$ and $\op{L}^\dagger$ have not yet been applied to provide a {\it nonunitary} transformation \eqref{eq:T-operator} to express the interacting Bose gas in a cluster-expansion friendly form, i.e.~in the excitation picture. Therefore, it clearly is interesting to study the implications of the excitation picture and its connections with the ``standard'' number-conserving approaches. We show in Secs.~\ref{sec:EP-relations} and \ref{sec:H-excitation} that the excitation picture provides a suitable platform to perform a nonperturbative 
cluster-expansion analysis of the strongly interacting Bose gas.

\subsection{Central relations of the excitation-picture}
\label{sec:EP-relations}

We can start with a simple relation $ \absN{\Phi} = \op{T}_{\rm ex}^\dagger \, \absNC{0} \otimes  \absNN{\Phi_T}$ that follows from Eqs.~\eqref{eq:Phi-MC2}--\eqref{eq:T-operator}. In the same way, any microcanonical density matrix $\op{\rho}$ can be transformed into the {\it excitation picture} via
\begin{eqnarray}
  \op{\rho} = 
  \op{T}_{\rm ex}^\dagger \; \op{\rho}_{\rm ex} \; \op{T}_{\rm ex},\, 
  \qquad \op{\rho}_{\rm ex} \equiv \absCC{0}{0} \otimes  \op{\rho}_{{\rm N},T}
\label{eq:rho-T}\;.
\end{eqnarray}
We adopt a notation that quantities in the excitation picture are denoted by a subindex ``ex''. When using the excitation picture, the BEC state is reduced into the vacuum state while $ \op{\rho}_{{\rm N},T}$ contains only the normal-component degrees of freedom, in full analogy to identification \eqref{eq:Phi-MC2}. Since the vacuum has vanishing particle correlations, the {\it BEC properties are trivial in the excitation picture}, which establishes a major simplification for describing the interacting Bose gas. 
In a sense, the excitation picture contracts the problem into a format where all nontrivial aspects involve {\it only} the normal component atoms excited by the quantum depletion, hence, the name {\it excitation picture}. The remaining normal-component contribution, i.e.~$\absNN{\Phi_T}$ or $\op{\rho}_{{\rm N},T}$, can obviously contain nontrivial atom clusters. However, since the normal component hosts a continuum of states within the same energy, it is clear that the quantum depletion does not excite atoms 
to a 
normal-component ``BEC'', but to a continuum of states. Consequently, one can expect that the excited atoms are far less correlated than they initially are inside the BEC. We next develop the formalism to express the many-body quantum kinetics entirely with the  excitation picture that converts the interacting Bose gas into a cluster-expansion-friendly format. 

To separate the highly correlated BEC state from the normal-component dynamics in interacting Bose gas, we utilize properties \eqref{eq:T-unitarity}--\eqref{eq:rho-T} to introduce the {\it excitation picture} for the many-body state and operators,
\begin{eqnarray}
  \op{\rho}_{\rm ex} = \op{T}_{\rm ex} \; \op{\rho} \; \op{T}_{\rm ex}^\dagger\,, \qquad
  \op{O}_{\rm ex} = \ex{\op{T}} \; \op{O} \; \ex{\op{T}}^\dagger
\label{eq:rho-T-and-O}\;,
\end{eqnarray}
respectively. Notice that $\op{\rho}$ and $\ex{\op{\rho}}$ appear to have a unitary connection, but this follows because $\op{\rho}$ contains more than $(\NC-1)$ condensed atoms, which makes $\ex{\op{T}}^\dagger \ex{\op{T}} = \ident$ according to Eq.~\eqref{eq:T-unitarity}. The validity of the $\op{\rho}$ identifications \eqref{eq:rho-T}--\eqref{eq:rho-T-and-O}  is further verified in \ref{app:T-properties}. 
For any other operator, the consequences of nonunitarity must be carefully examined. For example, the transformation of an operator product is not necessarily a product of individually transformed operators because
\begin{eqnarray}
  \ex{\left( \op{A} \, \op{B} \right)} \equiv \ex{\op{T}} \; \op{A} \, \op{B}  \; \ex{\op{T}}^\dagger
  \neq \ex{\op{A}} \, \ex{\op{B}}
\label{eq:AB-product}\;,
\end{eqnarray}
as shown in   \ref{app:T-properties}. Especially, one must be cautious when treating any normally ordered products of $\op{T}_{\rm ex}^\dagger$ and $\op{T}_{\rm ex}$.

Despite this complication, the excitation picture yields a set of extremely useful exact relations that simplify the many-body analysis considerably. For example, all expectation values can be computed completely within the excitation picture because we have
\begin{eqnarray}
  \ex{\ave{\ex{\op{O}}}} \equiv \trace{\ex{\op{O}}\,\ex{\op{\rho}}} =
  \trace{\ex{\op{T}} \; \op{O} \; \ex{\op{T}}^\dagger \,\ex{\rho}} = \trace{ \op{O} \; \ex{\op{T}}^\dagger \,\ex{\rho} \, \ex{\op{T}}} 
=
   \trace{ \op{O} \; \op{\rho} }  = \ave{\op{O}}
\label{eq:ex-average}\;,
\end{eqnarray}
where we have permuted  $\ex{\op{T}}$ under the trace, used property \eqref{eq:rho-T}, and identified the usual expression for the expectation value in the last step. 

In general, the evaluation of relevant expectation values can be simplified further by inserting identification \eqref{eq:rho-T} into definition \eqref{eq:ex-average}, yielding
\begin{eqnarray}
  \ex{\ave{\ex{\op{O}}}}
  =
  \trace{\ex{\op{O}}\,\absCC{0}{0} \otimes  \op{\rho}_{{\rm N},T}}
  =
  \traceN{{}_{\rm C}\langle 0 | \ex{\op{O}} \absNC{0}  \; \op{\rho}_{{\rm N},T}}
  \equiv
  \traceN{\exN{\op{O}}  \; \op{\rho}_{{\rm N},T}}
\label{eq:ave-next}\;,
\end{eqnarray}
after the trace $\traceN{\cdots}$ is performed over the normal-component degrees of freedom. Once we project the BEC part out of the remaining operator,
\begin{eqnarray}
  \exN{\op{O}} \equiv {}_{\rm C}\langle 0 | \ex{\op{O}} \absNC{0}
\label{eq:O-Normal}\;,
\end{eqnarray}
the resulting operator depends only on the normal-component degrees of freedom. Furthermore, we show in   \ref{app:T-properties} that the quantum dynamics of operators can be solved with Heisenberg equations of motion evaluated completely within the excitation picture, i.e.~
\begin{eqnarray}
  i \hbar \frac{\partial}{\partial t} \ex{\ave{\ex{\op{O}}}} 
  = \ex{\ave{\comm{\ex{\op{O}}}{\ex{\op{H}}}}}
\label{eq:HEM-ex}\;,
\end{eqnarray}
where $\ex{\op{H}} = \ex{\op{T}} \; \op{H} \; \ex{\op{T}}^\dagger$ stands for the Hamiltonian in the excitation picture; the explicit form of $\ex{\op{H}}$ is worked out in Sec.~\ref{sec:H-excitation}. 

Equations \eqref{eq:AB-product} and \eqref{eq:HEM-ex} guide us how a successful BEC analysis is performed in the excitation picture. The complication \eqref{eq:AB-product} means that it is not useful to transform the elementary boson operators $B_{\bf k}$ and $B^\dagger_{\bf k}$ to the excitation picture because the transformed $B_{{\bf k},{\rm ex}}$ and  $B^\dagger_{{\bf k},{\rm ex}}$ do not satisfy the bosonic commutation relations anymore, unlike for unitary transformations. This may seem a major setback, but the product-form transformation yield the correct $\ex{\left(\comm{B_{\bf k}}{B^\dagger_{{\bf k}'}}\right)} = \delta_{{\bf k},{\bf k}'}$ and the Heisenberg equation of motion has the usual form \eqref{eq:HEM-ex} under expectation value. 
Especially, one can apply a strategy where one only transforms the relevant operators $\op{O}$ and the Hamiltonian to the excitation picture. In practice, one starts with $\ex{\op{H}}$ and $\ex{\op{O}}$ and expresses them in terms of the usual boson operators. 

The resulting $\ex{\op{H}}$ and $\ex{\op{O}}$ become then some products of $B_{\bf k}$ and $B^\dagger_{\bf k}$, and $\ex{\ave{\comm{\ex{\op{O}}}{\ex{\op{H}}}}}$ can be efficiently be evaluated if commutators
\begin{eqnarray}
  \comm{B_{\bf k}}{\ex{\op{H}}} \equiv i \hbar \frac{\partial}{\partial t} B_{\bf k} \,,
  \qquad
  \equiv
  \comm{B^\dagger_{\bf k}}{\ex{\op{H}}} \equiv i \hbar \frac{\partial}{\partial t} B^\dagger_{\bf k} 
\label{eq:HEM-BB-comm}
\end{eqnarray}
are known. Strictly speaking, the identified differentials do not produce quantum kinetics of $B_{\bf k}$ and $B^\dagger_{\bf k}$ in the excitation picture, but they always produce their contribution when evaluated within the expectation value \eqref{eq:HEM-ex}. For example, the commutator of a product $\ex{\op{O}} = B_{\bf k} B_{{\bf k}'}$ yields $\comm{\ex{\op{O}}}{\ex{\op{H}}}=\comm{B_{\bf k}}{\ex{\op{H}}} B_{{\bf k}'} + B_{\bf k} \comm{B_{{\bf k}'}}{\ex{\op{H}}} \equiv \left[\ihddt B_{\bf k}\right] B_{{\bf k}'} + B_{\bf k} \left[\ihddt B_{{\bf k}'}\right]$. The same result is obtained by applying the product rule of differentiation $\ihddt B_{\bf k} B_{{\bf k}'}=\left[\ihddt B_{\bf k}\right] B_{{\bf k}'} + B_{\bf k} \left[\ihddt B_{{\bf k}'}\right]$. In other words, any $\comm{\ex{\op{O}}}{\ex{\op{H}}}$ follows by combining known commutators (dynamics) \eqref{eq:HEM-BB-comm} with product rule of differentiation, which makes identification \eqref{eq:HEM-BB-comm} extremely useful.

Provided with that the operator dynamics (commutator) \eqref{eq:HEM-BB-comm} is simple enough, the excitation picture yields major benefits from the point of view of the cluster expansion. We obviously can solve the quantum kinetics entirely in the excitation picture where the condensate remains as a vacuum  state for all times. Since a vacuum has no correlations, the excitation picture avoids the unnecessary tracking of the originally highly correlated BEC. Instead, the excitation picture exclusively follows how the normal-component excitations evolve around the BEC. Therefore, the excitation picture indeed converts the strongly interacting Bose gas into a cluster-expansion friendly format, as shown in Sec.~\ref{sec:Cluster-kinetics}. The cluster expansion can also be implemented directly to the original 
picture\cite{Kohler:2002,Kohler:2003} to access the dynamics of the lowest order clusters, but a general formulation requires further considerations to make cluster expansion efficient, as discussed in Sec.~\ref{sec:QS-basics}. Here, we attempt to develop a generic platform for all clusters.

\section{Basic quantum-statistical properties}
\label{sec:QS-basics}

The results in Sec.~\ref{sec:Excitation-pic} provide clear guidelines how to solve the quantum dynamics of interacting Bose gases. Since the quantum dynamics of {\it all} properties can be solved completely within the excitation picture,  according to Eq.~\eqref{eq:HEM-ex}, we first convert the relevant $\op{O}$ operators to the excitation picture. Once the explicit $\ex{\op{O}}$ form is known, we construct the corresponding $\exN{\op{O}}$ using Eq.~\eqref{eq:O-Normal} to determine whether the related property can exist in the interacting Bose gas. With these steps, we can classify which quantities are relevant for the BEC, even before any actual many-body computations are performed. 

For later identification, we categorize $\op{O}$ to be a {\it microcanonical operator} if it contains an equal number of creation and annihilation operators; the remaining operators are not microcanonical. In   \ref{app:O-elemntary}, we show  that {\it only the microcanonical operators produce a nonzero} $\exN{O}$ and $\ave{\op{O}}$ whenever the Bose gas has a fixed {\it total} particle number. Conversely, if $\op{O}$ is not a microconanical operator, the corresponding  $\ave{\op{O}}$ can exist only if the {\it total} particle number of the system is allowed to change. 

An operator $\left(\op{L}^\dagger\right)^J \, \left(\op{L}\right)^K \, \op{O}\left(J',K'\right)$, which contains $J'$ creation and $K'$ annihilation operators for the normal-component atoms, is microcanonical only if $J+J'$ is equal to $K+K'$; one can count the number of boson operators of $\op{L}$ and $\op{L}^\dagger$ with the help of identification \eqref{eq:L-operators-with-B}. In   \ref{app:O-elemntary}, 
we present the technical steps needed to produce a transformation
\begin{eqnarray}
  \ave{\left(\op{L}^\dagger\right)^J \, \left(\op{L}\right)^K \, \op{O}\left(J',K'\right)}
  \toX
  \avex{\op{O}\left(J',K'\right)}\,, \qquad
  J+J' = K+K'
\label{eq:org2exci}\;,
\end{eqnarray}
from the original to the excitation picture. Interestingly, $\op{O}\left(J',K'\right)$ is not changed and the transformed expectation value is not bound to be number conserving anymore. For example, the process related to $\ave{\op{L} \, \op{L} \, B^\dagger_{\bf k} B^\dagger_{-{\bf k}}}$ is microcanonical and yet its transformation $\avex{B^\dagger_{\bf k} B^\dagger_{-{\bf k}}}$ describes an amplitude of a process that creates two atoms into the normal component. This explicit example shows that the particle number is not conserved in the excitation picture because it focuses the investigation on the properties of the normal component alone. 

Also the Hartree-Fock Bogoliubov approximation\cite{Baranger:1961,Goodman:1974,Mang:1975,Zaremba:1999,Bender:2003, Milstein:2003, Wuster:2005} introduces expectation values of type $\ave{B^\dagger B^\dagger}$ as anomalous density, see discussion in Sec.~\ref{sec:Bogloliubov} for further details; this identification appears anomalous only because no change of picture is explicitly performed. When excitation picture is applied, ``anomalous'' quantities follow quite naturally because relation \eqref{eq:org2exci} connects, e.g., $\avex{B^\dagger_{\bf k} B^\dagger_{-{\bf k}}}$  with a number conserving transition amplitude $\ave{\op{L} \, \op{L} \, B^\dagger_{\bf k} B^\dagger_{-{\bf k}}}$ that is not anomalous as such. Therefore, the excitation picture gives $\avex{B^\dagger_{\bf k} B^\dagger_{-{\bf k}}}$ a natural interpretation as the elementary transition amplitude of the quantum depletion.

\begin{figure}[ht]
\includegraphics*[scale=0.7]{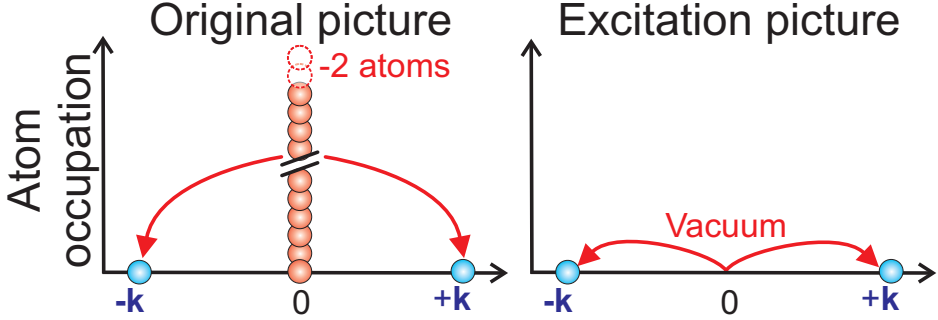}
\caption{(Color online) Quantum depletion in original vs.~excitation picture. 
Atomic occupation is schemetically represented as function of the atomic wave vector ${\bf k}$. Condensate atoms (red spheres) appear at zero ${\bf k}$ while normal-component atoms (blue filled circles) reside at ${\bf k}\neq 0$.  In the original picture (left), two BEC atoms scatter to normal-component $\pm{\bf k}$ atoms; the removed atoms are indicated as dashed circles. In the excitation picture (right), only the transtions appear (arrows). 
}\label{fig:EXCpicture}
\end{figure}

Figure \ref{fig:EXCpicture} illustrates the transitions related to $\ave{\op{L} \, \op{L} \, B^\dagger_{\bf k} B^\dagger_{-{\bf k}}}$  (left) and its excitation-picture equivalent $\avex{B^\dagger_{\bf k} B^\dagger_{-{\bf k}}}$ (right). The BEC is depicted as piled-up red spheres and the created normal-component atoms are symbolized by blue spheres. In the original picture, two BEC atoms are removed (dashed circles) to create two normal-component atoms (arrows), which represents the simplest process that initiates the quantum depletion. In the excitation picture, this process creates two atoms are out of the particle vacuum and therefore appears to be anomalous even though it is not in the original picture. This simple example illustrates nicely how the transformation to the excitation picture indeed focuses the investigation on the relevant excitation processes around the BEC.

\subsection{Excitation picture and BEC quantum statistics}
\label{sec:BECQS}

Based on the formulation of Sec.~\ref{sec:Excitation-pic}, the expectation-value representation\cite{Kira:2008} of BEC quantum statistics reduces to microcanonical combinations:
\begin{eqnarray}
   \ave{[B^\dagger_0]^J [B_0]^L}
   = 
   \delta_{J,L} \ave{[B^\dagger_0]^J [B_0]^J}
\label{eq:I-diagonal}\;.
\end{eqnarray}
These have a one-to-one connection to a density matrix
\begin{eqnarray}
   \op{\rho}_{\rm cond} = \sum_{n=0}^{\cal N}
   \absN{n}\, p_n \, \absNS{n}
\label{eq:rho--diagonal}
\end{eqnarray}
that is diagonal in the Fock-state representation due to the diagonality \eqref{eq:I-diagonal}; see also Sec.~\ref{app:Coherence} for further discussion. In this context, $p_n$ is positive definite and it describes the probability of finding exactly $n$ atoms in the BEC. Therefore, $p_n$ defines the {\it BEC statistics} in analogy to photon statistics\cite{Walls:2008} that determines the photon-occurrence probability for quantized light fields. 

Using the standard photon-statistics formulation\cite{Walls:2008,Kira:2008}, the normally-ordered expectation values uniquely define the BEC statistics via
\begin{eqnarray}
   p_n = {\textstyle \ave{:\frac{\left(\op{N}_0\right)^n}{n!}\, e^{-\op{N}_0} :}}
\label{eq:p_n-eval}\;,
\end{eqnarray}
where $: \cdots :$ enforces the normal order of the BEC operators and
\begin{eqnarray}
   \op{N}_0 \equiv B^\dagger_0 B_0
\label{eq:N0}
\end{eqnarray}
is the BEC number operator.  The normally-ordered BEC operators are connected with $\op{N}_0$ via
\begin{eqnarray}
   : \op{N}_0^J : \equiv  [B^\dagger_0]^J [B_0]^L =
  \op{N}_0 (\op{N}_0 -1) \cdots (\op{N}_0-J+1) \equiv 
  {\textstyle \frac{\op{N}_0!}{(\op{N}_0-J)!}}
\label{eq:Cond-op}
\end{eqnarray}
that follows after applying the bosonic commutation relations \eqref{eq:boson-comm} several times.

In   \ref{app:T-properties}, we show  how the expectation values \eqref{eq:I-diagonal} can be evaluated with the help of an exact substitution $\op{N}_0 \toX \opNC$ into the excitation picture. The resulting $\opNC$ operator is expressible entirely in terms of normal-component operators defined by Eq.~\eqref{eq:C-number-op}. More specifically, these $\op{N}_0$-based expectation values become
\begin{eqnarray}
   \ave{: \op{N}_0^J :} &\toX& \avex{ \frac{\opNC!}{(\opNC-J)!} } = \avex{\opNC (\opNC -1) \cdots (\opNC-J+1)}\,,
\nonumber\\
   \ave{\op{N}_0^J} &\toX& \avex{ \opNC^J} 
\label{eq:N_0-trafo}\;,
\end{eqnarray}
in the excitation picture. 

Most important, the BEC statistics \eqref{eq:p_n-eval} can be evaluated exactly from the normal-component properties when the system is microcanonical. In other words, BEC statistics is defined uniquely by the quantum statistics of the normal component after we have introduced the excitation picture.  As the major benefit, the quantum statistics of the normal component typically involves only low-rank clusters, which makes the cluster-expansion approach an attractive and efficient method for describing strongly interacting Bose gas. The simplest example involves a vanishing normal component because then each of the $\opNC$ operators can be replaced by ${\cal N}$. It is straightforward to see that this replacement in Eq.~\eqref{eq:N_0-trafo} reproduces the quantum-statistical results presented already in Sec.~\ref{sec:PLE_correlations}.

\subsection{Relevant doublets in the excitation picture}
\label{sec:Doublets}

To assess the principal influence of the quantum depletion on BEC statistics, we consider a homogeneous system where the normal component exhibits only clusters up to doublets. In this situation, expectation values are nonvanishing  only when the total momentum of creation and annihilation operators is equal\cite{Kira:2011}; Neither $\avex{B_{\bf k}}$ nor $\avex{B^\dagger_{\bf k}}$ can satisfy this condition for a normal component that necessarily has ${\bf k} \neq 0$. Therefore, all normal-component singlets must vanish; also the BEC has vanishing singlets $\ave{B_0}=\ave{B^\dagger_0}=0$, according to Eq.~\eqref{eq:I-diagonal}.

Based on the cluster-expansion representation \eqref{eq:chi_N}--\eqref{eq:DI-rep}, the doublets are defined by the difference of two-atom expectation values and its single-particle factorization. For example, $\avex{B^\dagger_{\bf k} B_{\bf k}}$ can be factored to a singlet product
$\avex{B^\dagger_{\bf k}} \avex{B_{\bf k}}$ according to the Wick's theorem\cite{Wick:1950}. For homogeneous excitations, doublets
\begin{eqnarray}
   f_{\bf k} 
   &\equiv& 
   \ave{B^\dagger_{\bf k} B_{\bf k}} - \avex{B^\dagger_{\bf k}} \avex{B_{\bf k}}
   =
   \avex{B^\dagger_{\bf k} B_{\bf k}} \,,
\nonumber\\
   s_{\bf k} &\equiv& \avex{B_{\bf k} B_{-{\bf k}}} - \avex{B_{\bf k}} \avex{B_{-{\bf k}}} = \avex{B_{\bf k} B_{-{\bf k}}} 
\label{eq:N-Doublets}\;,
\end{eqnarray}
are equal to the two-particle expectation values because singlets vanish as discussed above.\cite{Kira:2008} Naturally, also the complex-conjugated $s^\star_{\bf k} = \avex{B^\dagger_{-{\bf k}} B^\dagger_{\bf k} }$ may exist in homogeneous systems. Physically, $f_{\bf k}$ defines the occupation of normal-component component ${\bf k}$ while $s_{\bf k}$ is the transition amplitude identified in connection with Fig.~\ref{fig:EXCpicture}.

For later use, we define a total number operator for the normal-component atoms 
\begin{eqnarray}
 \opNN \equiv \sum_{\bf k}{}' B^\dagger_{\bf k} B_{\bf k} \toX \sum_{\bf k}{}' B^\dagger_{\bf k} B_{\bf k} 
\label{eq:N_N}\;.
\end{eqnarray}
This remains unchanged when transformed into the  excitation picture, based on properties \eqref{eq:app-B4H_ex}. For the microcanonical systems studied here, $\opNN$, $\opNC$, and the total atom number ${\cal N}$ are furthermore connected through
\begin{eqnarray}
  \opNN + \opNC = {\cal N} \quad \Leftrightarrow \quad \opNC = {\cal N} - \opNN
\label{eq:N_connect}\;,
\end{eqnarray}
based on the sum of Eqs.~\eqref{eq:C-number-op} and \eqref{eq:N_N}. The average number of BEC and normal-component atoms follows from
\begin{eqnarray}
   \NC &\equiv& \avex{\opNC} = {\cal N} - \NN
\label{eq:NCave}\,,
\\
   \NN &\equiv& \avex{\opNN}= 
   \sum_{\bf k}{}'f_{\bf k}
\label{eq:NNave}\;,
\end{eqnarray}
defined here via the excitation-picture expectation values, respectively.

The singlet-doublet (SD) clusters always correspond to a physical state\cite{Kira:2006b,Kira:2008} and they define a simple correlation generating function in terms of quadratic $\alpha$ contributions,
\begin{eqnarray}
  \xi_{\rm SD}(\left\{ \alpha \right\}) 
  \equiv 
 -\sum_{\bf k} f_{\bf k} |\alpha_{\bf k}|^2
    + 
  {\textstyle \frac{1}{2}}
   \sum_{\bf k} 
   \left(
      \alpha_{\bf k} \alpha_{-{\bf k}} s_{\bf k}
      +
      \alpha^\star_{\bf k} \alpha^\star_{-{\bf k}} s_{\bf k}
   \right)
\label{eq:xi_SD}\;,
\end{eqnarray}
when the singlets (S) vanish; see Ref.~\cite{Kira:2008} for an explicit derivation. This $\xi_{\rm SD}(\left\{ \alpha \right\})$ is expressed in the excitation picture and formally follows from Eqs.~\eqref{eq:chi_N} and \eqref{eq:xi_N} by setting $\alpha_0$ and $\alpha^\star_0$ to zero and taking the average in the Fock space of the excitation picture. Since we include only clusters up to the doublets, we have added the subscript ``SD'' to $\xi$. It is straightforward to show that Eq.~\eqref{eq:xi_SD} indeed produces the correct singlet--doublet factorization \eqref{eq:N-Doublets} when the reduction formula \eqref{eq:DI-rep} is applied.  The corresponding characteristic function is defined by
\begin{eqnarray}
  \chi_{\rm SD}(\left\{ \alpha \right\}) 
  &\equiv& 
  e^{\xi_{\rm SD}(\left\{ \alpha \right\})}
\label{eq:chi_SD}\;,
\end{eqnarray}
based on connection \eqref{eq:xi_N}, whenever the normal component contains clusters up to doublets \eqref{eq:xi_SD}. Strictly speaking, $e^{\xi_{\rm SD}(\left\{ \alpha \right\})}$ should be Taylor expanded up to the order ${\cal N}$ because all expectation values containing more than ${\cal N}$ boson annihilation operators should vanish. Since realistic atom trap experiments typically have a very large atom number, we use the full $e^{\xi_{\rm SD}(\left\{ \alpha \right\})}$ instead of the Taylor-expanded form.

\subsection{Shape of quantum depletion vs.~BEC quantum statistics}
\label{sec:shape}

Next, we will study how a normal component, whose quantum statistics is described by $\xi_{\rm SD}(\left\{ \alpha \right\})$, modifies the quantum statistics of the BEC. Since the transformation \eqref{eq:N_0-trafo} into the excitation picture allows us to express BEC properties exactly in terms of normal-component operators \eqref{eq:N_connect}, we also may use the normal-component $\chi_{\rm SD}(\left\{ \alpha \right\})$ to evaluate any property of the BEC as well.  Based on Eq.~\eqref{eq:I-diagonal}, we only need to consider those expectation values that have an equal amount of creation and annihilation operators. We adopt a strategy where we first identify the correlations in the original picture and then transform them with Eq.~\eqref{eq:N_0-trafo} into the excitation picture to evaluate them explicitly by using only the normal-component properties.
 
The BEC number $\ave{B^\dagger_0 B_0}$ is the lowest-order contribution we need to consider, followed by the four-atom expectation value $\ave{B^\dagger_0 B^\dagger_0 B_0 B_0}$. Cluster expansion essentially implements the Wick's theorem\cite{Wick:1950} and expresses any expectation value in terms of all possible factorizations into atom clusters. For example, $\ave{B^\dagger_0 B_0}$ follows from the sum of its singlet factorization $\ave{B^\dagger_0}\ave{B_0}$ and doublet correlation $\Delta\ave{B^\dagger_0 B_0}$. Since the singlets vanish, we find
\begin{eqnarray}
  \ave{B^\dagger_0 B_0} = \Delta \ave{B^\dagger_0 B_0} \toX \NC
\label{eq:NC-SD}\;,
\end{eqnarray}
where we have applied transformation \eqref{eq:N_0-trafo} as well as the BEC-number relation \eqref{eq:NCave}. 

To investigate nontrivial quantum-statistical aspects, we analyze next the four-atom expectation value $\ave{B^\dagger_0 B^\dagger_0 B_0 B_0}$ that contains the factorization into products of doublets $\Delta\ave{B^\dagger_0 B_0}\,\Delta\ave{B^\dagger_0 B_0}$. To count each factorization possibility only once, we formally label each boson operator by its position, associating $\ave{B^\dagger_0 B^\dagger_0 B_0 B_0}$ with $\ave{B^\dagger_1 B^\dagger_2 B_3 B_4}$ where indices 1, 2, 3, and 4 denote the position of each BEC operator. With this notation, the doublet factorization\cite{Kira:2011} into all possible $\Delta \ave{B^\dagger_j B_k}$ pairs yields
\begin{eqnarray}
  \ave{B^\dagger_1 B^\dagger_2 B_3 B_4} 
  = 
  \Delta \ave{B^\dagger_1 B_4} \Delta \ave{B^\dagger_2 B_3}
  + 
  \Delta \ave{B^\dagger_1 B_3} \Delta \ave{B^\dagger_2 B_4} + \Delta\ave{B^\dagger_1 B^\dagger_2 B_3 B_4} 
\label{eq:NC-4-SD}\;.
\end{eqnarray}
In case singlets,  $\Delta \ave{B^\dagger_j B^\dagger_k}$,  $\Delta\ave{B_j B_k}$, or triplets exist, one also needs to include the corresponding factorizations; they do not exist for the microcanonical system studied here. In general, the explicit evaluation of more complicated cluster-based factorizations can be realized most efficiently by expressing characteristic functions in terms of the correlation-generating function and by following a derivation similar to that performed in Sec.~\ref{sec:PLE_correlations}. 

By setting all indices of factorization \eqref{eq:NC-4-SD} identical, we can identify the four-atom BEC correlation to be
\begin{eqnarray}
  \Delta\ave{B^\dagger_0 B^\dagger_0 B_0 B_0} 
  &=&
  \ave{B^\dagger_0 B^\dagger_0 B_0 B_0} 
  - 
  2\Delta \ave{B^\dagger_0 B_0} \Delta \ave{B^\dagger_0 B_0} 
\nonumber\\
  &\toX&
  \avex{\opNC\,(\opNC-1)} 
  - 
  2\NC^2
  =
  \avex{\opNC^2}
  -\NC
  - 
  2\NC^2
\label{eq:NC-quad-corr1}\;,
\end{eqnarray}
where we have transferred the expressions into the excitation picture with relations \eqref{eq:N_0-trafo} and \eqref{eq:NC-SD}. In the last step, we have used the property $\avex{\opNC\,(\opNC-1)} = \avex{\opNC^2-\opNC}$ and the identification \eqref{eq:NCave}. By combining the first and the last line of Eq.~\eqref{eq:NC-quad-corr1}, we may also express the four-atom expectation value,
\begin{eqnarray}
  \ave{B^\dagger_0 B^\dagger_0 B_0 B_0} 
  \toX
  \avex{\opNC\,(\opNC-1)} = \avex{\opNC^2} - \NC
\label{eq:NC-quad-exp1}\;,
\end{eqnarray}
in the excitation picture.

To determine either $\Delta\ave{B^\dagger_0 B^\dagger_0 B_0 B_0}$ or $\ave{B^\dagger_0 B^\dagger_0 B_0 B_0}$, we obviously need to evaluate $\avex{\opNC^2}$ explicitly. We start by inserting connection \eqref{eq:N_connect} into it, yielding
\begin{eqnarray}
  \avex{\opNC^2} 
  =
  \avex{({\cal N}-\opNN)^2}
  =
  \avex{{\cal N}^2-2\opNN\,{\cal N} + \opNN^2}
  =
  {\cal N}^2-2\NN\,{\cal N} + \avex{\opNN^2}
\label{eq:NC-opNC2}\;,
\end{eqnarray}
when we apply identification \eqref{eq:NNave} to be able to use $\chi_{\rm SD}$ and $\xi_{\rm SD}$ later on. We then substitute definition \eqref{eq:N_N} into the remaining expectation value $\avex{\opNN^2}$ term, producing
\begin{eqnarray}
 \avex{\opNN^2} 
 =
 \sum_{{\bf k},{\bf k}'}{}' 
 \avex{B^\dagger_{\bf k} B_{\bf k} B^\dagger_{{\bf k}'} B_{{\bf k}'}}
 =
 \sum_{{\bf k},{\bf k}'}{}' 
 \avex{B^\dagger_{\bf k} B^\dagger_{{\bf k}'} B_{{\bf k}'} B_{\bf k} }
 +
 \sum_{{\bf k}}{}'
 \avex{B^\dagger_{\bf k} B_{\bf k}}
\label{eq:aveN_N2a}\;,
\end{eqnarray}
after having normally ordered the operators. The second contribution produces the average number of normal-component atoms while the first contribution can be computed from the characteristic function \eqref{eq:chi_SD} by applying the reduction formula \eqref{eq:I-rep}. We then find
\begin{eqnarray}
  \avex{\opNN^2} 
  &=&
  \sum_{{\bf k},{\bf k}'}{}' 
  \partial_{{\bf k}} \partial_{{\bf k}'}
  \partial^\star_{{\bf k}'}
  \partial^\star_{{\bf k}}  
  \left.
  \chi_{\rm SD}(\left\{ \alpha \right\}) \right|_{\{ \alpha =0 \}}
  + \NN
  =
  \sum_{{\bf k},{\bf k}'}{}'
  \left(
    f_{\bf k} \,f_{{\bf k}'}
    +
    \delta_{{\bf k}',{\bf k}} \,f_{\bf k}^2
    +
    \delta_{{\bf k}',-{\bf k}} \,|s_{\bf k}|^2
  \right)
  +
  \NN
\nonumber\\
  &=&
  \NN^2
  +
  \sum_{{\bf k}}{}'
  \left(
    f_{\bf k}^2
    +
    |s_{\bf k}|^2
  \right)
  +
  \NN
\label{eq:aveN_N2b}\;,
\end{eqnarray}
as a result of straightforward differentiation. 

In general, $\sum_{{\bf k}}{}'\left(f_{\bf k}^2+|s_{\bf k}|^2\right)$ scales like $\NN$ and the proportionality is determined by the exact shape of the $(f_{\bf k},\,s_{\bf k})$ excitation. We show in Sec.~\ref{sec:Bogloliubov} that $f_{\bf k}^2+|s_{\bf k}|^2$ produces $f_{\bf k}$ to leading order for low levels of quantum depletion, which reduces the sum approximatively to $\NN$. Nevertheless, when quantum depletion becomes strong enough, $f_{\bf k}^2+|s_{\bf k}|^2$ deviates from  $f_{\bf k}$  due to the excitation-specific shape of the quantum depletion. To quantify the shape of quantum depletion with a single number, we identify a shape correction
\begin{eqnarray}
  c_{\rm shape} 
  \equiv 
  \frac{1}{2}
   + \frac{1}{2\NN}
    \sum_{\bf k}{}'\left(f_{\bf k}^2+|s_{\bf k}|^2 \right)
\label{eq:c_shape}\;.
\end{eqnarray}
For $c_{\rm shape}=1$, $f_{\bf k}^2+|s_{\bf k}|^2$ behaves like $f_{\bf k}$ under a sum. Its value is computed to be $c_{\rm shape}\approx1.1781$ for the Bogoliubov excitations studied in more detail in Sec.~\ref{sec:Bogloliubov}. The actual $c_{\rm shape}$ depends sensitively on the quantum-depletion details such that one can characterize how strongly the quantum depletion deviates from Bogoliubov excitations by monitoring $c_{\rm shape}$, as shown in Ref.~\cite{Kira:HBE}.

By substituting result \eqref{eq:c_shape} into Eqs.~\eqref{eq:NC-opNC2} and \eqref{eq:aveN_N2b}, we find
\begin{eqnarray}
  \avex{\opNN^2} = \NN^2 + 2 \,c_{\rm shape}\NN\,,
  \qquad
  \avex{\opNC^2} = \NC^2 + 2 \,c_{\rm shape}\NN
\label{eq:aveNNC}\;.
\end{eqnarray}
If $c_{\rm shape}$ is exactly one, $\avex{\opNN^2}$ reduces to the well-known form for a single-mode thermal state\cite{Book:2011}, i.e.~$\avex{\opNN^2} = \NN^2 + 2 \,\NN$. For $c_{\rm shape}>1$, the added fluctuations are larger than for an ideal thermal state. At the same time, the normal component adds the contribution $2 \,c_{\rm shape}\NN$ also to the BEC $\avex{\opNC^2}$ such that the atom--atom correlation \eqref{eq:NC-quad-corr1} becomes
\begin{eqnarray}
  \Delta\ave{B^\dagger_0 B^\dagger_0 B_0 B_0} 
  =
  - 
  \NC^2
  -
  \NC
  +
  2 \,c_{\rm shape}\NN  
\label{eq:NC-quad-x}\;.
\end{eqnarray}
Compared with the normal-component-free relation \eqref{eq:dI_Fock}, the total atom number is now replaced by the BEC number, i.e.~$-{\cal N} \left( {\cal N} +1\right) \rightarrow -\NC \left(\NC -1\right)$. The quantum depletion then opposes the pure BEC part via the $2 \,c_{\rm shape}\NN$ contribution. Therefore, the normal component adds additional fluctuations that depend on the shape of the quantum depletion, i.e.~$c_{\rm shape}$. In Sec.~\ref{sec:complementaryQS}, we will study how this shape can be characterized in correlation measurements.

To fully resolve the connection of quantum depletion and quantum fluctuations, it is often useful to examine a {\it set} of complementary quantities that depend on the atom--atom correlations. In analogy to photon counting, the results of BEC-atom counting are characterized by the $J$-th order moments of the counts,
\begin{eqnarray}
  \left[ n^J \right] \equiv \sum_{n=0}^\infty n\, p_n
\label{eq:ac-moments}
\end{eqnarray}
that follow directly from the density matrix \eqref{eq:rho--diagonal}. These averages are denoted within brackets $\left[ \cdots \right]$ to distinguish them from the usual expectation values $\ave{\cdots}$. With the help of definition \eqref{eq:p_n-eval}, we find that the first- and second-order moments of atom counts produce
\begin{eqnarray}
  \left[ n \right] = \ave{\op{N}_0}\,, \qquad 
  \left[ n^2 \right] = \ave{\op{N}_0^2}
\label{eq:ac-moments12a}\;.
\end{eqnarray}
The transformation \eqref{eq:N_0-trafo} converts these into the excitation picture
\begin{eqnarray}
  \left[ n \right] = \avex{\opNC} = \NC\,,
  \qquad 
  \left[ n^2 \right] = \avex{\opNC^2} = \NC^2+ 2\,c_{\rm shape} \NN
\label{eq:ac-momentFIN}\;,
\end{eqnarray}
after having combined the results \eqref{eq:NC-opNC2} and \eqref{eq:aveNNC}. The fluctuations of the BEC-number counts $\Delta \NC$ around the $\NC$ average then become
\begin{eqnarray}
  \Delta \NC^2 \equiv \left[ n^2 \right] - \left[ n \right]^2 = 2\,c_{\rm shape} \NN
\label{eq:ac-FIN}\;,
\end{eqnarray}
when the normal component is a singlet--doublet state. If the BEC counts were Poisson distributed, the number fluctuations would be $\Delta N_{C,{\rm Poiss}}^2 \equiv \NC$; analogously, a perfect laser has a photon statistics that is Poisson distributed\cite{Walls:2008,Book:2011}. Interestingly, the fluctuations of BEC number scale with the normal-component atom number, not with BEC-atom number. This implies that {\it quantum depletion generally produces a non-Poissonian BEC}, also meaning that the atomic BECs are only partial analogous to lasers\cite{Bloch:1999,Hagley:1999,Ketterle:2002} in the quantum-statistical sense.

In quantum optics, the second-order coherence\cite{Glauber:1963,Walls:2008} is characterized by
\begin{eqnarray}
  g^{(2)} \equiv \frac{\ave{B^\dagger_0 B^\dagger_0 B_0 B_0}}{\ave{B^\dagger_0 B_0}\ave{B^\dagger_0 B_0}}
\label{eq:g(2)-def}
\end{eqnarray}
that is proportional to the conditional probability of detecting another atom when one atom is already present. With the help of results \eqref{eq:NC-quad-exp1}--\eqref{eq:NC-opNC2} and \eqref{eq:aveNNC}, we find
\begin{eqnarray}
  g^{(2)} = 1 + \frac{2\,c_{\rm shape} \NN-\NC}{\NC^2}
\label{eq:g(2)-FIN}\;.
\end{eqnarray}
In quantum optics, $g^{(2)}$ is typically measured with a coincidence measurement using the so-called 
Hanbury Brown--Twiss setup.\cite{Hanbury:1956} For Poissonian fields, $g^{(2)}$ is equal to unity. In case $g^{(2)}$ is below one, the detection of another atom is lower than for Poissonian fields such that the boson field shows antibunching in the detection events. Atom bunching is indicated by $g^{(2)}$ greater than one. For photons, realizing perfect antibunching is a central research topic in the important effort to construct stable single-photon sources\cite{Michler:2000,Yuan:2002,Santori:2002,Peyronel:2012} for quantum-information processing\cite{Nielsen:2010}. 

\subsection{Complementary characterization of BEC's quantum statistics}
\label{sec:complementaryQS}

\begin{figure}[t]
\includegraphics*[scale=0.38]{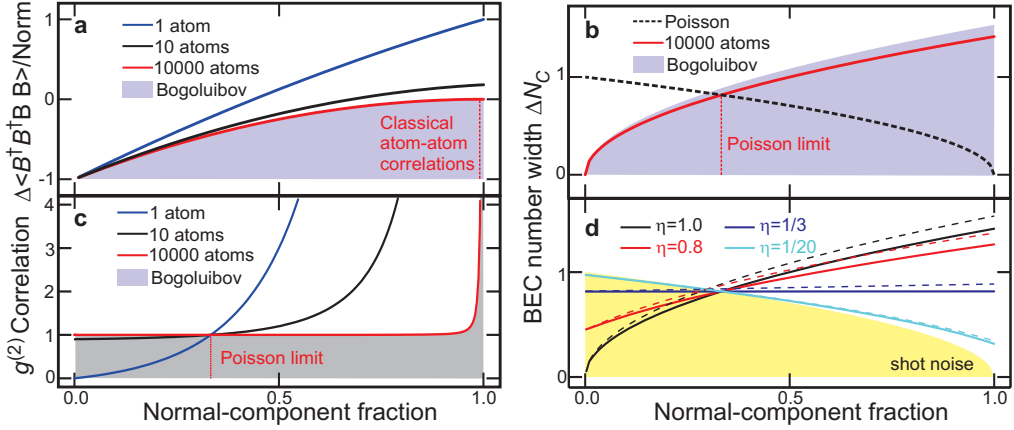}
\caption{(Color online) Connection of quantum statistics and shape of quantum depletion. {\bf a} Normalized atom--atom correlation $\Delta\ave{B^\dagger_0 B^\dagger_0 B_0 B_0}$ [with ${\rm Norm}={\cal N} \left( {\cal N} +1\right)$] is plotted as function of normal-component fraction $\nuN$. Computations with ${\cal N}=1$ (blue line), ${\cal N}=10$ (black line), and  ${\cal N}=10^4$ (red line) atoms contain no shape effects ($c_{\rm shape}=1$); shaded area shows atom--atom correlations with shape effects ($c_{\rm shape}=1.1781$) and  ${\cal N}=10^4$. The solid line denotes the shot-noise limit (Poisson limit). {\bf b} Condensate's number fluctuations for ${\cal N}=10^4$ atoms, with (shaded area) and without (red line) shape effects. {\bf c} Second-order coherence of BEC is presented with the same line styles as in frame {\bf a}. The vertical dashed lines identify the special points vanishing fluctuations for ${\cal N}=10^4$  and $c_{\rm shape}=1$. 
{\bf d} Effect of quantum efficiency on $\Delta \NC$ for ${\cal N}=10^4$ atoms. The quantum efficiency is $\eta=1$ (black), $\eta=0.8$ (red), $\eta=0.33$ (dark blue), and $\eta=0.05$ (light blue). Solid lines are computed with $c_{\rm shape}=1$ while dashed lines use $c_{\rm shape}=1.1781$ corresponding to the Bogoliubov excitations. The shaded area shows the shot-noise limit  \eqref{eq:Poisson}.
}\label{fig:CondensateQS}
\end{figure}

It is insightful to study BEC correlations via three complementary four-atom correlations: its atom--atom correlations \eqref{eq:NC-quad-x}, atom-number fluctuations \eqref{eq:ac-FIN}, and second-order coherence \eqref{eq:g(2)-FIN}. Figure \ref{fig:CondensateQS}{\bf a} shows the normalized  $\frac{\Delta\ave{B^\dagger_0 B^\dagger_0 B_0 B_0} }{{\cal N} \left( {\cal N} +1\right)}$ as function of a normal-component fraction 
\begin{eqnarray}
  \nuN \equiv \frac{\NN}{\NC}
\label{eq:Normal-fraction}\;,
\end{eqnarray}
for ${\cal N}=1$ (blue line), ${\cal N}=10$ (black line), and  ${\cal N}=10^4$ (red line) atoms by neglecting the shape contributions, i.e.~$c_{\rm shape}=1$. The shaded area shows how the shape of the quantum depletion effects the atom--atom correlations when the system contains Bogoliubov excitations ($c_{\rm shape}=1.1781$) and ${\cal N}=10^4$ atoms. As discussed in connection with Eq.~\eqref{eq:chi_SD}, the  analysis used is strictly speaking valid only for ${\cal N} \gg 1$ (actually ${\cal N} \ge 2$ for the four-atom correlations). Nevertheless, we also use the ${\cal N}=1$ limit of the correlation expression \eqref{eq:NC-quad-x} in order to illustrate the functional limit of the  correlations studied.

As a general tendency, the atom--atom correlation dips to its maximal negative value for a vanishing normal component $\nuN =0$; its normalized value is exactly -1. We also observe that the atom correlation increases monotonically as function of $\nuN$. Based on Eq.~\eqref{eq:NC-quad-x}, the extremal values $\Delta\ave{B^\dagger_0 B^\dagger_0 B_0 B_0}$ are $-{\cal N} \left( {\cal N} +1\right)$ at $\nuN=0$ and $+2\,c_{\rm shape} \NN$ at $\nuN=1$. As expected, the magnitude of atom--atom correlations for low $\nuN$ is significantly larger than it is for the case with a dominant normal component ($\nuN \rightarrow 1$), whenever the atom number is substantial. The normalized $\Delta\ave{B^\dagger_0 B^\dagger_0 B_0 B_0}$ approaches  $-(1-\nuN)^2$ for elevated atom numbers, which is demonstrated by the similarity of ${\cal N}=10$ and ${\cal N}=10^4$ results. Especially, the shape of quantum depletion seems to be indistinguishable because ($c_{\rm shape}=1$, red line) and ($c_{\rm shape}=1.1781$, shaded area) 
appear 
to be 
identical for the large-${\cal N}$ cases analyzed here. However, we show below that the shape of quantum depletion produces a detectable difference when the BEC quantum fluctuations are studied via complementary correlations.

We  may also conclude that the normal-component contributions eventually reverse the sign of atom--atom correlations once $\nuN$ becomes large enough. The root of Eq.~\eqref{eq:NC-quad-x} yields a zero crossing of  $\Delta\ave{B^\dagger_0 B^\dagger_0 B_0 B_0}$ at
\begin{eqnarray}
  \nuN^{\rm zero} = 
  {\textstyle
    1 + \frac{1+2\,c_{\rm shape}}{2\,{\cal N}}
    - 
    \sqrt{\frac{2 \,c_{\rm shape}}{{\cal N}}+\left( \frac{1+2\, c_{\rm shape}}{2\, {\cal N}} \right)^2}
  }
\label{eq:NC-zeroX}\;.
\end{eqnarray}
More specifically, we find $\nuN^{\rm zero} = 0.4384$ for  ${\cal N}=1$,  $\nuN^{\rm zero} = 0.6783$ for ${\cal N}=10$, and $\nuN^{\rm zero} = 0.9860$ without ($c_{\rm shape}=1$, vertical red-dashed line) or at $\nuN^{\rm zero} = 0.9848$ ($c_{\rm shape}=1.1781$, not shown) with the Bogoliubov excitations for ${\cal N}=10^4$. In other words, the shape of quantum depletion slightly modifies the zero-crossing value for the case studied here, such that it does modify BEC statistics even though it is not apparent in Fig.~\ref{fig:CondensateQS}{\bf a}. As a general tendency, the zero-crossing \eqref{eq:NC-zeroX} approaches
\begin{eqnarray}
  \nuN^{\rm zero} \rightarrow
  {\textstyle
    1 
    - 
    \sqrt{\frac{2 \,c_{\rm shape}}{{\cal N}}}
  }\,,
  \qquad
  {\cal N} \gg 1
\label{eq:NC-zeroXlim}\;,
\end{eqnarray}
for a large enough atom number. This verifies that the zero-crossing always depends on the shape of the quantum depletion. However, the sign reversal of atom correlations becomes less dramatic for larger than for lower atom numbers, which often makes characterizing $\nuN^{\rm zero}$ rather insensitive to the specific details of the quantum depletion. Even though $\Delta\ave{B^\dagger_0 B^\dagger_0 B_0 B_0}$ vanishes at $\nuN^{\rm zero}$, this does not yet mean that the BEC fluctuations trivially vanish, because $\Delta\ave{B^\dagger_0 B^\dagger_0 B_0 B_0}$ describes just one aspect of quantum statistics, as we show below.

To analyze BEC statistics through a complementary correlation, Fig.~\ref{fig:CondensateQS}{\bf b} presents the normalized width $\Delta \bar{N}_{\rm C} = \frac{\Delta \NC}{\sqrt{\cal N}}$ of BEC-number counts, computed from Eq.~\eqref{eq:ac-FIN} as function of $\nuN$. We have assumed here that the system has ${\cal N}=10^4$ atoms. The results with (shaded area) and without (red line) shape effects are compared with the normalized Poisson distributed fluctuations  
\begin{eqnarray}
  \Delta \bar{N}_{C,{\rm Poiss}} \equiv \sqrt{\frac{\NC}{\cal N}}
\label{eq:Poisson}\;,
\end{eqnarray}
defining also the shot-noise limit\cite{Walls:2008}, i.e.~border to classical behavior, plotted as dashed line. We observe that the functional dependence of $\Delta \bar{N}_{\rm C}$ is far from $\Delta \bar{N}_{C,{\rm Poiss}}$ because $\Delta \bar{N}_{\rm C}$ and $\Delta \bar{N}_{C,{\rm Poiss}}$ exhibit opposite monotonic dependencies. We also may conclude that the shape of quantum depletion, assumed here to follow from the Bogoliubov excitation ($c_{\rm shape}=1.1781$, shaded area), produces a visible change in BEC number fluctuations compared to the case without them ($c_{\rm shape}=1$, red line). 
Therefore, a BEC-atom counting measurement characterizes the details of quantum depletion much more directly than a $\Delta \ave{B^\dagger_0 B^\dagger_0 B_0 B_0}$ measurement, although both $\Delta\ave{B^\dagger_0 B^\dagger_0 B_0 B_0}$ and $\Delta \NC$ describe correlations up to the four-atom level. 

Atom-number fluctuation \eqref{eq:ac-FIN} approaches the border to classical behavior, i.e.~$\Delta \NC = \sqrt{\NC}$, when the normal-component fraction becomes
\begin{eqnarray}
  \nuN^{\rm class} = \frac{1}{1+2 \,c_{\rm shape} }
\label{eq:NC-class}\;.
\end{eqnarray}
This fraction is realized roughly at $\nuN = \frac{1}{3}$ that is generally different from the limit of vanishing four-atom correlations \eqref{eq:NC-zeroXlim}. The actual $\nuN^{\rm class}$ are marked in Fig.~\ref{fig:CondensateQS}{\bf} as red-dashed vertical line for $c_{\rm shape}=1$. This position clearly differs from the classical limit of $\Delta\ave{B^\dagger_0 B^\dagger_0 B_0 B_0}$ correlation, shown also as red-dashed line in Figs.~\ref{fig:CondensateQS}{\bf a}. In other words, $\Delta \NC$ deviates from the classical-border limit $\sqrt{\NC}$ even when $\Delta \ave{B^\dagger_0 B^\dagger_0 B_0 B_0}$ vanishes. Therefore, $\Delta \ave{B^\dagger_0 B^\dagger_0 B_0 B_0}$ and $\Delta \NC$ cannot simultaneously approach the classical border whenever the atom number is substantial. This shows that a {\it single correlation qualifier cannot conclusively determine whether or not the BEC correlations become simple}. 

To extend our complementary study, Fig.~\ref{fig:CondensateQS}{\bf c} shows $g^{(2)}$ as function $\nuN$ for ${\cal N}=1$ (blue line), ${\cal N}=10$ (black line), and  ${\cal N}=10^4$ without (red line, $c_{\rm shape}=1$) and with (shaded area, $c_{\rm shape}=1.1781$) shape effects. Only ${\cal N}=1$ atom case yields strong antibunching for low $\nuN$, i.e.~a large BEC fraction. For single-atom ``BEC', antibunching has  a trivial explanation because single atom systems cannot produce detection of two atoms, which forces $g^{(2)}$ to vanish for all $\nuN$. Vanishing $g^{(2)}$ for all $\nuN$ follows if we properly Taylor expand $\chi_{\rm SD}(\left\{ \alpha \right\})$, as discussed subsequent to Eq.~\eqref{eq:chi_SD}. We do not perform that here because realistic BECs have ${\cal N} \gg 1$ and the ${\cal N} \rightarrow 1$ used provides the mathematical limit of large-${\cal N}$ $g^{(2)}$ formula.

The second-order coherence still shows residual antibunching even for a large ${\cal N}$, but it comes infinitesimal due to the $-\frac{1}{\cal N}$ proportionality in Eq.~\eqref{eq:g(2)-FIN}. More specifically, large ${\cal N}$ and small $\nuN$ produce a $g^{(2)}$ that approaches one from below, implying essentially perfect second-order coherence as long as the BEC has a macroscopically large atom occupation. In other words, the antibunching level is so low that it usually is below the experimental sensitivity. However, an elevated $\nuN$ eventually produces a diverging $g^{(2)}$, i.e.~BEC exhibits massive bunching of BEC atoms when the BEC fraction approaches zero. The transition to large bunching emerges at larger $\nuN$ when ${\cal N}$ is increased. It is also clearly interesting to determine when the residual antibunching 
turns into bunching at the vertical-dashed line, establishing the classical boundary $g^{(2)}=1$. Based on Eq.~\eqref{eq:g(2)-FIN}, we find exactly the same classical limit \eqref{eq:NC-class} as from the atom-number fluctuations. Consequently, $g^{(2)}$ and $\Delta \NC$ identify the same classical border.

However, $g^{(2)}$ detects deviations from classical aspects much less sensitively than $\Delta \NC$ does, analyzed in Figs.~\ref{fig:CondensateQS}{\bf c} and \ref{fig:CondensateQS}{\bf b}, respectively. For example, $g^{(2)}$ stays close to one over an extended $\nuN$ range for all cases shown, exceeding ${\cal N}=10$. Also the shape of the quantum depletion seems not to have a noticeable effect on $g^{(2)}$ because $c_{\rm shape}=1$ (red solid line) appears to be indistinguishable from $c_{\rm shape}=1.1781$ (shaded area); the shape effects as well as a strong $\nuN$ dependence are clearly resolved by the $\Delta \NC$. Therefore, measurement of BEC's number fluctuations resolves the effect of quantum depletion on a BEC most sensitively.

\subsection{Quantum efficiency in the BEC characterization}
\label{sec:eta-effect}

An actual measurement cannot be ideally precise, which inevitably distorts the BEC characterization. For example, an individual atom may be detected with probability $\eta$ that is smaller than one. In other words, a realistic measurement may count $\eta \NC$ atoms when in reality the system contains $\NC$ BEC atoms; the quantity $\eta$ is commonly referred to as the quantum efficiency. To estimate $\eta$ effect on BEC characterization, we follow the standard photon-counting formulation with $\eta <1$. According to Ref.~\cite{Walls:2008}, $\eta$-deteriorated measurement detects 
\begin{eqnarray}
   p_n(\eta) =  {\textstyle \ave{:\frac{\left(\eta\,\op{N}_0\right)^n}{n!}\, e^{-\eta\,\op{N}_0} :} }
\label{eq:p_n-eta}\;,
\end{eqnarray}
instead of the ideal relation \eqref{eq:p_n-eval}. This generates $\eta$-deteriorated moments of BEC counts via 
$\left[n^J\right]_\eta \equiv \sum_{n=0}^\infty n^J\, p_n(\eta)$.

The resulting $\eta$-deteriorated BEC number and number fluctuations follow straightforwardly from Eq.~\eqref{eq:p_n-eta}, yielding
\begin{eqnarray}
   \left[ n \right]_\eta = \eta\, \NC\,,
   \qquad
   \left[\Delta \NC^2\right]_\eta = (1-\eta)\, \eta\, \NC + 2 c_{\rm shape} \, \eta^2 \NN
\label{eq:eta_NC}\;.
\end{eqnarray}
The same detection scheme measures $\eta {\cal N}$ as the total number of atoms. As a result, an $\eta$-deteriorated measurement detects 
\begin{eqnarray}
   \left.\Delta \NC\right|_\eta 
   = 
   \sqrt{(1-\eta)\frac{\NC}{\cal N}+  2 c_{\rm shape} \, \eta \, \frac{\NN}{\cal N}}
\label{eq:eta_dNC}\;,
\end{eqnarray}
as the normalized BEC number fluctuations.

Figure \ref{fig:CondensateQS}{\bf d} compares $\left.\Delta \NC\right|_\eta$ for an ideal $\eta=1$ (black),  $\eta=0.8$ (red), $\eta=0.33$ (dark blue), and $\eta=0.05$ (light blue) as function of the normal-state fraction $\nuN$; the system has ${\cal N} = 10^4$ atoms and the solid (dashed) lines are computed with (without) the Bogoliubov excitations and the shaded area corresponds to the shot-noise limit \eqref{eq:Poisson}. We observe that 
 the BEC statistics tends to approach the shot-noise limit for a decreasing $\eta$. The direction of monotonic normal-fraction dependence is changed at $\eta=\frac{1}{3}$; at this value BEC fluctuations are essentially independent of $\nuN$. As another $\eta$ property, the effects of quantum-depletion shape becomes weaker as $\eta$ is reduced. The shape effects remain clearly visible for $\eta=\frac{1}{3}$ while one can hardly distinguish $c_{\rm shape}=1$ (solid line) from $c_{\rm shape}=1.1781$ at $\eta=\frac{1}{20}$.

\section{System Hamiltonian in the excitation picture}
\label{sec:H-excitation}

We may now return to determining the system Hamiltonian in the excitation picture, based on the transformation properties discussed in Sec.~\ref{sec:EP-relations}. All operators that appear in the system Hamiltonian \eqref{eq:H-in-B} are microcanonical and their transformation into the excitation picture is derived in   \ref{app:T-properties}. The resulting explicit transformations, given by Eq.~\eqref{eq:app-B4H_ex}, yield a excitation-picture Hamiltonian:
\begin{align}
  \op{H}_{\rm ex} 
   &= 
    \sum_{\bf k}{}' E_{\bf k} B^\dagger_{\bf k} B_{\bf k} 
    + 
    \frac{V_0}{2} 
    \left( 
      \opNC\,\left(\opNC - 1 \right)
      + 2 \opNC \sum_{\bf k}{}' B^\dagger_{\bf k} B_{\bf k} 
      + \sum_{{\bf k},{\bf k}'}{}' B^\dagger_{\bf k} B^\dagger_{{\bf k}'} B_{{\bf k}'} \, B_{\bf k}
    \right)
\nonumber\\
  &+
  \sum_{\bf k}{}' V_{\bf k} \opNC  B^\dagger_{\bf k} B_{\bf k} 
  +\frac{1}{2}\sum_{\bf k}{}' V_{\bf k}  
  \left[ 
      B_{\bf k} \, B_{-{\bf k}} \, \sqrt{(\opNC+1)(\opNC+2)}
    +
    \sqrt{(\opNC+1)(\opNC+2)} \, B^\dagger_{-{\bf k}} B^\dagger_{\bf k}  
  \right]
\nonumber\\
  &+
  \sum_{{\bf k},{\bf k}'}{}' V_{\bf k} 
  \left[ 
     B^\dagger_{{\bf k} + {\bf k}'} B_{{\bf k}'}\,  B_{\bf k} \,\sqrt{\opNC}
    +
    \sqrt{\opNC}\, B^\dagger_{\bf k} \, B^\dagger_{{\bf k}'}  B_{{\bf k} + {\bf k}'} 
  \right]
  +
  \frac{1}{2} \sum_{{\bf q}} \sum_{{\bf k}\neq {\bf k}'}{}' \, V_{{\bf k}-{\bf k}'} 
  B^\dagger_{{\bf k}} B^\dagger_{{\bf q}-{\bf k}} B_{{\bf q}-{\bf k}'}\,  B_{{\bf k}'}
\label{eq:H-in-ex-start}\;,\quad
\end{align}
which is exact while it is not anymore number conserving because the excitation picture focuses the analysis on the physics of the normal-component atoms alone. The $B_{\bf k}$ and $B^\dagger_{\bf k}$ operators are still the usual bosonic operators which also determine the condensate-number operator $\opNC$ through connection
\begin{eqnarray}
  \opNC \equiv {\cal N} - \sum_{\bf k}{}' B^\dagger_{\bf k} B_{\bf k}
\label{eq:opNC}\;,
\end{eqnarray}
based on identifications \eqref{eq:N_N}--\eqref{eq:N_connect}.

Alternatively, one can replace $\opNC$ by${\cal N} - \opNN$ based on connection \eqref{eq:N_N}. We can use this exact substitution to simplify the second contribution of Eq.~\eqref{eq:H-in-ex-start} to
\begin{eqnarray}
  \op{H}^{\rm 2nd}_{\rm ex} 
  &\equiv& 
    \frac{V_0}{2} 
    \left( 
      \opNC\,\left(\opNC - 1 \right)
      + 2 \opNC \opNN
      + \sum_{{\bf k},{\bf k}'}{}' B^\dagger_{{\bf k}'} B_{{\bf k}'} \, B^\dagger_{\bf k}  B_{\bf k}
      - \sum_{\bf k}{}' B^\dagger_{\bf k} \, B_{\bf k}
    \right)
\nonumber\\
  &=&
      \frac{V_0}{2} 
    \left( 
      \opNC\,\left(\opNC - 1 \right)
      + \opNC \opNN + \opNN \opNC
      + \opNN \opNN
      - \opNN
    \right)
\nonumber\\
  &=&
      \frac{V_0}{2} 
    \left( \opNC + \opNN \right)
      \left(\opNC + \opNN - 1 \right) 
      = \frac{V_0}{2}
    {\cal N}
    \left( 
      {\cal N} - 1 \right)
\label{eq:H-2nd-ex}\,
\end{eqnarray}
which is obtained after we reorganize the terms by commuting, before we use identification \eqref{eq:N_N}. 
This describes a constant energy shift produced by interactions among ${\cal N}$ interacting bosons. As we apply relation \eqref{eq:sqrt-term} and insert identification \eqref{eq:H-2nd-ex} into system Hamiltonian \eqref{eq:H-in-ex-start}, we obtain
\begin{align}
  \op{H}_{\rm ex} 
  &= 
  \frac{V_0}{2}
  {\cal N} \left( {\cal N} - 1 \right)
  +
  \sum_{\bf k}{}' \left( E_{\bf k} + V_{\bf k} \opNC\, \right) \, B^\dagger_{\bf k} B_{\bf k} 
  +\frac{1}{2}\sum_{\bf k}{}' V_{\bf k}  
  \left[ 
      \sqrt{(\opNC+1)(\opNC+2)} \, B^\dagger_{-{\bf k}} B^\dagger_{\bf k} +{\rm h.c.} 
  \right]
\nonumber\\
  &
  +\sum_{{\bf k},{\bf k}'}{}' V_{\bf k} 
  \left[ 
     B^\dagger_{{\bf k} + {\bf k}'} B_{{\bf k}'}\,  B_{\bf k} \,\sqrt{\opNC}
    +
    \sqrt{\opNC}\, B^\dagger_{\bf k} \, B^\dagger_{{\bf k}'}  B_{{\bf k} + {\bf k}'} 
  \right]
  +
  \frac{1}{2} \sum_{{\bf q}} \sum_{{\bf k}\neq {\bf k}'}{}' \, V_{{\bf k}-{\bf k}'} 
  B^\dagger_{{\bf k}} B^\dagger_{{\bf q}-{\bf k}} B_{{\bf q}-{\bf k}'}\,  B_{{\bf k}'}
\label{eq:H-in-ex}\;,
\end{align}
after we also have organized the $B^\dagger_{\bf k} B_{\bf k}$ contributions together.

The excitation-picture Hamiltonian \eqref{eq:H-in-ex} has many recognizable connections with alternative formulations. For example, the number-conserving approach\cite{Girardeau:1956,Gardiner:1997,Girardeau:1998,Castin:1998,Morgan:2003,Gardiner:2007,Gardiner:2012,Billam:2012,Mora:2003},
combined with a unitary transformation, yields the standard Bogoliubov Hamiltonian that is nearly the same as the first line of Eq.~\eqref{eq:H-in-ex}; the last contribution is simply replaced by $\sqrt{(\opNC+1)(\opNC+2)} \, B^\dagger_{-{\bf k}} B^\dagger_{\bf k} \, \op{L}^2 + {\rm h.c.}$, as shown by Eq.~(14) in Ref.~\cite{Girardeau:1998} where $B_{\bf k}$ is denoted by $\hat{a}_{\bf k}$ and $\op{L}$ by $\hat{\beta}_0$. The $B^\dagger_{-{\bf k}} B^\dagger_{\bf k} \, \op{L}^2$ operator clearly corresponds to the number-conserving transition exciting two atoms from the BEC to the normal component, as presented in the left part of Fig.~\ref{fig:EXCpicture}. 
At the same time, $\ex{\op{H}}$ contains the corresponding excitation-picture transition $B^\dagger_{-{\bf k}} B^\dagger_{\bf k}$ which does not conserve the particle number, as illustrated in the right part of Fig.~\ref{fig:EXCpicture}. In other words, the condensate-lowering (raising) operators $\op{L}$ ($\op{L}^\dagger$) are missing from the excitation-picture form because the state is transformed with the nonunitary transformation \eqref{eq:rho-T}--\eqref{eq:rho-T-and-O}. Nevertheless, the physical atom number is still fully conserved when the excitation picture results are transformed back to the original picture using relation \eqref{eq:rho-T}, which connects the excitation-picture analysis directly to the standard number-conserving approaches.

The Beliaev approach\cite{Leggett:2001,Griffin:2012,Wright:2012} introduces coherence by substituting $B_0 \rightarrow \beta$ and $B^\dagger_0 \rightarrow \beta^\star$ with a complex number, as discussed in Sec.~\ref{app:Coherence}. According to e.g.~in Ref.~\cite{Yang:2011}, this substitution to the original Hamiltonian \eqref{eq:H-in-B} 
produces the first line of Eq.~\eqref{eq:H-in-ex} where also $\opNC$ is replaced by a number, which connects the Beliaev approach with the excitation picture.
Both the Beliaev approach and the excitation picture yield a Hamiltonian that does not conserve the atom number anymore.
However, the excitation picture and the Beliaev approach violate the number conservation for very different reasons --- introduction of coherence vs.~nonunitary transformation, $\ex{T}$, respectively. The number conservation is irreversibly lost in the Belieav approach because one cannot retrieve the actual quantum statistics of the BEC from the complex-valued amplitude. In contrast to this, the excitation picture can uniquely be transformed back to the original picture where the atom number is fully conserved. In this sense, the excitation picture still provides a fully number conserving approach for all atoms.

The connectivity of the excitation picture to both number-conserving and Beliaev approaches may reconcile some fundamental differences of these approaches debated e.g.~in Refs.~\cite{Griffin:2012,Wright:2012}. We do not pursue this line of investigations further here, but concentrate on the cluster-expansion aspects of the excitation picture. Clearly, the excitation picture is different from many traditional approaches because it describes strongly interacting Bose gas from an alternative point of view where only some aspects agree with the traditional results, as shown above. Most importantly, BEC properties become trivial in the excitation picture because the BEC remains a vacuum state at all times, which is a major benefit for any cluster-expansion based approach. The actual BEC properties must then be analyzed through properties of $\opNC$, as is done in Sec~\ref{sec:QS-basics}. Additionally, full many-body study of strongly interacting Bose gas should keep the new contributions in the 
second line of \eqref{eq:H-in-ex}, often dropped in Bogoliubov-type approaches, because one needs to account for the many-body effects among the normal component atoms created by the quantum depletion.

\subsection{Condensate occupation and fluctuations}
\label{sec:Coccupation}

From the cluster-expansion point of view, Hamiltonian \eqref{eq:H-in-ex} has one more problematic issue; the square-root terms formally lead to contributions where $B^\dagger_{\bf k} B_{\bf k}$ appear to all orders, which formally produces a direct coupling of a single-particle boson operators to all particle orders. However, this problematic issue can be completely removed by inspecting how $\opNC$ must behave in a strongly interacting Bose gas. In other words, we show that the square-root terms can be linearized with respect to $\opNC$. In general, these square-root terms define how strongly the BEC changes the normal component and we will analyze how the linearization alters the coupling.

As a starting point, $\NC \equiv \avex{\opNC}$ defines the average macroscopic occupation of the BEC, as discussed in Sec.~\ref{sec:QS-basics}. We may then compare this with a zero-momentum occupation obtained as an extrapolation from the normal-component
$f_0 \equiv \lim_{{\bf k} \rightarrow 0} f_{\bf k}$.
Since $f_0$ does not correspond to a macroscopic occupation, it cannot scale with the quantization volume ${\cal L}^3$. Nevertheless, we are studying dense enough Bose gas where we can expect $f_0$ to be appreciable, i.e.~$f_0 \gg 1$ when the quantum depletion is strong. The actual occupation of the zero-momentum state can be defined via
\begin{eqnarray}
  \NC \equiv f_0 + N_{\rm add}
\label{eq:N_add}
\end{eqnarray}
because the presence of the BEC adds a macroscopic occupation $N_{\rm add}$ to the zero-momentum state. Since the BEC results from a macroscopic occupation, $N_{\rm add}$ must scale with ${\cal L}^3$ such that also the $\NC$ dominantly scales with the volume. Clearly $N_{\rm add}$ must be positive while $f_0$ remains large, which makes $\NC \gg 1$ even when the BEC vanishes (implying $N_{\rm add}=0$) as long as we study a dense enough Bose gas.
In other words, $\NC$ must remain large (although not macroscopic) even when the BEC is annihilated by the quantum depletion since $\NC$ still contains the $f_0$ part of the uncondensed Bose gas.
This seemingly innocent conclusion, allows us to linearize \eqref{eq:H-in-ex} without a real loss of generality in the many-body analysis of the strongly interacting Bose gas. 

As the very nature of any atom BEC, $\NC$ is not only macroscopically large but the BEC number also does not fluctuate much on the scale of the total atom number ${\cal N}$, as shown in Fig.~\ref{fig:CondensateQS}{\bf b}. Therefore, it is meaningful to identify a number fluctuation operator
\begin{eqnarray}
  \delta \opNC \equiv \opNC - \NC
\label{eq:deltaNC}\;,
\end{eqnarray}
for the BEC.
As shown in Secs.~\ref{sec:shape}--\ref{sec:complementaryQS},
the quantum depletion creates BEC atom-number fluctuations, and the mean fluctuations of the BEC number $\Delta\NC^2 = \avex{\delta \opNC^2}$ defines their overall magnitude. The actual value of $\Delta\NC$ is given by Eq.~\eqref{eq:ac-FIN}, assuming that the normal component is a singlet--doublet state. In general, the BEC fluctuations can become large only if also the normal component has large fluctuations, due to the overall number conservation during the quantum depletion. The number fluctuations of individual normal-component atoms can become large when they follow a thermal state because it maximizes the entropy for a fixed average boson number\cite{Walls:2008,Book:2011}. Since the singlet--doublet form of the normal component includes the possibility to form a thermal state\cite{Kira:2008} within the normal component, the singlet-doublet form indeed describes strong number fluctuations for individual normal-component atoms. Therefore, already the singlet--doublet analysis gives a good 
estimate how large $\Delta\NC$ can become due to quantum depletion as function of the BEC fraction $F_{\rm BEC} \equiv \frac{\NC}{\cal N}$.

To determine an explicit estimate, we determine the relative BEC fluctuations
\begin{eqnarray}
  \frac{\Delta \NC}{\NC} = \sqrt{\frac{2 \, c_{\rm shape}}{\cal N}}\,\sqrt{\frac{1-F_{\rm BEC}}{F_{\rm BEC}^2}}
\label{eq:RELdeltaNC}\;,
\end{eqnarray}
which is obtained directly from Eq.~\eqref{eq:ac-FIN} by expressing  the normal-component atom number as $\NN = (1 - F_{\rm BEC}) \, {\cal N}$. We see now that the relative BEC fluctuations scale with $\frac{1}{\sqrt{\cal N}}$, which tends to make them small even when the normal-component fluctuations are large. To get a reasonable estimate for the $\Delta \NC$, we assume that the system has ${\cal N}=10^4$ atoms, $c_{\rm shape}=1.1781$ (as in Fig.~\ref{fig:CondensateQS}{\bf b}), and that the quantum depletion is very strong, leaving only $\NC=200$ atoms to the BEC, i.e.~$F_{\rm BEC}=0.02$. With these inputs, Eq.~\eqref{eq:RELdeltaNC} produces $\frac{\Delta \NC}{\NC} = 0.76$. In other words, even when the quantum depletion is very strong, $\Delta\NC$ remains smaller than the limit of thermal fluctuations, i.e.~$\Delta \NC = 2 \NC$. The same calculation produces only $\frac{\Delta \NC}{\NC} = 0.069$ for $F_{\rm BEC}=0.2$ which also implies a significant quantum depletion. 

Next, we study how the square-root expressions of $\ex{\op{H}}$ can be accurately linearized for a broad range of conditions, covering BEC fluctuations $-0.76 \NC \le \delta \op{\NC} \le +0.76 \NC$ (or $-0.069 \NC \le \delta \op{\NC} \le +0.069 \NC$) estimated above as a reasonable range for the strongly interacting Bose gas; in this inequality (as well as in the following discussion), $\delta \op{\NC}$ should be perceived as the number it generates when it acts upon the relevant many-body state. The $\sqrt{(\opNC+1)(\opNC+2)}$ contribution of Hamiltonian \eqref{eq:H-in-ex} reduces to $Y_C \equiv \sqrt{(\NC+1)(\NC+2)}$ for  vanishing number fluctuations. A Taylor expansion around $Y_C$ yields
\begin{eqnarray}
  \sqrt{(\opNC+1)(\opNC+2)} 
  &=& 
  Y_C 
  +
  {\textstyle 
    \sqrt{1+\frac{1}{4 Y_C^2}}
  } \, \delta \opNC
  + 
  {\cal O}
  \left( 
    {\textstyle \left[ \frac{\delta \opNC}{Y_C} \right]^2}
  \right) 
\nonumber\\
  &=&
  {\textstyle 
    \sqrt{1+\frac{1}{4 Y_C^2}}
  } \, 
  \left( \opNC + {\textstyle \frac{3}{2}} \right)
  -
  {\textstyle \frac{1}{4 Y_C}}
  +
  {\cal O}
  \left( 
    {\textstyle \left[ \frac{\delta \opNC}{Y_C} \right]^2}
  \right) 
\label{eq:sqrt-term}\;,
\end{eqnarray}
after applying definition \eqref{eq:deltaNC} and reorganizing the terms. Just like  $\delta \op{\NC}$, also $\opNC$ can be perceived as number it generates as it acts upon a state. We introduce $x \equiv \frac{\opNC}{\NC}$ that remains within interval 
\begin{eqnarray}
  \begin{array}{ll}
    0.24 \le x \le 1.76\,, & {\rm for}\;F_{\rm BEC} = 0.02
    \\
    0.931 \le x \le 1.069\,, & {\rm for}\;F_{\rm BEC} = 0.2
  \end{array}
\label{eq:x-limits}\;,
\end{eqnarray}
based on discussion following Eq.~\eqref{eq:RELdeltaNC}. Since $F_{\rm BEC} = 0.02$ corresponds to virtually a collapsed BEC while $F_{\rm BEC} = 0.2$ implies strongly reduced BEC, $x$ ranges \eqref{eq:x-limits} exemplify well the extreme limits of the quantum-depletion.

\begin{figure}[t]
\includegraphics*[scale=0.55,angle=0]{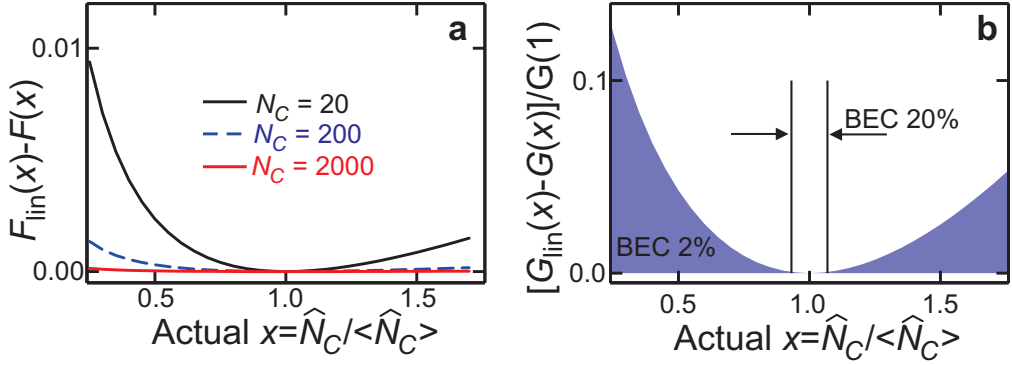}
\caption{(Color online) Effect of linerizing $\opNC$ under a square root. {\bf a} The accuracy of linearized $\sqrt{(\opNC+1)(\opNC+2)}$ is presented through a difference of $F_{\rm lin}(x)$ and $F(x)$ for $\NC=20$ (black line), $\NC=200$ (dashed line), and $\NC=20$ (red line); $x$ is the scaled BEC number.
{\bf b} The accuracy of linearized $\sqrt{\opNC}$ is analyzed via the relative deviation (shaded area) of $G_{\rm lin}(x)$ and $G(x)$; this deviation does not depend on $\NC$. In both frames, the $x$ range is given by $F_{\rm BEC}=20\%$ in Eq.~\eqref{eq:x-limits}; the vertical lines indicate the $F_{\rm BEC}=2\%$ range.
}\label{sqrtNCs}
\end{figure}

To check how well the linear part of Eq.~\eqref{eq:sqrt-term} describes the full expression within reasonable ranges \eqref{eq:x-limits}, we define the original function and its linearized version
\begin{eqnarray}
  F(x) \equiv \sqrt{(\NC\,x+1)(\NC\,x+2)}\,,
  \qquad
  F_{\rm lin}(x) \equiv 
    {\textstyle 
    \sqrt{1+\frac{1}{4 Y_C^2}}
  } \, 
  \left( \NC\,x + {\textstyle \frac{3}{2}} \right)
  -
  {\textstyle \frac{1}{4 Y_C}}
\label{eq:sqrt-term2}\;,
\end{eqnarray}
respectively, based on expansion~\eqref{eq:sqrt-term} and identification $\opNC \equiv \NC\,x$. Figure \ref{sqrtNCs}{\bf a} shows the difference $F_{\rm lin}(x)-F(x)$ as function of $x$ within the relevant fluctuation range \eqref{eq:x-limits}, when the system has $\NC=20$ (dashed line), $\NC=200$ (black line), and $\NC=2000$ (red line) atoms within the BEC. We observe that $F_{\rm lin}(x)$ remains always above the actual $F(x)$ (this property is valid all $x\ge 0$), and the accuracy is better than 1\% even for the extreme range \eqref{eq:x-limits} with $F_{\rm BEC} = 0.02$. This accuracy improves by orders of magnitude as as $\NC$ grows from 20 to 2000. Consequently, the linear contribution of Eq.~\eqref{eq:sqrt-term} provides always a very accurate upper limit
\begin{eqnarray}
  \sqrt{(\opNC+1)(\opNC+2)} 
  \le
  {\textstyle 
    \sqrt{1+\frac{1}{4 Y_C^2}}
  } \, 
  \left( \opNC + {\textstyle \frac{3}{2}} \right)
  -
  {\textstyle \frac{1}{4 Y_C}}
\label{eq:sqrt-termFIN}\;,
\end{eqnarray}
for the BEC effects. The inequality should be understood to involve the operator-related number that results when the operator acts upon a many-body state. 

Since the overestimate remains extremely small, we may replace $\sqrt{(\opNC+1)(\opNC+2)}$ by its linearized form \eqref{eq:sqrt-term} because {\it the effect of the BEC on the many-body dynamics is only infinitesimally overestimated by the linearization}. In general, we are studying dense Bose gas with $\NC \gg 1$ and $Y_C \gg 1$ even when the macroscopic BEC occupation vanishes, which makes both $\frac{1}{4 Y_C^2}$ and $\frac{1}{4 Y_C}$ too negligible to have any practical relevance for the interacting Bose gas. Therefore, we can drop these contributions and use a simple linearization
\begin{eqnarray}
  \sqrt{(\opNC+1)(\opNC+2)} 
  \rightarrow
  \opNC = \NC + \delta \opNC
\label{eq:sqrt-termLIN}\;,
\end{eqnarray}
which provides an accurate description of BEC effects on the strongly interacting Bose gas we are studying here.  

We perform a similar analysis for the other square-root contribution within the Hamiltonian \eqref{eq:H-in-ex}. A straightforward Taylor expansion around $\opNC$ around $\NC$ produces 
\begin{eqnarray}
  \sqrt{\opNC}
  &=& 
  \sqrt{\NC} + 
  {\textstyle \frac{1}{2} \frac{\delta\opNC}{\sqrt{NC}} 
  + 
  \sqrt{\NC} \,  {\cal O}
  \left( 
    \left[\frac{\delta \opNC}{\NC}\right]^2
  \right) 
  }
=
  {\textstyle   
  \frac{\NC+\opNC}{2\sqrt{\NC}} 
  + 
  \sqrt{\NC} \,  {\cal O}
  \left( 
    \left[\frac{\delta \opNC}{\NC}\right]^2
  \right) 
  }
\label{eq:sqrtB1}\;,
\end{eqnarray}
after the terms have been reorganized. An alternative derivation of linearization \eqref{eq:sqrtB1} is presented in   \ref{app:Commutators}. To check the validity range of the linearization, we use $\opNC = \NC\,x$ and identify the original and linearized square-root expression,
\begin{eqnarray}
  G(x) \equiv \sqrt{\NC\,x} = \sqrt{x}\,\sqrt{\NC}\,,
  \qquad
  G_{\rm lin}(x) \equiv 
    {\textstyle   
  \frac{1+x}{2}
  }\sqrt{\NC}
\label{eq:sqrt-termB2}\;,
\end{eqnarray}
respectively, based on expansion \eqref{eq:sqrtB1}. Figure \ref{sqrtNCs}{\bf b} shows a normalized difference $(G_{\rm lin}(x)-G(x))/G(1)$ as function of $x$ for range \eqref{eq:x-limits} with $F_{\rm BEC}=0.02$; the limits of $F_{\rm BEC}=0.2$ are indicated by the vertical lines. This normalization produces the same curve for all $\NC$ values and  always {\it overestimates} the actual square-root expression. Even with the almost completely annihilated BEC ($F_{\rm BEC}=0.02$), the linear approximation yields maximally a 12\% overestimate. For $F_{\rm BEC}=0.2$, the overestimate is below $6.2\times10^{-4}$, which is extremely small. In other words, the linearization \eqref{eq:sqrtB1} is virtually exact even for cases with a very small BEC fraction. Therefore, also $\sqrt{\opNC}$ can be accurately linearized within the Hamiltonian \eqref{eq:H-in-ex} for quantum-depletion studies.

Most important, this linearization produces an upper limit for the BEC effects on the interacting Bose gas because the difference is always positive for all $x \ge 0$; in other words, replacing $\sqrt{\opNC}$ by its linearized form,
\begin{eqnarray}
  \sqrt{\opNC}
  \rightarrow
  \sqrt{\NC} + 
  {\textstyle \frac{1}{2} \frac{\delta\opNC}{\sqrt{\NC}} 
  }
\label{eq:sqrtBFIN}\;,
\end{eqnarray}
at most overestimates the strength of the BEC effects caused by the $\sqrt{\opNC}$ contribution to the $\ex{\op{H}}$. Only the limit of vanishing BEC ($F_{\rm BEC} \rightarrow 0$) may produce an appreciable overestimate for the BEC effects, while any appreciable $F_{\rm BEC}$ yields a virtually exact linearization. At the same time, the $\sqrt{\NC}$ part of Eq.~\eqref{eq:sqrtBFIN} correctly yields vanishing BEC effects when the BEC ceases to exist. It may seem that the $\frac{\delta\opNC}{\sqrt{\NC}}$ part overestimates the BEC effects by diverging at the $\NC \rightarrow 0$ limit. However, we will explain in Sec.~\ref{sec:occuANDfluctH} why even this part yields vanishing BEC effects to the many-body dynamics at the limit $\NC \rightarrow 0$. Therefore, linearization \eqref{eq:sqrtBFIN} provides an accurate description of the strongly interacting Bose gas even at very large quantum-depletion levels. 

To improve linearization \eqref{eq:sqrtBFIN}, one can either include the full square-root expressions or quadratic corrections to them rather straightforwardly in order to extend the validity range. However, the linear approximation covers a very broad range of conditions relevant for the quantum-depletion studies of the strongly interacting Bose gas. In other words, the linearization of the square-root terms does not yield a perturbative description of the many-body effects in the traditional sense because it rather provides an accurate overestimate of the role of BEC. A traditional perturbation theory relies on including effects in terms of powers of the interaction-matrix element $V_{\bf k}$. We apply the linearization because it also simplifies the cluster-expansion analysis, as explained in Sec.~\ref{sec:Cluster-kinetics}, and then apply cluster expansion to include $V_{\bf k}$ effects nonperturbatively; cf.~Ref.~\cite{Book:2011} for a textbook discussion why the cluster-expansion approach is 
fundamentally a systematic nonperturbative 
approach.

\subsection{Condensate occupation and fluctuations in the system Hamiltonian}
\label{sec:occuANDfluctH}

The linearizations \eqref{eq:sqrt-termLIN} and \eqref{eq:sqrtBFIN} clarify the role of BEC in the excitation-picture system Hamiltonian \eqref{eq:H-in-ex} and provide an accurate nonperturbative description of the strongly interacting Bose gas, as shown in Sec.~\ref{sec:Coccupation}. As we apply them to Eq.~\eqref{eq:H-in-ex} and \eqref{eq:H-2nd-ex}, we find a straightforward separation 
\begin{eqnarray}
  \op{H}_{\rm ex} \equiv \op{H}^{\rm occ.}_{\rm ex} + \delta\op{H}_{\rm ex}
\label{eq:H-occ-delta}
\end{eqnarray}
that contains the BEC occupation part
\begin{align}
  \op{H}^{\rm occ.}_{\rm ex} 
  &= 
    \frac{V_0}{2} 
    {\cal N} \left( {\cal N} -1 \right)
    +
    \sum_{\bf k}{}' E^C_{\bf k} B^\dagger_{\bf k} B_{\bf k} 
    + 
    \sum_{\bf k}{}'  \frac{\NC \, V_{\bf k}}{2}  
  \left[ 
      B_{\bf k} \, B_{-{\bf k}} 
    +
      B^\dagger_{-{\bf k}} B^\dagger_{\bf k}  
  \right]
\nonumber\\
  &
  +\sum_{{\bf k},{\bf k}'}{}' \sqrt{\NC}\, V_{\bf k} 
  \left[ 
     B^\dagger_{{\bf k} + {\bf k}'} B_{{\bf k}'}\,  B_{\bf k} 
    +
     B^\dagger_{\bf k} \, B^\dagger_{{\bf k}'}  B_{{\bf k} + {\bf k}'} 
  \right]
  +
  \sum_{{\bf q}} \sum_{{\bf k}\neq {\bf k}'}{}' \, \frac{V_{{\bf k}-{\bf k}'}}{2} 
  B^\dagger_{{\bf k}} B^\dagger_{{\bf q}-{\bf k}} B_{{\bf q}-{\bf k}'}\,  B_{{\bf k}'}
\label{eq:H-in-ex-occup}\;.
\end{align}
describing how pairwise interactions convert the BEC $\NC$ to normal-component atoms. To simplify the notation, we have identified
\begin{eqnarray}
  E^C_{\bf k} \equiv E_{\bf k} + N_C \, V_{\bf k}
\label{eq:E-renormalized}\;,
\end{eqnarray}
as the BEC-renormalized kinetic energy. Since the BEC number has fluctuations, they also induce small contribution to  $\op{H}_{\rm ex}$ described by
\begin{eqnarray}
  \delta\op{H}_{\rm ex} 
  &=& 
  \sum_{\bf k}{}' V_{\bf k} \delta\opNC  B^\dagger_{\bf k} B_{\bf k} 
  +\frac{1}{2}\sum_{\bf k}{}' V_{\bf k}  
  \left( 
      B_{\bf k} \, B_{-{\bf k}} \, \delta\opNC
    +
    \delta\opNC \, B^\dagger_{-{\bf k}} B^\dagger_{\bf k}  
  \right)
\nonumber\\
  &&
  +\sum_{{\bf k},{\bf k}'}{}' V_{\bf k}  
  \left(
     B^\dagger_{{\bf k} + {\bf k}'} B_{{\bf k}'}\,  B_{\bf k} \, \frac{\delta\opNC}{\sqrt{\NC}}
     +
     \frac{\delta\opNC}{\sqrt{\NC}}\, 
    B^\dagger_{\bf k} \, B^\dagger_{{\bf k}'}  B_{{\bf k} + {\bf k}'} 
  \right) 
\label{eq:H-in-ex-fluct}\;.
\end{eqnarray}
In general, $\op{H}^{\rm occ.}_{\rm ex}$ describes the dominant part of the many-body interactions because $\delta\opNC$ scales with $\frac{1}{\sqrt{\cal N}}$, according to Eq.~\eqref{eq:RELdeltaNC}.

\begin{figure}[t]
\includegraphics*[scale=0.44]{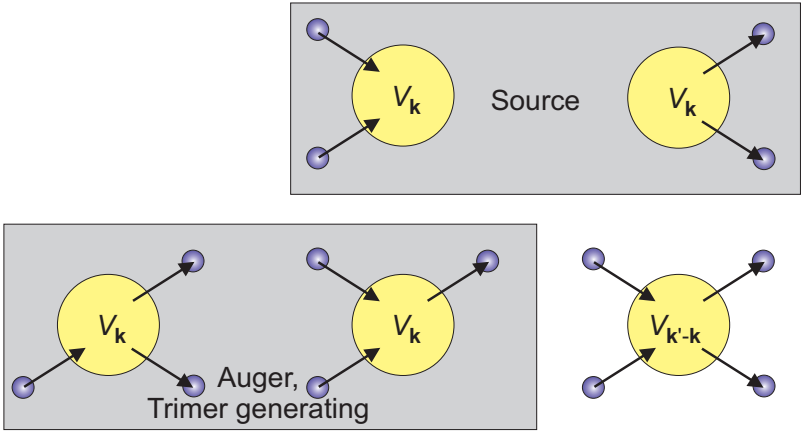}
\caption{(Color online) Diagrammatic representation of atom--atom interactions in the excitation picture. The diagrammatic rules are the same as in Fig.~\ref{Vdiagams}.}\label{VexDiagams}
\end{figure}

The dominant $\op{H}^{\rm occ.}_{\rm ex}$ part of  Hamiltonian \eqref{eq:H-occ-delta} is shown diagrammatically in Fig.~\ref{VexDiagams} with the same rules as in Fig.~\ref{Vdiagams}. We see that it contains only five diagrams out of the nine  the original $\op{H}$ has, shown in Fig.~\ref{Vdiagams}. In particular, the four first diagrams of $\op{H}$ (Fig.~\ref{Vdiagams}) become an energy renormalization \eqref{eq:H-2nd-ex} in the excitation picture (Fig.~\ref{VexDiagams}). The $\NC$-dependent contributions within $\op{H}^{\rm occ.}_{\rm ex}$ either create or annihilate two normal-component atoms because the number of in- and out-going arrows is not the same. In the original picture, they describe processes where BEC atoms are either converted to or created from the  normal-component atoms. As the major advantage of the excitation picture, these contributions identify the elementary process of quantum depletion as excitation and annihilation of normal component as $B^\dagger_{\bf k} B^\dagger_{-{\bf k}}$ and 
$B_{\bf k} B_{-{\bf k}}$, respectively. Physically, these terms inject only two-atom clusters into the system, as illustrated in Fig.~\ref{fig:EXCpicture}, which acts as a source to the quantum depletion. 
Once the normal-component atoms are created, in (out) scattering of an atom can create (annihilate) two atoms, as indicated by diagrams with three arrows. These represent Auger-type processes, also observed in semiconductors\cite{Klimov:2000,Hader:2005,Robel:2009} when either photon or phonon absorption/emission triggers transitions among three carrier states. In the strongly interacting Bose gas, the BEC takes the role of the photons/phonons. I will show in Ref.~\cite{Kira:HBE} that such Auger processes also contribute to the formation of Efimov trimers.\cite{Stecher:2009,Castin:2010} The remaining contribution with four arrows describes pairwise atom--atom interactions among the normal-component atoms. They can, e.g., bind two atoms to a molecular dimer state, as shown in Ref.~\cite{Kira:HBE}; in semiconductors, analogous processes are responsible for the formation of excitons\cite{Damen:1990,Kira:2001,Kira:2006}, i.e.~Coulomb bound electron--hole pairs.

Hamiltonian $\delta\op{H}_{\rm ex}$ describes the effect of BEC-number fluctuations on these processes as well as on the energy renormalization (not shown in Fig.~\ref{VexDiagams}). To isolate the orders of fluctuations from $\delta\op{H}_{\rm ex}$, we introduce three fluctuation operators
\begin{align}
  &\delta\left[B^\dagger_{\bf k} B_{\bf k}\right]
  \equiv
  B^\dagger_{\bf k} B_{\bf k}-f_{\bf k}\,,
\qquad
  &\delta\left[B_{\bf k} B_{-{\bf k}} \right]
  \equiv
  B_{\bf k} B_{-{\bf k}}- s_{\bf k}\,,
\nonumber\\
  &
  \delta\left[B^\dagger_{{\bf k}+{\bf k}'} B_{\bf k} B_{{\bf k}'}\right]
  \equiv
  B^\dagger_{{\bf k}+{\bf k}'} B_{\bf k} B_{{\bf k}'}-
  T_{{\bf k},{\bf k}'}\,,
  \qquad 
  &T_{{\bf k},{\bf k}'} \equiv \avex{B^\dagger_{{\bf k}+{\bf k}'} B_{\bf k} B_{{\bf k}'}}
\label{eq:fsT-fluct}\,,
\end{align}
where  $T_{{\bf k},{\bf k}'}$ describes the transition amplitude among three different atoms, hence it is a three-particle quantity. As we insert definitions \eqref{eq:fsT-fluct} into Eq.~\eqref{eq:H-in-ex-fluct}, we find
\begin{eqnarray}
  \delta\op{H}_{\rm ex} 
  &=& 
  \sum_{\bf k}{}' V_{\bf k} 
  \left(f_{\bf k}+{\rm Re} \left[s_{\bf k} +\frac{1}{\sqrt{\NC}} \sum_{{\bf k}'}{}' T_{{\bf k},{\bf k}'} \right] \right) \delta \opNC
\nonumber\\
  &&+
  \delta\opNC  \sum_{\bf k}{}' V_{\bf k}\, \delta\left[B^\dagger_{\bf k} B_{\bf k} \right]
  +\sum_{\bf k}{}'  \frac{V_{\bf k}}{2} \, \delta\left[B_{\bf k} \, B_{-{\bf k}} \right]\, 
  \delta\opNC
  +
  \delta\opNC 
  \sum_{\bf k}{}' \frac{V_{\bf k}}{2}  
  \, \delta\left[B^\dagger_{-{\bf k}} B^\dagger_{\bf k} \right] 
\nonumber\\
  &&
  +\sum_{{\bf k},{\bf k}'}{}' \frac{V_{\bf k}}{\sqrt{\NC}} 
     \delta\left[B^\dagger_{{\bf k} + {\bf k}'} B_{{\bf k}'}\,  B_{\bf k} \right]\, \delta\opNC
  +
  \delta\opNC\,
  \sum_{{\bf k},{\bf k}'}{}' \frac{V_{\bf k}}{\sqrt{\NC}} 
     \delta\left[B^\dagger_{{\bf k} + {\bf k}'} B_{{\bf k}'}\,  B_{\bf k} \right]
  +{\cal O}\left( {\cal L}^0 \right)
\label{eq:H-in-ex-fluct2}\;,\quad
\end{eqnarray}
where only the first term is linear in fluctuations. Since the summed fluctuations, such as $\delta\opNC$ and $\sum_{\bf k}{}' \delta\left[B_{\bf k} B_{-{\bf k}} \right]$, scale with $\frac{1}{\sqrt{\cal N}}$, we conclude that the quadratic fluctuation are not extensive whereas the linear term is. Therefore, the leading order contribution of BEC fluctuations reduces to
\begin{eqnarray}
  \delta\op{H}_{\rm ex} 
  = 
  E_{\rm fluct} \,
  \delta \opNC
\label{eq:H-in-ex-fluctFIN}\;,
\end{eqnarray}
after we have introduced a BEC fluctuation-induced energy shift
\begin{eqnarray}
  E_{\rm fluct}
  \equiv
  \sum_{\bf k}{}' V_{\bf k} \, f_{\bf k}
  +
  \sum_{\bf k}{}' V_{\bf k} \, {\rm Re} \left[s_{\bf k}\right] 
  +
  \frac{1}{\sqrt{\NC}} \sum_{{\bf k},{\bf k}'}{}' 
  V_{\bf k} \,
  {\rm Re} 
  \left[
    T_{{\bf k},{\bf k}'}
  \right] 
\label{eq:fluct-energy}\;.
\end{eqnarray}
Since $\delta\opNC$ is a two-particle operator, fluctuations of the BEC number do not induce a hierarchy problem. The potential divergence of linearization \eqref{eq:sqrtBFIN} enters only the $T_{{\bf k},{\bf k}'}/\sqrt{\NC}$ term. However, the three-atom correlation $T_{{\bf k},{\bf k}'}$ vanish fasters than $\sqrt{\NC}$ at the limit $\NC \rightarrow 0$ as shown in Ref.~\cite{Kira:HBE}, which makes $E_{\rm fluct}$ and linearization effects nondivergent.

\subsection{Bogoliubov excitations}
\label{sec:Bogloliubov}

Next, we seek for simple structures within the excitation-picture Hamiltonian \eqref{eq:H-occ-delta}--\eqref{eq:H-in-ex-occup} and \eqref{eq:H-in-ex-fluctFIN}. From all contributions, the constant and single-particle contributions to $\ex{\op{H}}^{\rm occ.}$ constitute a Hamiltonian
\begin{eqnarray}
  \op{H}^{\rm BG}_{\rm ex} 
  = \frac{V_0}{2} 
    {\cal N} \left( {\cal N} -1 \right)
    +\sum_{\bf k}{}' E^C_{\bf k} B^\dagger_{\bf k} B_{\bf k}    
  +\sum_{\bf k}{}'  \frac{\NC \, V_{\bf k}}{2}  
  \left[ 
      B_{\bf k} \, B_{-{\bf k}} 
    +
     `B^\dagger_{-{\bf k}} B^\dagger_{\bf k}  
  \right]
\label{eq:H-Bogo}\;
\end{eqnarray}
that can be diagonalized by introducing the standard Bogoliubov transformation:
\begin{eqnarray}
  B_{\bf k} \equiv u_{\bf k} D_{\bf k} - v_{-{\bf k}} D^\dagger_{-{\bf k}}\,,
\qquad
  B^\dagger_{\bf k} \equiv u_{\bf k} D^\dagger_{\bf k} - v_{-{\bf k}} D_{-{\bf k}} 
\label{eq:D-Bogo}\;.
\end{eqnarray}
We have chosen $u_{\bf k}$ and $v_{-{\bf k}}$ to be real-valued coefficients. As long as they satisfy the normalization $u^2_{\bf k} - v^2_{-{\bf k}} =1$, $D_{\bf k}$ and $D^\dagger_{\bf k}$ are bosonic operators defining the Bogoliubov excitations. By parameterizing the Bogoliubov transformation \eqref{eq:D-Bogo} via hyperbolic functions, we find that a specific Bogoliubov transformation,
\begin{eqnarray}
  {\rm tanh} \, 2 \beta_{\bf k} = \frac{\NC\,V_{\bf k}}{E_{\bf k}^C}\,,
  \qquad
  u_{\bf k} = {\rm cosh} \, \beta_{\bf k}\,, \quad v_{{\bf k}} = {\rm sinh} \, \beta_{\bf k}
\label{eq:D-Bogo-state}\;,
\end{eqnarray}
diagonalizes the Hamiltonian \eqref{eq:H-Bogo}, yielding 
\begin{eqnarray}
  \op{H}^{\rm BG}_{\rm ex} 
  = \frac{V_0}{2} 
    {\cal N} \left( {\cal N} -1 \right)
    +\sum_{\bf k} \left( E^{\rm BG}_{\bf k} - E^C_{\bf k} \right)
    +\sum_{\bf k}{}' E^{\rm BG}_{\bf k} D^\dagger_{\bf k} D_{\bf k} 
\label{eq:H-BogoFIN}\;.
\end{eqnarray}
The $E^{\rm BG}_{\bf k}$ identified defines the excitation energy of the Bogoliubov excitations\cite{StamperKurn:1999,Jin:1996,Mewes:1996,Utsunomiya:2008}
\begin{eqnarray}
 E^{\rm BG}_{\bf k} \equiv \sqrt{(E^C_{\bf k})^2 -(\NC\,V_{\bf k})^2 } = \sqrt{E^2_{\bf k} + 2 E_{\bf k} \NC\, V_{\bf k}}
\label{eq:E-Bogo}\;,
\end{eqnarray}
where we have used definition \eqref{eq:E-renormalized} to simplify the expression. 

Since the atom--atom interaction has a very short range in the real space, $V_{\bf k}$ is nearly constant for relevant momenta. The resulting 
\begin{eqnarray}
 \NC\, V_{\bf k} \rightarrow 8\pi a_{\rm scatt}  \frac{\hbar^2}{2 m} \frac{\NC}{{\cal L}^3}= 8\pi a_{\rm scatt}  \frac{\hbar^2\, \rhoC}{2 m}
\label{eq:V-contact}
\end{eqnarray}
is often parametrized\cite{Leggett:2001,Giorgini:2008,Bloch:2008} in terms of scattering length $a_{\rm scatt}$ and the density of BEC atoms $\rhoC \equiv \frac{\NC}{{\cal L}^3}$. Mathematically, $a_{\rm scatt}$ is the scattering length of the contact potential. With this approximation, the constant part of $\op{H}^{\rm BE}_{\rm ex}$ reduces to the famous Lee-Huang-Yang energy\cite{Lee:1957} that describes the system energy at the ground state of the Bogoliubov excitations.

To determine the effect of quantum depletion on correlation measurements in Sec.~\ref{sec:complementaryQS}, we evaluate the doublets at the Bogoliubov ground state, i.e.~$\avex{D^\dagger_{\bf k} D_{{\bf k}'}}=0$ and $\avex{D_{\bf k} D_{{\bf k}'}}=0$. With the help of the Bogoliubov transformation \eqref{eq:D-Bogo} and condition \eqref{eq:D-Bogo-state}, we find
\begin{eqnarray}
 \avex{B^\dagger_{\bf k} B_{{\bf k}'}}^{\rm BG} &=& \delta_{{\bf k}',{\bf k}}\, n^{\rm BG}_{\bf k}\,, 
 \qquad 
  n^{\rm BG}_{\bf k} \equiv \avex{B^\dagger_{\bf k} B_{{\bf k}}}^{\rm BG} = \frac{1}{2} \frac{E^C_{\bf k}-E^{\rm BG}_{\bf k}}{E^{\rm BG}_{\bf k}}\,,
\nonumber\\
\avex{B_{\bf k} B_{{\bf k}'}}^{\rm BG} &=& \delta_{{\bf k}',-{\bf k}} \, s^{\rm BG}_{\bf k}\,, 
 \qquad 
  s^{\rm BG}_{\bf k} \equiv \avex{B_{\bf k} B_{-{\bf k}}}^{\rm BG} = -\frac{1}{2} \frac{\NC\, V_{\bf k}}{E^{\rm BG}_{\bf k}}
\label{eq:nANDs4bogo}\;,
\end{eqnarray}
where $\avex{\cdots}^{\rm BG}$ the average is evaluated using the ground state of Bogoliubov excitations. Whenever, the BEC has a macroscopic density, i.e.~$\rho_{\rm BEC} \neq 0$, atoms at the normal component become excited even at vanishing temperature, provided with that the atoms are interacting.

It is straightforward to show that Eq.~\eqref{eq:nANDs4bogo} yields property
\begin{eqnarray}
  \left(n^{\rm BG}_{\bf k}\right)^2 + \left|s^{\rm BG}_{\bf k} \right|^2 =
  n^{\rm BG}_{\bf k}  + \left(n^{\rm BG}_{\bf k}\right)^2
\label{eq:Bogo-property}\;.
\end{eqnarray}
When inserted into Eq.~\eqref{eq:c_shape}, we find find a shape correction $c_{\rm shape} = \frac{3\pi}{8} \approx1.1781$ for any fixed $a_{\rm scatt}$. When $(f_{\bf k}, s_{\bf k})$ deviates from the Bogoliubov excitations, $c_{\rm shape}$ can have any value from one to infinity. Therefore, $c_{\rm shape}$ provides a convenient measure of how close the quantum depletion is to the Bogoliubov excitation, see analysis in Sec.~\ref{sec:complementaryQS}.

It is clear that the full Hamiltonian \eqref{eq:H-in-ex-occup}--\eqref{eq:H-in-ex-fluct} contains the Bogoliubov excitations as a subset, and thus, describes the quantum depletion related to them. However, the nonlinear parts of Eqs.~\eqref{eq:H-in-ex-occup} and \eqref{eq:H-in-ex-fluct} cannot be diagonalized with the Bogoliubov transformation \eqref{eq:D-Bogo}. Therefore, the nonlinear contributions introduce interactions among Bogoliubov excitations, such that strong atom--atom coupling can considerably modify the approximative results \eqref{eq:nANDs4bogo}. In particular, when the dimer- and trimer-forming contributions become relevant, it is not beneficial to convert the system with the Bogoliubov transformation, but to study the quantum dynamics of the relevant particle clusters in terms of $B_{\bf k}$ and $B^\dagger_{\bf k}$. We will formally develop the theory into this direction in Sec.~\ref{sec:Cluster-kinetics} while I will develop the explicit cluster-kinetics formalism in Ref.~\cite{Kira:HBE}.

\subsection{Quantum dynamics in a cluster-expansion friendly form}
\label{sec:Cluster-kinetics}

The excitation picture expresses the quantum statistics of the BEC itself in terms of occupation $\NC$ as well as its fluctuation operator $\dopNC$. For the microcanonical systems studied here, both of these can be presented exactly using {\it only} the normal-component operators, based on relations \eqref{eq:N_N}--\eqref{eq:N_connect} and \eqref{eq:deltaNC}. Therefore, the quantum dynamics 
\eqref{eq:HEM-BB-comm}
of $B_{{\bf k}\neq 0}$ and $B^\dagger_{{\bf k} \neq 0}$ defines the quantum kinetics of both the normal-component and BEC atoms, which makes the excitation picture very useful. To simplify the notation for $B_{\bf k}$ and $B^\dagger_{\bf k}$, we assume that ${\bf k}$ implicitly refers to the normal-component atoms with ${\bf k} \neq 0$.

The quantum dynamics of any observable can then be evaluated straightforwardly from the Heisenberg equation of motion \eqref{eq:HEM-ex} with the help of Eq.~\eqref{eq:HEM-BB-comm}. In practice, we start from the excitation-picture Hamiltonian that is a sum of Eqs.~\eqref{eq:H-in-ex-occup} and \eqref{eq:H-in-ex-fluctFIN}; the $\delta\op{H}_{\rm ex}$ contains the fluctuation operator $\dopNC$ whose commutation relations with $B_{\bf k}$ and $B^\dagger_{\bf k}$ are simple,
\begin{eqnarray}
  \comm{B_{\bf k}}{\delta \opNC} = -\, B_{\bf k}\,,
  \qquad
  \comm{B^\dagger_{\bf k}}{\delta \opNC} = B_{\bf k}^\dagger
\label{eq:NC-commutator}\;,
\end{eqnarray}
based on definitions \eqref{eq:N_N}--\eqref{eq:N_connect}. As we use these and bosonic commutation relations several times, the elementary normal-component operators evolve according to
\begin{eqnarray}
  \ihddt B_{\bf k}
  &=&
  \left(E^C_{\bf k}-E_{\rm fluct} \right) B_{\bf k} 
  + \sum_{{\bf q},{\bf k'}}{}' V_{{\bf k}'-{\bf k}} B^\dagger_{{\bf q}-{\bf k}} B_{{\bf q}-{\bf k}'} B_{{\bf k}'}
\nonumber\\
  &&+\NC V_{\bf k} B^\dagger_{-{\bf k}}
  +
  \sqrt{\NC} \sum_{{\bf k}'}{}'
  \left[
    \left(V_{\bf k} + V_{{\bf k}'-{\bf k}} \right) B^\dagger_{{\bf k}'-{\bf k}} B_{{\bf k}'}
    +
    V_{{\bf k}'} B_{{\bf k}-{\bf k}'} B_{{\bf k}'}
  \right]
\label{eq:BoccuD}\,,
\\
  \ihddt B^\dagger_{\bf k}
  &=&
  -\left(E^C_{\bf k}-E_{\rm fluct} \right)  B^\dagger_{\bf k} 
  - \sum_{{\bf q},{\bf k'}}{}' V_{{\bf k}'-{\bf k}} 
   B^\dagger_{{\bf k}'} B^\dagger_{{\bf q}-{\bf k}'} B_{{\bf q}-{\bf k}}
\nonumber\\
  &&-\NC V_{\bf k} B_{-{\bf k}}
  -
  \sqrt{\NC} \sum_{{\bf k}'}{}'
  \left[
    \left(V_{\bf k} + V_{{\bf k}'-{\bf k}} \right) B^\dagger_{{\bf k}'} B_{{\bf k}'-{\bf k}} 
    +
    V_{{\bf k}'}  B^\dagger_{{\bf k}'} B^\dagger_{{\bf k}-{\bf k}'}
  \right]
\label{eq:B+occuD}\,.
\end{eqnarray}
Strictly speaking, these differential equations represent commutators \eqref{eq:HEM-BB-comm}. This seemingly innocent detail could prevent us from determining a general $\ex{\op{O}}$ dynamics because the functional dependence of  $\ex{\op{O}}$ on boson operators can have a nontrivial form in the excitation picture, as shown in \ref{app:O-elemntary}. However, we have another strong result --- Eq.~\eqref{eq:HEM-ex} --- that expresses the quantum kinetics via an ordinary commutation relation $\ihddt \avex{\ex{\op{O}}} = \avex{\comm{\ex{\op{O}}}{\ex{\op{H}}}}$. 
As discussed at the end of Sec.~\ref{sec:EP-relations}, this simplification allows us to always use to apply Eqs.~\eqref{eq:BoccuD}--\eqref{eq:B+occuD} to generate any $\ihddt \avex{\ex{\op{O}}} = \avex{\comm{\ex{\op{O}}}{\ex{\op{H}}}}$ using the usual differentiation rules where  Eqs.~\eqref{eq:BoccuD}--\eqref{eq:B+occuD} are the elementary differentiations. Therefore, Eqs.~\eqref{eq:BoccuD}--\eqref{eq:B+occuD} indeed determine the bosonic quantum 
kinetics when evaluated in the excitation picture.

We have also found another set of strong results in Secs.~\ref{sec:EP-relations}--\ref{sec:QS-basics}; the BEC remains a vacuum state in the excitation picture for all times, which eliminates {\it all} normally ordered $\avex{\ex{\op{O}}}$ containing BEC operators. Consequently, the full quantum dynamics of the interacting Bose gas (including the BEC properties) follows exclusively from the atom correlations excited to the normal component; explicit examples given in Sec.~\ref{sec:QS-basics}. When analyzing quantum depletion, even $\avex{\ex{\op{O}}}$ containing only normal-component operators vanish before the onset of quantum depletion. 
This scenario corresponds to semiconductor excitations that vanish before, e.g., an optical excitation is applied. In this situation, an optical field generates excited clusters sequentially, which can be efficiently described with very few clusters as shown in Ref.~\cite{Book:2011}. 
I will show in Ref.~\cite{Kira:HBE} that Eqs.~\eqref{eq:BoccuD}--\eqref{eq:B+occuD} and the excitation-picture describe quantum depletion via an analogous sequential build up of atom clusters. In other words, the excitation picture Eqs.~\eqref{eq:BoccuD}--\eqref{eq:B+occuD} set up a clusters-friendly description for a strongly interacting Bose gas.

From the structural point of view, Eqs.~\eqref{eq:BoccuD}--\eqref{eq:B+occuD} still produce the BBGKY hierarchy problem.
As a general classification, a product of $n$ boson operators belongs to the class of $n$-particle operators as explained in Sec.~\ref{sec:CE_representation}; for a textbook discussion cf.~Ref.~\cite{Book:2011}. Therefore, the first line of $B_{\bf k}$ and $B^\dagger_{\bf k}$ dynamics couples a single-particle operator with three-particle operators, which yields the standard BBGKY hierarchy problem, also observed in semiconductors.
The BEC produces two new classes of contributions; the part that is proportional to $\NC$ does not yield a hierarchy problem because the single-particle contribution is coupled with a conjugated single-particle operator. This pure single-particle dynamics can be solved exactly by introducing the Bogoliubov excitations, discussed in Sec.~\ref{sec:Bogloliubov}. 
The second BEC contribution is proportional to $\sqrt{\NC}$ and it couples the single-particle dynamics to two-particle operators. This hierarchy problem is less severe than the standard one, in the first line. Therefore, also it can be efficiently treated with the same cluster-expansion-based approach\cite{Wyld:1963,Fricke:1996,Kira:2006,Book:2011} as the standard hierarchy problem.

\section{Hartree-Fock Bogoliubov (HFB) approximation}
\label{sec:HFB}

Condensates are often described with the Gross-Pitaevskii equation (GPE)\cite{Dalfovo:1999,Pethick:2002,Leggett:2001} and its generalizations\cite{Gardiner:1997,Castin:1998,Milstein:2003,Andersen:2004,Wuster:2005,Wuster:2007}. Therefore, it is interesting to compare how the excitation-picture relates to such standard methods. We start from the original system Hamiltionian \eqref{eq:Hamiltonian} and compute the Heisenberg equation of motion for the field operator \eqref{eq:Field-ops}, yielding straightforwardly
\begin{eqnarray}
  \ihddt \hat{\Psi}({\bf r}) = H_0({\bf r}) \, \hat{\Psi}({\bf r}) 
  + 
  \int d^3x \, V({\bf x}-{\bf r}) \, \hat{\Psi}^\dagger({\bf x}) \hat{\Psi}({\bf x}) \hat{\Psi}({\bf r}) 
\label{eq:Psi-dyn}\,.
\end{eqnarray}
This form still describes the full many-body dynamics, including the hierarchy problem induced by the three field-operator contribution.

Besides its plane-wave representation \eqref{eq:Field-ops}, we may use any other orthonormal basis of single-particle wave functions $\phi_\nu({\bf r})$, to express the field operator $\hat{\Psi}({\bf r}) = \sum_\nu \phi_\nu({\bf r}) B_\nu$. Since BEC and normal-component refer to orthogonal states, $\hat{\Psi}({\bf r})$ can be separated {\it exactly} into pure BEC ($\nu=c$) and normal-component ($\nu\neq c$) contributions $\hat{\Psi}_{\rm c}$ and $\hat{\Psi}_{\rm n}  \equiv \sum_{\nu\neq c} \phi_\nu({\bf r}) B_\nu$, respectively. As shown in Sec.~\ref{sec:complementaryQS}, strong BECs have very small number fluctuations, which supports Bogoliubov's original idea\cite{Bogoliubov:1947} to replace BEC operators $B_c$ and $B^\dagger_c$ by complex numbers instead of their full operator form
\begin{eqnarray}
  \hat{\Psi}({\bf r}) =  \hat{\Psi}_{\rm c}({\bf r}) + \hat{\Psi}_{\rm n}({\bf r})
  \rightarrow
  \phi({\bf r}) + \hat{\Psi}_{\rm n}({\bf r})
\label{eq:Psi-sepration}\,,
\end{eqnarray}
where the identification of the complex-valued wave function $\phi({\bf r})$ introduces coherence as an approximation, as pointed out, e.g., after Eq.~(3.14) in Ref.~\cite{Leggett:2001}. Such a treatment assumes that the BEC remains as a pure coherent state\cite{Kira:2011}, parametrized by the coherent amplitude $\ave{\hat{\Psi}({\bf r})} = \phi({\bf r})$, whereas the normal component is assumed to represent an incoherent fluctuation field with $\ave{\hat{\Psi}_{\rm n}({\bf r})}=0$. For sake of generality, we have not defined the explicit spatial dependence for $\phi({\bf r})$. Procedure \eqref{eq:Psi-sepration} was first generalized for BEC by Beliaev\cite{Beliaev:1958} and, nowadays, there are multiple strategies\cite{Zaremba:1999,Blakie:2008,Griffin:2012,Wright:2012} to implement it to the actual computations. Instead of analyzing a specific method, we study how the assumption of coherence \eqref{eq:Psi-sepration} is connected with the excitation-picture approach.

As approximation \eqref{eq:Psi-sepration} is inserted to Eq.~\eqref{eq:Psi-dyn}, we obtain
\begin{align}
  &\ihddt \left[ \phi({\bf r}) + \hat{\Psi}_{\rm n}({\bf r}) \right] 
  = 
  H_0({\bf r}) \,  \left[ \phi({\bf r}) + \hat{\Psi}_{\rm n}({\bf r}) \right] 
\nonumber\\
  &
  \qquad+ 
  \int d^3x \, V({\bf x}-{\bf r}) 
  \left[
    |\phi({\bf x})|^2 \phi({\bf r}) 
    + \hat{\Psi}^\dagger_{\rm n}({\bf x})\hat{\Psi}_{\rm n}({\bf x}) \phi({\bf r}) 
    + \hat{\Psi}_{\rm n}({\bf x})\hat{\Psi}_{\rm n}({\bf r}) \phi^\star({\bf x}) 
    + \hat{\Psi}^\dagger_{\rm n}({\bf x})\hat{\Psi}_{\rm n}({\bf r}) \phi({\bf x}) 
  \right]
\nonumber\\
  &
  \qquad+ 
  \int d^3x \, V({\bf x}-{\bf r}) 
  \left[
    \phi^\star({\bf x}) \phi({\bf r}) \hat{\Psi}_{\rm n}({\bf r})
    + \phi({\bf x}) \phi({\bf r}) \hat{\Psi}^\dagger_{\rm n}({\bf r})
    + |\phi({\bf x})|^2 \hat{\Psi}_{\rm n}({\bf r}) 
    + \hat{\Psi}^\dagger_{\rm n}({\bf x})\hat{\Psi}_{\rm n}({\bf x}) \hat{\Psi}_{\rm n}({\bf r})
  \right]  
\label{eq:Psi-dyn-Bogo1}\,,\quad
\end{align}
where we have organized the even and odd orders of the normal-component operators to the second and third line, respectively. When the $\hat{\Psi}_{\rm n}({\bf r})$ and $\hat{\Psi}^\dagger_{\rm n}({\bf r})$ are perceived as pure fluctuation operators, any expectation that contains an odd number of them can be assumed to vanish. Using this constraint and taking an expectation value of Eq.~\eqref{eq:Psi-dyn-Bogo1}, we and up with the modified GPE
\begin{align}
  \ihddt \phi({\bf r})
  &= 
  H_0({\bf r}) \, \phi({\bf r})
   + 
  \int d^3x \, V({\bf x}-{\bf r}) 
  \left[
    |\phi({\bf x})|^2 
    + 
    f({\bf x},{\bf x})
  \right] 
  \phi({\bf r}) 
\nonumber\\
  &
  +
  \int d^3x \, V({\bf x}-{\bf r}) 
  \left[
    s({\bf x},{\bf r}) \phi^\star({\bf x}) 
    + f({\bf x},({\bf r}) \phi({\bf x}) 
  \right]
\label{eq:Gross-Pitaevskii}\,,\quad
\end{align}
where we have identified a density $f({\bf x},{\bf r}) \equiv \ave{\hat{\Psi}^\dagger_{\rm n}({\bf x})\hat{\Psi}_{\rm n}({\bf r})}$ and $s({\bf x},{\bf r}) \equiv \ave{\hat{\Psi}_{\rm n}({\bf x})\hat{\Psi}_{\rm n}({\bf r})}$ as the anomalous density, following many previous identifications\cite{Milstein:2003,Wuster:2005,Wuster:2007}. When the atom--atom interaction is replaced by a contact potential, $V({\bf x}-{\bf r}) = U_0 \, \delta({\bf x}-{\bf r})$, Eq.~\eqref{eq:Gross-Pitaevskii} reduces to the same GPE as in Refs.~\cite{Milstein:2003,Wuster:2005,Wuster:2007}. In case the we want to include the possibility to form three-atom coherences to the normal component, we also need to add a coherent 
$\int d^3x \, V({\bf x}-{\bf r}) \ave{ \hat{\Psi}_{\rm n}^\dagger({\bf x})\hat{\Psi}_{\rm n}({\bf x}) \hat{\Psi}_{\rm n}({\bf r})}$ contribution to the GPE. However, this contribution represents the build up of three-atom coherences to the normal component, which is beyond the standard Hartree-Fock Bogoliubov (HFB) approach.

In general, the three-atom coherences are generated by the quadratic operators in Eq.~\eqref{eq:Psi-dyn-Bogo1}. Omitting the corresponding coherences from the GPE can be formally achieved by replacing these quadratic terms with expectation values. With this approximation, Eqs.~\eqref{eq:Psi-dyn-Bogo1}--\eqref{eq:Gross-Pitaevskii} uniquely determine the dynamics of the normal-component operators
\begin{align}
  \ihddt &\hat{\Psi}_{\rm n}({\bf r}) 
  = 
  H_{\rm eff}({\bf r}) \,   \hat{\Psi}_{\rm n}({\bf r})
\nonumber\\
  &+ 
  \int d^3x \, V({\bf x}-{\bf r}) 
  \left[
   \phi^\star({\bf x}) \phi({\bf r}) \hat{\Psi}_{\rm n}({\bf r})
   +
    \phi({\bf x}) \phi({\bf r}) \hat{\Psi}^\dagger_{\rm n}({\bf r})
    + \hat{\Psi}^\dagger_{\rm n}({\bf x})\hat{\Psi}_{\rm n}({\bf x}) \hat{\Psi}_{\rm n}({\bf x})
  \right]\,,
\nonumber\\
  \ihddt &\hat{\Psi}^\dagger_{\rm n}({\bf r}) 
  = 
  -H_{\rm eff}({\bf r}) \,   \hat{\Psi}^\dagger_{\rm n}({\bf r}) 
\nonumber\\
  &- 
  \int d^3x \, V({\bf x}-{\bf r}) 
  \left[
   \phi^\star({\bf r}) \phi({\bf x}) \hat{\Psi}^\dagger_{\rm n}({\bf r})
   +
    \phi^\star({\bf x}) \phi^\star({\bf r}) \hat{\Psi}_{\rm n}({\bf r})
    + 
    \hat{\Psi}^\dagger_{\rm n}({\bf x})
    \hat{\Psi}^\dagger_{\rm n}({\bf x}) 
    \hat{\Psi}_{\rm n}({\bf x})
  \right]
\label{eq:Psi-normal}\,,\qquad
\end{align}
where $H_{\rm eff} ({\bf r}) \equiv  H_0({\bf r}) + \int d^3x \, V({\bf x}-{\bf r})\, |\phi({\bf x})|^2$ 
is an effective single-particle Hamiltonian. It is straightforward to show that operator dynamics \eqref{eq:Psi-normal}, the contact potential, and implementation of Wick's theorem\cite{Wick:1950} produces a $f({\bf x},{\bf r})$ and  $s({\bf x},{\bf r})$ dynamics that us identical to those applied in several investigations\cite{Milstein:2003,Wuster:2005,Wuster:2007}. The resulting $f({\bf x},{\bf r})$, $s({\bf x},{\bf r})$, and $\phi({\bf r})$ dynamics forms then a closed set of dynamical HFB equations that have been extremely successful in explaining the intriguing properties of weakly interacting Bose gases. However, the need to extend the HFB approach has become apparent in strongly interacting Bose gas, as pointed out in Ref.~\cite{Wuster:2007}. 
The excitation-picture result \eqref{eq:BoccuD}--\eqref{eq:B+occuD} establishes a systematic cluster-expansion platform for pragmatic generalizations, as I show in Ref.~\cite{Kira:HBE}. An alternative approach, based on projection operators, has also been developed and discussed in Ref.~\cite{Sahlberg:2013}.

\subsection{Extending HFB approach with excitation-picture analysis}
\label{sec:HFBextend}

The HFB approach is founded on the idea of separating the coherent BEC contribution from the incoherent normal-component fluctuations, which cannot be rigorously motivated\cite{Girardeau:1956,Gardiner:1997,Leggett:2001} even though it works superbly in many cases, as discussed in Sec.~\ref{app:Coherence}. Consequently, a direct extensions of the HFB approach becomes unambiguous because one must {\it a priori} decide which part of the higher-order correlations belong to the coherent vs.~incoherent many-body dynamics. The excitation picture introduced in this paper removes this unambiguity because it does not rely on sorting out coherences of the interacting Bose gas. Therefore, it is instructive to check which aspects of the excitation-picture approach are already included to the standard HFB approach.

To perform this comparison, we assume homogeneous excitation and use the plane-wave basis \eqref{eq:Field-ops}, 
$\op{\Psi}_{\rm n}({\bf r}) \equiv \frac{1}{{\cal L}^{3/2}} \sum_{\bf k}' e^{i{\bf k} \cdot {\bf r}} \, B_{\bf k}$ because $\op{\Psi}_c \equiv B_0\, \phi_0({\bf r})$ represents the BEC. In the HFB analysis and homogeneous conditions, $\ave{\hat{\Psi}^\dagger_c({\bf r}) \hat{\Psi}_c({\bf r})} \equiv |\phi({\bf r})|^2$ is a constant BEC density $\frac{\NC}{{\cal L}^3}$; approximation \eqref{eq:Psi-sepration} implies that the BEC wave function $\phi({\bf r})$ is replaced by $\sqrt{\frac{\NC}{{\cal L}^3}}$. Projecting the plane-wave ${\bf k}$ component from HFB relation~\eqref{eq:Psi-normal} yields
\begin{eqnarray}
  \left.\ihddt B_{\bf k}\right|_{\rm HFB}
  &=&
  \left(E^C_{\bf k}+V_0 \NC \right) B_{\bf k} 
  + \sum_{{\bf q},{\bf k'}}{}' V_{{\bf k}'-{\bf k}} B^\dagger_{{\bf q}-{\bf k}} B_{{\bf q}-{\bf k}'} B_{{\bf k}'}
  +\NC V_{\bf k} B^\dagger_{-{\bf k}}
\label{eq:B-HFB}\,,
\\
  \left. \ihddt B^\dagger_{\bf k}\right|_{\rm HFB}
  &=&
  -\left(E^C_{\bf k}+V_0 \NC \right)  B^\dagger_{\bf k} 
  - \sum_{{\bf q},{\bf k'}}{}' V_{{\bf k}'-{\bf k}} 
   B^\dagger_{{\bf k}'} B^\dagger_{{\bf q}-{\bf k}'} B_{{\bf q}-{\bf k}}
  -\NC V_{\bf k} B_{-{\bf k}}
\label{eq:B+HFB}\,,
\end{eqnarray}
which is structurally very close to the excitation-picture results \eqref{eq:BoccuD}--\eqref{eq:B+occuD}. As major differences, the HFB analysis does not contain the two-atom operator contributions, which induce higher order coherences to the normal component. Additionally, the BEC-number fluctuations \eqref{eq:fluct-energy} differ from the $V_0 \NC$ contribution of the HFB model.

Since the HFB approach approximates the BEC properties, it is understandable that it cannot fully describe the subtle aspects of many-body coherences or BEC fluctuations beyond the coherent-state approximation. Nevertheless, the HFB approach is so close to the systematic excitation-picture computations that the resulting many-body physics must agree as long as BEC fluctuation are not appreciable and high-order coherences are not building up to the normal component. This condition should be well valid in weakly interacting Bose gas whereas excitation picture is more appropriate for the strongly interacting Bose gas. I will derive the full correlation dynamics systematically using the excitation picture in Ref.~\cite{Kira:HBE}. The HFB become then replaced by a more general set of equations.

\subsection{Coherence aspects of BEC}
\label{app:Coherence}

The exact transformation rule \eqref{eq:N_0-trafo} has fundamental implications how the coherence properties of the BEC have to be perceived for microcanonic systems. We analyze the exact field operators,
\begin{eqnarray}
  \op{\Psi}({\bf r}) =  \hat{\Psi}_{\rm c}({\bf r}) + \hat{\Psi}_{\rm n}({\bf r})
   = \phi_0({\bf r})\, B_0 + \sum_{\nu\neq 0} \phi_{\nu}({\bf r})\, B_\nu
\label{eq:app-new-Psi}\;,
\end{eqnarray}
where we have separated the BEC state $\phi_0({\bf r})$ from the normal components $\phi_\nu({\bf r})$ with $\nu \neq 0$. These states just need to form an orthonormal set of single-particle wave functions. In case of a trap, the basis set can be chosen to optimally describe the confinement effects. Approximation \eqref{eq:Psi-sepration} follows by replacing $B_0$ with an amplitude $\beta_0$ and by identifying $\phi_0({\bf r})\, B_0 \rightarrow \phi_0({\bf r})\, \beta_0 \equiv \phi({\bf r})$. However, this phenomenological identification is not rigorous for microcanonical systems; Since $\op{\Psi}({\bf r})$ is not a microcanonical operator, $\ave{\op{\Psi}({\bf r})}$ must rigorously vanish for micorcanonical systems studied here, based on Eq.~\eqref{eq:org2exci}. In other words, the identification of $\ave{\op{\Psi}_c({\bf r})}\equiv\phi({\bf r})$ coherence cannot strictly speaking be valid, as also pointed out in Refs.~\cite{Girardeau:1956,Gardiner:1997,Leggett:2001}.

To recover the usual GPE \eqref{eq:Gross-Pitaevskii} despite the lack of coherence, we follow the derivation of a single-particle density matrix, as in Ref.~\cite{Leggett:2001}, in order to identify whether  coherence emerges in the excitation-picture treatment. In other words, we start by studying the generic atom density
\begin{eqnarray}
  \rho({\bf r},{\bf r}')
  &\equiv& \ave{\op{\Psi}^\dagger({\bf r})\,\op{\Psi}({\bf r}')}
\nonumber\\
  &=& \phi^\star_0({\bf r})\,\phi_0({\bf r}') \ave{B^\dagger_0 B_0}
  + \sum_\nu{}' 
  \left[\phi^\star_0({\bf r})\,\phi_\nu({\bf r}') \ave{B^\dagger_0 B_\nu}
    +
    \phi^\star_\nu({\bf r})\,\phi_0({\bf r}') \ave{B^\dagger_\nu B_0}
  \right]
\nonumber\\
  &&+
  \sum_{\nu,\nu'}{}'
  \phi^\star_\nu({\bf r})\,\phi_{\nu'}({\bf r}') \ave{B^\dagger_\nu B_{\nu'}}
\label{eq:app-rho11}\;.
\end{eqnarray}
As we implement exact transformations \eqref{eq:app-B4H_ex}, this quantity becomes
\begin{eqnarray}
  \rho({\bf r},{\bf r}')
  &\toX& \phi^\star_0({\bf r})\,\phi_0({\bf r}') \avex{\opNC}
  + \sum_\nu{}' 
  \left[\phi^\star_0({\bf r})\,\phi_\nu({\bf r}') \avex{\sqrt{\opNC}\, B_\nu}
    +
    \phi^\star_\nu({\bf r})\,\phi_0({\bf r}') \avex{B^\dagger_\nu \sqrt{\opNC}}
  \right]
\nonumber\\
  &&+
  \sum_{\nu,\nu'}{}'
  \phi^\star_\nu({\bf r})\,\phi_{\nu'}({\bf r}') \avex{B^\dagger_\nu B_{\nu'}}
\label{eq:app-rho22}\;.
\end{eqnarray}
The first expectation value produces directly the average number of BEC atoms that is $N_C$. In the same way, all the remaining $\opNC$ contributions can be replaced by $\NC$ because the corrections to this approximation scale like $\frac{1}{\NC}$. At this point, we can identify a macroscopic wave function
\begin{eqnarray}
  \phi({\bf r}) \equiv \phi_0({\bf r}) \sqrt{\NC}
\label{eq:app-macroPSI}\;,
\end{eqnarray}
in analogy to the phenomenological identification \eqref{eq:Psi-sepration}. With these steps, Eq.~\eqref{eq:app-rho22} reduces into
\begin{eqnarray}
  \rho({\bf r},{\bf r}')
  &=& \phi^\star({\bf r})\,\phi({\bf r}') 
  + \sum_\nu{}' 
  \left[\phi^\star({\bf r})\,\phi_\nu({\bf r}') \avex{ B_\nu}
    +
    \phi^\star_\nu({\bf r})\,\phi({\bf r}') \avex{B^\dagger_\nu}
  \right]
  + f({\bf r},{\bf r}')
\label{eq:app-rho3}\;,
\end{eqnarray}
where $f({\bf r},{\bf r}')$ is the normal-component density identified already in connection with the GP Eq.~\eqref{eq:Gross-Pitaevskii}.

As discussed in the beginning of Sec.~\ref{sec:Doublets}, neither $\avex{ B_\nu}$ nor $\avex{B^\dagger_\nu}$ coherences can build up to the normal component. Therefore, Eq.~\eqref{eq:app-rho3} becomes
\begin{eqnarray}
  \rho({\bf r},{\bf r}')
  &=& \phi^\star({\bf r})\,\phi({\bf r}')  + \rho_N({\bf r},{\bf r}')
\label{eq:app-rho4}\;.
\end{eqnarray}
Consequently, the BEC part has the same form as the density matrix of a pure state, identified by wave function $\phi({\bf r})$. At the same time, the normal component $ \rho_N({\bf r},{\bf r}')$ cannot generally be reduced to a pure state. This separation into a pure-state $\phi({\bf r})$ and a normal-component density matrix appears generally in all expectation values having equal number of creation and annihilation operators. Therefore, all properties involving the BEC degrees of freedom show the coherence properties of a single macroscopic wave function $\phi({\bf r})$, exactly as predicted by the phenomenological substitution \eqref{eq:Psi-sepration}, even when true coherence does not exist.

In particular, these coherences are approximatively described by the GPE \eqref{eq:Gross-Pitaevskii}, provided with that the third-order normal-component coherences are not formed. However, the existence of macroscopic wave function does not require the BEC to be a true coherent state, implemented phenomenologically by the substitution $B_0 \rightarrow \beta_0$ or $\ave{B_0} \rightarrow \beta_0$; The corresponding expectation value simply does not exist in microcanonical systems, and yet all the relevant coherence properties behave as if this substitution were true. In other words, introducing the phenomenological substitution $B_0 \rightarrow \beta_0$ is 'convenient fiction', as put by M{\o}lmer\cite{Molmer:1997} who explained why coherent state provides such a good description of a laser despite it rigorously does not exist in the experiments. Personally, I believe a very similar connection exists between a BEC and its coherence properties; despite the coherent state cannot exist in the BEC experiments, 
$\phi({\bf r})$ describes the BEC properties excellently, making the GPE very useful indeed.

To understand the requirements for inducing a coherent amplitude to the BEC, we consider a simplified situation where all atoms are condensed. The corresponding system can be generally described using a single-boson density matrix
\begin{eqnarray}
  \op{\rho}
  \equiv 
  \sum_{N,N'=0}^\infty \absN{N} \, \rho_{N,N'} \, \absNS{N'}
\label{eq:app-coh-rho}\;.
\end{eqnarray}
Alternatively, we may think that $\op{\rho}$ is the density matrix of the system after all other degrees of freedom are traced out from it. The expectation value of the coherent amplitude becomes then
\begin{eqnarray}
  \ave{B_0}
  \equiv 
  \sum_{N,N'=0}^\infty \absNS{N'} B_0 \absN{N} \, \rho_{N,N'} 
  =
  \sum_{N=0}^\infty \sqrt{N} \, \rho_{N,N-1} 
\label{eq:app-coh-amp}
\end{eqnarray}
that follows after applying property \eqref{eq:app-B_rules} and orthogonality of the Fock states. In order to have a non-vanishing coherent amplitude, at least one of the density-matrix elements $\rho_{N,N-1}$ must be nonzero. The coherence-generating parts of $\op{\rho}$ must, therefore, look like 
\begin{eqnarray}
  \op{\rho}^{\rm coh}
  \equiv 
   \absN{N} \, \rho_{N,N-1} \, \absNS{N-1} + {\rm h.c.}
\label{eq:app-coh-part}
\end{eqnarray}
which is a combination of states with $N$ and $N-1$ atoms. Due to this mixture, the resulting $\op{\rho}$ is not microcanonical anymore. In other words, coherent amplitude can exist in the system only if the system does not have a fixed particle number. Therefore, the {\it BEC in a microcanocical Bose gas cannot have a coherent amplitude}. We have actually shown in   \ref{app:O-elemntary} that an even stronger statement holds: any operator that is not microcanonical has a vanishing expectation value for an interacting Bose gas with a fixed atom number.

To understand how amplitude coherence could be generated to the BEC, we consider the properties of a coherent state\cite{Glauber:1963} $\absN{\alpha}$. Coherent states are frequently used in quantum optics \cite{Walls:2008,Kira:2011} because they satisfies the eigen-value problem of $B_0 \absN{\alpha} = \alpha \absN{\alpha}$ where $\alpha$ is a complex-value amplitude. In other words, any normally ordered expectation value, $\ave{[B^\dagger_0]^J [B_0]^K} = (\alpha^\star)^J \alpha^K$ can then be evaluated directly by a formal substitution $B_0 \rightarrow \alpha$. This treats the boson operators classically, which makes coherent states the most classical representation of bosons. Therefore, the coherent state has coherence to all orders. At the same time, this substitution is identical to the common-wisdom substitution $\hat{\Psi}_c({\bf r}) \rightarrow \phi({\bf r})$, also used in Eq.~\eqref{eq:Psi-sepration}, and the coherent state can always be represented in terms of Fock states\cite{Walls:2008,Kira:2011}:
\begin{eqnarray}
  \absN{\alpha} = \sum_{N=0}^\infty \frac{\alpha^N}{\sqrt{N!}}\, \absN{N} \,e^{-\frac{|\alpha|^2}{2}}
\label{eq:app-coherent-state}\;.
\end{eqnarray}
We see that the coherent state is a superposition of many number states. In the context of BEC, it means that the system is indeed not microcanonical because different {\it total} atom numbers emerge in the superposition; note that we have assumed that the normal component vanishes in the present discussion.

Typically, the atom traps can change their atom number, e.g., via evaporative cooling or three-body loss \cite{Burt:1997,Esry:1999,Kim:2004,Brodsky:2006,Daley:2009,Castin:2010,Laurent:2013}. Obviously, such processes are dissipative, which tends not to generate coherent superposition states with a coherent amplitude $\alpha = e^{i\theta}|\alpha|$ having a specific direction $\theta$. Instead, such processes should randomize at least the direction $\theta$, such that the dissipation may generate a phase randomized coherent state,
\begin{eqnarray}
  \op{\rho}_{|\alpha|} 
  \equiv \frac{1}{2 \pi} \int_0^{2 \pi} d\theta\, \absN{e^{i\theta}|\alpha| }\absNS{e^{i\theta}|\alpha| } = 
  \sum_{N=0}^\infty   \absN{N} \frac{|\alpha|^{2N}}{N!}\,e^{-|\alpha|^2} \, \absNS{N} 
\label{eq:app-rho-coherent-state}\;,
\end{eqnarray}
at its best. The resulting $\op{\rho}_{|\alpha|}$ does not have a constant atom number, yet it does not connect the different atom-number levels as in Eq.~\eqref{eq:app-coh-part}, which is required for the amplitude coherence. In addition, $\op{\rho}_{|\alpha|}$ can well be described by studying the quantum kinetics of each of the fixed-number components $\absN{N}\absNS{N}$ components with the microcanonical theory developed here, and by weighting separate microcanonical computations with the probability weights $p_N \equiv \absNS{N} \op{\rho} \absN{N}$ of the initial states. We can also introduce loss of particles through dynamics of $p_N$, which directly generalizes the excitation-picture analysis to be applicable even when the atom trap looses particles. In other words, the fundamental interaction properties of Bose gas can be described using the microcanonical ensemble. This is not surprising because the concept of condensation is itself based on the idea that particle number must be conserved. 
 

\section{Conclusions}
\label{sec:Conclusions}

The standard many-body Hamiltonian of an interacting Bose gas yields excessive cluster-correlations among the BEC atoms, which makes a direct application of the cluster expansion inefficient. Here, I have introduced an excitation picture that formally eliminates {\it all} clusters within the Bose-Einstein condensate (BEC) by expressing the entire many-body system with a few atom clusters {\it excited} to the normal component alone. In other words, the excitation picture focuses the analysis exclusively to the normal-component excitations, generated from the BEC by the atom--atom interactions.

As the main result, the excitation picture yields a cluster-expansion friendly formulation of a strongly interacting Bose gas. On this basis, one can straightforwardly start applying the existing coupled-cluster knowhow  in quantum optics\cite{Kira:2006,Kira:2006b,Kira:2008,Kira:2011,Jahnke:2012,Hunter:2014} as well as
in semiconductor optics\cite{Schmitt-Rink:1989,Shah:1999,Aoki:1999,Khitrova:1999,Chemla:2001,Rossi:2002,Koch:2006,Meier:2007,Haug:2009,Smith:2010,Cundiff:2012,Malic:2013}
to solve the many-body dynamics in a strongly interacting Bose gas. In particular, the presented elementary operator dynamics \eqref{eq:BoccuD}--\eqref{eq:B+occuD} serves as a general starting point to derive the quantum dynamics of the clusters. As an ultimate goal, the explicit quantum kinetics of {\it all} 
clusters can be efficiently solved in the excitation picture. I will complete this final step of investigations in Ref.~\cite{Kira:HBE} because one needs significant extension of existing formalisms to execute that. More specifically, Ref.~\cite{Kira:HBE} introduces the implicit-notation formalism to unravel the quantum dynamics of {\it all} clusters with one derivation. In this connection, also an extended comparison of semiconductors vs.~strongly interacting Bose gas will be continued.

One may view a strongly interacting Bose gas as a prototype of a highly correlated system in the original picture. The success of this work suggests that there may be a more general approach to treat highly correlated systems with the cluster-expansion approach by introducing an appropriate excitation picture. 
Like for the interacting Bose gas, the transformation will most likely be nontrivial and only partially unitary because it aims to represent the physics via excitations around the highly correlated state. 
Therefore, I expect that one needs to define the transformation and its specific properties on a case-by-case basis. Nonetheless, a success to find a suitable transformation significantly clarifies how many-body effects evolve in highly correlated systems.

For BECs, the structure of the excitation picture alone reveals intriguing BEC properties. The presented analysis exposes several unexpected limitations and connections between interactions and Bogoliubov excitations, BEC's coherent amplitude, second-order coherence, and atom-number fluctuations. For example, the order parameter becomes rigorously identified in the excitation picture where
the BEC operators are substituted by the square root of the BEC number operator. Physically, the existence of order parameter does not require the presence of coherent amplitude, but only a singular BEC occupation,  as shown in   \ref{app:Coherence}. Additionally, BEC quantum statistics cannot be reduced to a single number even for relatively simple properties. For example, atom--atom correlations, BEC number fluctuation, and second-order coherence are shown to highlight different complementary aspects of the BEC correlations.

The conversion process of BEC to normal component is called quantum depletion, which can have many forms 
influencing the quantum statistics of the BEC through interactions. I have identified the shape correction $c_{\rm shape}$ to classify how the shape of quantum depletion influences measurements that are capable of resolving four-atom clusters. Among such measurements, BEC's number fluctuation $\Delta\NC$ is singled out as the most sensitive  because it is directly proportional to $\sqrt{c_{\rm shape}}$, as shown by Eq.~\eqref{eq:ac-FIN}. For Bogoluibov excitations, $c_{\rm shape}$ was shown to be $\frac{3\pi}{8} \approx 1.1781$, but there is no upper limit how large $c_{\rm shape}$ can be as the atom--atom interactions become stronger.
Such modifications are expected in a strongly interacting Bose gas because
the system Hamiltonian contains Bogoluibov excitation only as a subset, and the full excitation-picture Hamiltonian introduces new interaction possibilities among all excitations. This observation suggests that experimental detection of $\Delta\NC$ can quantitatively determine how much the quantum depletion deviates from the Bogoluibov 
excitations as function of excitation level and interaction strength.

Once the atom-cluster dynamics becomes explored further, I expect many intriguing cross comparisons between strongly interacting Bose gas and semiconductor many-body physics. Since the cluster-expansion approach is also used in quantum chemistry and nuclear-many-body physics, these studies can generate true synergy in producing new insights to the challenging problem of many-body quantum kinetics. Clearly, the developed excitation-picture formalism serves as a common platform to execute such investigations systematically. 

\section*{Acknowledgements}

I wish to thank Steven Cundiff for bringing to my attention the fact that experiments on BEC and quantum-degenerate Fermi gas seem to be converging toward the same set of many-body problems as the ultrafast spectroscopy on semiconductors. I am also indebted to Deborah Jin for explaining the experimentally relevant aspects in Fermi and Bose gases, and Chris Greene and Matthew Davis for insightful discussions about how many-body BEC theory is currently perceived.  I have enjoyed very much my cooperation with Stephan Koch concerning semiconductor many-body physics, over the years. I am also grateful to Renate Schmid for a very careful proof reading of THIS manuscript.

\appendix

\section{Consequences of the excitation transform}
\label{app:T-properties}

To access the dynamics in interacting Bose-gas, it is important to know how products of operators behave under excitation-picture transform \eqref{eq:rho-T-and-O}. For this purpose, we compute the excitation transformation of a general operator pair $\op{A}\,\op{B}$:
\begin{align}
  \ex{\left( \op{A} \, \op{B} \right)}
  &= \ex{\op{T}} \; \op{A} \, \op{B}  \; \ex{\op{T}}^\dagger
  = \ex{\op{T}} \; \op{A}   \left( \ex{T}^\dagger \ex{T} + \sum_{j=0}^{\opNC-1} \absCC{j}{j} \right) \op{B} \ex{\op{T}}^\dagger
\nonumber\\
   &= \ex{\op{T}} \; \op{A}  \ex{T}^\dagger \ex{T} \op{B} \ex{\op{T}}^\dagger
  +\ex{\op{T}} \; \op{A}   \sum_{j=0}^{\opNC-1} \absCC{j}{j}  \op{B} \ex{\op{T}}^\dagger
\nonumber\\
   &= \ex{\op{A}} \, \ex{\op{B}}
  +\ex{\op{T}} \; \op{A}   \sum_{j=0}^{\opNC-1} \absCC{j}{j}  \op{B} \ex{\op{T}}^\dagger
\label{eq:app_AB-product}\;
\end{align}
where we have used definition \eqref{eq:rho-T-and-O} and identity-relation \eqref{eq:T-unitarity} to get the expression inside the parentheses. The last step follows after we have applied definition \eqref{eq:rho-T-and-O}  the second time to identify the product of the excitation-pcture operators. Since the remaining therm is generally not vanishing, the excitation picture of operator product is not necessarily a product of transformed operators, as stated by Eq.~\eqref{eq:AB-product}.

The density operators $\op{\rho}$ and $\op{\rho}_{\rm ex}$ have a one-to-one connection based on identifications \eqref{eq:rho-T} and \eqref{eq:rho-T-and-O}. Next, we show that this connection can also be derived using the properties of the $\ex{\op{T}}$ and $\ex{\op{T}}^\dagger$ operators. The difficult part is to show that knowing $\ex{\op{\rho}}$ through definition \eqref{eq:rho-T-and-O} yields $\op{\rho}$ because the calculation that follows involves normally ordered transformation matrices. In short, we want to show that Eq.~\eqref{eq:rho-T} is the appropriate inverse of transformation \eqref{eq:rho-T-and-O}. Therefore, we insert $\op{\rho}_{\rm ex} =\op{T}_{\rm ex} \; \op{\rho} \; \op{T}_{\rm ex}^\dagger$ into Eq.~\eqref{eq:rho-T}, producing
\begin{eqnarray}
  \op{T}_{\rm ex}^\dagger \; \op{\rho}_{\rm ex} \ex{\op{T}}
  &=& 
  \op{T}_{\rm ex}^\dagger \op{T}_{\rm ex} \; \op{\rho} \; \op{T}_{\rm ex}^\dagger \; \ex{\op{T}}
\nonumber\\
  &=& 
  \left( \ident - \sum_{j=0}^{\op{N}_C-1} \absCC{j}{j}\right)  
  \; \op{\rho} \; 
  \left( \ident - \sum_{j=0}^{\op{N}_C-1} \absCC{j}{j}\right)
\label{eq:app-rho1}\;
\end{eqnarray}
after having used the property \eqref{eq:T-unitarity}. Here, $\op{N}_C$ projects the number of BEC atoms within $\op{\rho}$. We notice that $j$ remains always smaller than the BEC-atom number such that both  $\absCC{j}{j} \op{\rho}$ and  $\op{\rho} \absCC{j}{j} \op{\rho}$ contributions strictly vanish. Therefore, we find that only the identity terms within Eq.~\eqref{eq:app-rho1} remain, producing
\begin{eqnarray}
  \op{T}_{\rm ex}^\dagger \, \op{\rho}_{\rm ex} \, \op{T}_{\rm ex}
  =
  \op{\rho}
\label{eq:app-rho2}\;,
\end{eqnarray}
which verifies that $\op{\rho}$ follows uniquely from $\ex{\op{\rho}}$, as stated by Eq.~\eqref{eq:rho-T}. The inverse of \eqref{eq:app-rho2} produces relation \eqref{eq:rho-T-and-O} straightforwardly because we can use property $\op{T}_{\rm ex} \op{T}_{\rm ex}^\dagger = \ident$ to produce it, according to property \eqref{eq:T-unitarity}. As a result, relations \eqref{eq:rho-T} and \eqref{eq:rho-T-and-O} indeed provide a one-to-one mapping between $\op{\rho}$ and $\ex{\op{\rho}}$.

In the Schr\"{o}dinger picture, the quantum dynamics of the system follows from $i \hbar \ddt \op{\rho} = \comm{\op{H}}{\op{\rho}}$. We use this to solve the quantum dynamics of $\ex{\ave{\ex{\op{O}}}} = \ave{\op{O}}$ directly. By starting from Eq.~\eqref{eq:ex-average}, we obtain
\begin{eqnarray}
  i \hbar \ddt \ex{\ave{\ex{\op{O}}}} 
  &=&
  i \hbar \ddt \trace{\ex{\op{O}} \, \ex{\op{T}} \, \op{\rho}  \, \ex{\op{T}}^\dagger }
  = 
   \trace{\ex{\op{O}} \, \ex{\op{T}} \, \left(i \hbar \ddt \op{\rho}\right)  \, \ex{\op{T}}^\dagger }
\nonumber\\
  &=&
   \trace{\ex{\op{O}} \, \ex{\op{T}} \, \left(\op{H}\, \op{\rho} - \op{\rho}\,\op{H} \right)  \, \ex{\op{T}}^\dagger }
\label{eq:app-HEM1}\;,
\end{eqnarray}
in the Schr\"{o}dinger picture. We then replace the density matrix by relation \eqref{eq:app-rho2} and get
\begin{eqnarray}
  i \hbar \ddt \ex{\ave{\ex{\op{O}}}} 
  &=&
   \trace{\ex{\op{O}} \, 
   \left( 
      \ex{\op{T}} \, \op{H}\, \op{T}_{\rm ex}^\dagger \, \op{\rho}_{\rm ex} \, \op{T}_{\rm ex} \, \ex{\op{T}}^\dagger
      - \ex{\op{T}} \, \op{T}_{\rm ex}^\dagger \, \op{\rho}_{\rm ex} \, \op{T}_{\rm ex} \, \op{H} \, \ex{\op{T}}^\dagger \right)
      }
\nonumber\\
  &=&
   \trace{\ex{\op{O}} \, 
   \left( 
      \ex{\op{H}} \, \op{\rho}_{\rm ex} \, \op{T}_{\rm ex} \, \ex{\op{T}}^\dagger
      - \ex{\op{T}} \, \op{T}_{\rm ex}^\dagger \, \op{\rho}_{\rm ex} \, \ex{\op{H}} \right)  }
\label{eq:app-HEM2}\;.
\end{eqnarray}
The remaining transfer operators produce unity, yielding
\begin{eqnarray}
  i \hbar \ddt \ex{\ave{\ex{\op{O}}}} 
  &=&
   \trace{\ex{\op{O}} \, 
   \left( 
      \ex{\op{H}} \, \op{\rho}_{\rm ex} 
      - \op{\rho}_{\rm ex} \, \ex{\op{H}} \right)  }
  =
   \trace{
   \left(\ex{\op{O}} \, \ex{\op{H}} -\ex{\op{H}}\,\ex{\op{O}} \right)
      \op{\rho}_{\rm ex}  }
\nonumber\\
    &=& \trace{\comm{\ex{\op{O}}}{\ex{\op{H}}} \op{\rho}_{\rm ex}}
  =
  \ex{\ave{\comm{\ex{\op{O}}}{\ex{\op{H}}}}}
\label{eq:app-HEM_FIN}\;,
\end{eqnarray}
where we have cyclically permutated $\ex{\op{H}}$ under the trace to identify the commutator between $\ex{\op{O}}$ and $\ex{\op{H}}$. The final form shows that the quantum dynamics of interacting Bose gas can be evaluated using the Heisenberg equation of motion 
\begin{eqnarray}
  i \hbar \ddt \ave{\ex{\op{O}}}
  =
  \comm{\ex{\op{O}}}{\ex{\op{H}}}
\label{eq:app-HEM_OP}\;,
\end{eqnarray}
evaluated completely within the excitation picture. The last operator identification is valid only in the Hilbert space where the BEC state is unoccupied, i.e.~the density matrix is defined by $\ex{\op{\rho}}$.

\section{Elementary operators in the excitation picture}
\label{app:O-elemntary}

To evaluate the physical properties in the excitation picture, we need to determine how the physically relevant operators behave under transformation \eqref{eq:rho-T-and-O}. In Sec.~\ref{sec:QS-basics}, we have identified microcanonical operators as the only relevant ones. As an example, we start by transforming a microcanocical operator \eqref{eq:org2exci},
\begin{eqnarray}
  \left(\op{L}^\dagger\right)^J \, \left(\op{L}\right)^K \, \op{O}\left(J',K'\right)
  \equiv
  \left(\op{L}^\dagger\right)^J \, \left(\op{L}\right)^K \, \op{O}^{J'}_{K'}\,, \qquad
  J+J' = K+K'
\label{eq:app-microO}\;,
\end{eqnarray}
into the excitation picture. We denote the operator containing $J'$ ($K'$) normal-component creation (annihilation) operators by $\op{O}^{J'}_{K'}$ to shorten the notation. We also consider generic basis states
\begin{eqnarray}
  \absN{C,\,\group{n_{\bf k}}}
  \equiv
  \absNC{C} \otimes
  \absNN{\group{n_{\bf k}}}
\label{eq:app-basis}
\end{eqnarray}
that contains $C$ atoms at BEC and $\NN = \sum_{\bf k} n_{\bf k}$ normal-component atoms, according to Eq.~\eqref{eq:Normal-state}.

To evaluate the explicit excitation-picture form of operators, we calculate the product of transformation \eqref{eq:rho-T-and-O} and state \eqref{eq:app-basis}, yielding
\begin{align}
  \ex{\op{O}}\, \absN{C,\,\group{n_{\bf k}}} 
  &=
  {\op T} \op{O} \left( \op{L}^\dagger \right)^\opNC
  \, \absN{C,\,\group{n_{\bf k}}} 
  =
    {\op T} \op{O} \left( \op{L}^\dagger \right)^{{\cal N} -\NN}
  \, \absN{C,\,\group{n_{\bf k}}}
\nonumber\\
  &=
   {\op T} \op{O} 
  \, \absN{{\cal N}-\NN+C,\,\group{n_{\bf k}}}\,,
  \qquad \NN \le {\cal N}
\label{eq:app-O-help1}\;,
\end{align}
after expressing $\op{T}^\dagger$ with help of definition \eqref{eq:T-operator}, replacing $\opNC={\cal N}-\opNN$ by ${\cal N} -\NN$ (because $\absN{C,\,\group{n_{\bf k}}}$ contains $\NN$ normal-component atoms), and applying property \eqref{eq:L-property2}. In principle, $\NC={\cal N} -\NN$ must be positive definite, which sets up the condition. Since we are analyzing microcanonical systems with exactly ${\cal N}$ atoms, this condition is automatically satisfied for the relevant basis states; nevertheless, we have denoted this condition for the sake of completeness. As an other help relation, we evaluate
\begin{align}
  \op{O}^{J'}_{K'}
  \, \absN{C,\,\group{n_{\bf k}}} 
  =
  {\sf o}^{J'}_{K'} 
  \, \absN{C,\,\group{n_{\bf k}+J'-K'}}
\label{eq:app-O-help2}\,,
\end{align}
where $\absN{C,\,\group{n_{\bf k}+J'-K'}}$ is a normal-component state with $K'$ states removed and $J'$ states added by $\op{O}^{J'}_{K'}$. The prefactor ${\sf o}^{J'}_{K'}$ is defined such that it vanishes whenever  $\absN{C,\,\group{n_{\bf k}}}$ does not contain the states removed by $K'$ annihilation operators within $\op{O}^{J'}_{K'}$. 

In further derivations, we do not need to know the detailed structure of either ${\sf o}^{J'}_{K'}$ or $\absN{C,\,\group{n_{\bf k}}}$. However, they can be computed straightforwardly using the well-known rules\cite{Book:2011,Walls:2008}
\begin{eqnarray}
  (B)^j \absN{n} = \sqrt{\frac{n!}{(n-j)!}} \, \absN{n-j}\,, \qquad
  (B^\dagger)^k \absN{n} = \sqrt{\frac{(n+k)!}{n!}} \, \absN{n+k}
\label{eq:app-B_rules}\;
\end{eqnarray}
when boson operators act on a specific Fock state. Here, the first relation vanishes for $j>n$ because the factorial $(n-j)!$ has then a negative argument; when factorials are represented via the gamma function, this produces a diverging $(n-j)!$ and, thus, vanishing $\frac{1}{(n-j)!}$ for $j>n$. For later reference, we also identify a theta function
\begin{align}
  \theta_x =
  \left\{ 
    \begin{array}{cc}
      1\,, & x\ge 0
      \\
      0\,, & x<0
    \end{array}
  \right.
\label{eq:app-theta}
\end{align}
that exists only for positive-definite arguments $x$.

The explicit excitation-picture transformation of operator \eqref{eq:app-microO} produces
\begin{align}
  \ex{
   \left[
    \left(\op{L}^\dagger\right)^J \, \left(\op{L}\right)^K \, \op{O}^{J'}_{K'}
   \right]
   }
   \, \absN{C,\,\group{n_{\bf k}}} 
   &=
   \op{T}
    \left(\op{L}^\dagger\right)^J \, \left(\op{L}\right)^K \, \op{O}^{J'}_{K'}
   \, \absN{{\cal N}-\NN+C,\,\group{n_{\bf k}}}
\nonumber\\
   &=
   {\sf o}^{J'}_{K'}\, 
   \op{T}
    \left(\op{L}^\dagger\right)^J \, \left(\op{L}\right)^K \, 
   \, \absN{{\cal N}-\NN+C,\,\group{n_{\bf k}+J'-K'}}
\label{eq:app-microO-ex1}\;,
\end{align}
where we have used relations \eqref{eq:app-O-help1}--\eqref{eq:app-O-help2}, consecutively. Applying raising and lowering operator properties \eqref{eq:L-property2} to the state \eqref{eq:app-microO-ex1} leads to
\begin{align}
  &\ex{
   \left[
    \left(\op{L}^\dagger\right)^J \, \left(\op{L}\right)^K \, \op{O}^{J'}_{K'}
   \right]
   }
   \, \absN{C,\,\group{n_{\bf k}}} 
  =
  \theta_{\NC + C-K}\,
   {\sf o}^{J'}_{K'}\, 
   \op{T} 
   \, \absN{{\cal N}-\NN+C-K,\,\group{n_{\bf k}+J'-K'}}
\label{eq:app-microO-ex2}\;,
\end{align}
where the theta function appears because $\left(\op{L}\right)^K\, \absN{{\cal N}-\NN+C,\,\group{n_{\bf k}+J'-K'}}$ yields a vanishing state whenever $\NC + C-K$ is negative. Since $\absN{{\cal N}-\NN+C+J-K,\,\group{n_{\bf k}+J'-K'}}$ contains $\NN+J'-K'$ normal-component atoms, $\op{T}$ removes ${\cal N} -\NN-J'+K'$ atoms from its condensate part. This step converts \eqref{eq:app-microO-ex2} into
\begin{align}
  \ex{
   \left[
    \left(\op{L}^\dagger\right)^J \, \left(\op{L}\right)^K \, \op{O}^{J'}_{K'}
   \right]
   }
   \, \absN{C,\,\group{n_{\bf k}}} 
   =
    \theta_{\NC + C-K}\,
   {\sf o}^{J'}_{K'}\, 
   \absN{C+J+J'-K-K',\,\group{n_{\bf k}+J'-K'}}
\label{eq:app-microO-ex3}\;,
\end{align}
Since we are studying microcanonical operators, $J+J'=K+K'$, according to definition \eqref{eq:app-microO}. Consequently, result \eqref{eq:app-microO-ex3} becomes
\begin{align}
  \ex{
   \left[
    \left(\op{L}^\dagger\right)^J \, \left(\op{L}\right)^K \, \op{O}^{J'}_{K'}
   \right]
   }
   \, \absN{C,\,\group{n_{\bf k}}} 
   &=
    \theta_{\NC + C-K}\,
   {\sf o}^{J'}_{K'}\, 
   \absN{C,\,\group{n_{\bf k}+J'-K'}}
\nonumber\\
   &=
   \theta_{\NC + C-K}\,
   \op{O}^{J'}_{K'}\, 
   \absN{C,\,\group{n_{\bf k}}} 
   =
   \op{O}^{J'}_{K'}\, 
   \theta_{\opNC + C-K}\,
   \absN{C,\,\group{n_{\bf k}}} 
\label{eq:app-microO-ex4}\;,
\end{align}
where we have applied relation \eqref{eq:app-O-help2} in reversed direction. 
Since $\op{O}^{J'}_{K'}\, \theta_{\opNC + C-K}$ does not contain any condensate operators, the excitation-picture form of microcanonical operators contains only normal-component operators. For practical purposes, the theta function part can well be ignored because any existing BEC implies a macroscopic $\NC$ while $K$ is typically a small integer number. When this simplification is applied, we essentially limit the space of allowed states to minimum number of condensate atoms. This is not a necessary, but it will help bookkeeping and is justified for the BEC studies.

More specifically, result \eqref{eq:app-microO-ex4} implies
\begin{align}
  \ex{
   \left[
    \left(\op{L}^\dagger\right)^J \, \left(\op{L}\right)^K \, \op{O}^{J'}_{K'}
   \right]
   }
   &=
   \ex{
   \left[
    \left(\op{L}^\dagger\right)^J \, \left(\op{L}\right)^K \, \op{O}^{J'}_{K'}
   \right]
   }
   \sum_{C,\,\group{n_{\bf k}}}
   \, \absN{C,\,\group{n_{\bf k}}} \absNS{C,\,\group{n_{\bf k}}} 
\nonumber\\
   &=
   \sum_{C,\,\group{n_{\bf k}}}
   \ex{
   \left[
    \left(\op{L}^\dagger\right)^J \, \left(\op{L}\right)^K \, \op{O}^{J'}_{K'}
   \right]
   }
   \, \absN{C,\,\group{n_{\bf k}}} \absNS{C,\,\group{n_{\bf k}}} 
\nonumber\\
&=
   \sum_{C,\,\group{n_{\bf k}}}
   \op{O}^{J'}_{K'}\,\theta_{\opNC + C-K}\,
   \, \absN{C,\,\group{n_{\bf k}}} \absNS{C,\,\group{n_{\bf k}}} 
   =
   \op{O}^{J'}_{K'} \, \theta_{\opNC + C-K}\,
\label{eq:app-microO-FIN}\;,
\end{align}
where we have inserted the identity operator in the first step and have applied \eqref{eq:app-microO-ex4} as well as identified the identity operator in the last step. For realistic BEC studies, the theta function part can be eliminated producing
\begin{eqnarray}
  \left(\op{L}^\dagger\right)^J \, \left(\op{L}\right)^K \, \op{O}^{J'}_{K'}
  \toX
  \op{O}^{J'}_{K'}\,, \qquad
  J+J' = K+K'
\label{eq:app-microO2exci}\;,
\end{eqnarray}
which is the basis for transformation \eqref{eq:org2exci}.

A very similar derivation can be performed for microcanonical operators containing $B_0$ and $B_0^\dagger$ instead of $\op{L}$ and $\op{L}^\dagger$. The ones that appear in the Hamiltonian \eqref{eq:H-in-B} yield
\begin{align}
  &B^\dagger_{\bf k} B_{{\bf k}'} \toX B^\dagger_{\bf k} B_{{\bf k}'}\,,
  \quad
  B^\dagger_0 B_{\bf k} \toX \sqrt{\opNC} \, B_{\bf k} \,,
\qquad
  B^\dagger_{{\bf k}_1} B^\dagger_{{\bf k}_2} B_{{\bf k}_3} B_{{\bf k}_4}
    \toX B^\dagger_{{\bf k}_1} B^\dagger_{{\bf k}_2} B_{{\bf k}_3} B_{{\bf k}_4}\,, \qquad
\nonumber\\
  &B^\dagger_0 B_0 \toX \opNC\,,
    \quad
  B^\dagger_{\bf k} B_0 \toX  B_{\bf k}^\dagger \sqrt{\opNC} \,,  
\qquad
  B^\dagger_{{\bf k}_1} B^\dagger_{{\bf k}_2} B_{{\bf k}_3} B_{0}
    \toX B^\dagger_{{\bf k}_1} B^\dagger_{{\bf k}_2} B_{{\bf k}_3} \sqrt{\opNC}\,,
\nonumber\\
  &B^\dagger_{0} B^\dagger_{0} B_{{\bf k}} B_{{\bf k}'} 
    \toX B_{{\bf k}} B_{{\bf k}'}\,\sqrt{(\opNC+1)(\opNC+2)}, \qquad
\quad
  B^\dagger_{{\bf k}} B^\dagger_{{\bf k}'} B_{0} B_{0}
    \toX \sqrt{(\opNC+1)(\opNC+2)} \; B^\dagger_{{\bf k}} B^\dagger_{{\bf k}'}\,,
\nonumber\\
  &B^\dagger_{0} B^\dagger_{{\bf k}} B_{{\bf k}'} B_{0}
    \toX \opNC \, B_{{\bf k}} B_{{\bf k}'} \qquad
\qquad
  B^\dagger_{{\bf k}_1} B^\dagger_{{\bf k}_2} B_{{\bf k}_3} B_{0}
    \toX B^\dagger_{{\bf k}_1} B^\dagger_{{\bf k}_2} B_{{\bf k}_3}  \sqrt{\opNC}\,, 
\nonumber\\
  &B^\dagger_{0} B^\dagger_{{\bf k}_1} B_{{\bf k}_2} B_{{\bf k}_3}
    \toX  \sqrt{\opNC}\,B^\dagger_{{\bf k}_1} B_{{\bf k}_2} B_{{\bf k}_3}\,,
  \qquad
  B^\dagger_{0} B^\dagger_{0} B_{0} B_{0} 
    \toX \opNC \, \left( \opNC -1 \right)
\label{eq:app-B4H_ex}\;.
\end{align}
For these, the appearance of $\opNC$ operators replaces the theta operator observed in Eq.~\eqref{eq:app-microO-FIN}. To work out the quantum statistics of the BEC, we also need transformations
\begin{eqnarray}
  &&
  [B^\dagger_0 B_0]^J \toX \opNC^J 
\nonumber\\
  &&
  [B^\dagger_0]^J B_0^J  \toX \frac{\opNC!}{(\opNC-J)!} = \opNC (\opNC-1)\cdots(\opNC-J+1)  
\label{eq:app-Cond-QS}\;.
\end{eqnarray}
In the original picture, $\op{N}_0 \equiv B^\dagger_0 B_0$ defines the number operator of in the BEC. By comparing transformations \eqref{eq:app-Cond-QS} with relation \eqref{eq:Cond-op}, we conclude that excitation picture introduces a transformation
\begin{eqnarray}
  \op{N}_0\toX \opNC
\label{eq:app-N0-toX}
\end{eqnarray}
wherever a pure BEC number operator appears.

On a more general level, transformations \eqref{eq:app-microO2exci}--\eqref{eq:app-N0-toX} suggest that the microcanonical operators contain {\it exclusively  only} normal-component operators in the excitation picture, especially, $\ex{\op{H}}$ does [see Eq.~\eqref{eq:H-in-ex-start}]. As $\ex{\op{H}}$ is inserted into the Heisenberg equation of motion \eqref{eq:app-HEM_OP}, we can solve the quantum dynamics of {\it any} $\avex{\op{O}^{J'}_{K'}}$ in terms of the normal-component operators alone. In other words, the strongly correlated BEC part becomes formally eliminated, and the {\it physically relevant quantum kinetics is followed through the weakly correlated normal-component excitations}, which is the major benefit of the excitation picture.

\section{Properties of the $\sqrt{\opNC}$ operator}
\label{app:Commutators}

The original form of the Hamiltonian \eqref{eq:H-in-ex} contains a $\sqrt{\opNC}$ operator that is more difficult to treat than $\opNC$ due to the square root that appears. We use definition \eqref{eq:deltaNC} to express $\opNC$ in terms of the average BEC number $\NC$ and BEC-number fluctuation operator $\delta\opNC$, producing
\begin{eqnarray}
  \sqrt{\op{N}_C} = \sqrt{\NC + \delta\opNC} = \sqrt{\NC} \left({\textstyle 1 +\frac{\delta\opNC}{\NC}} \right)^{\frac{1}{2}}
\label{eq:app_NC}\;,
\end{eqnarray}
where the BEC number is moved to the front as a common factor of the summed terms. As shown in Sec.~\ref{sec:complementaryQS}, expectation values related to $\delta\opNC$ scale at most like $\sqrt{\NC}$. Therefore, $\frac{\delta\opNC}{\NC}$ remains a small number such that we can apply a converging Taylor expansion to express the operator part of Eq.~\eqref{eq:app_NC}. We then find
\begin{eqnarray}
  \sqrt{\op{N}_C} 
  = 
  \sqrt{\NC}  
  \left(
  {\textstyle
    1 
    + 
    \frac{1}{2\sqrt{\NC}} \frac{\delta\opNC}{\sqrt{\NC}} 
    + 
    {\cal O}\left( \frac{1}{\NC} \right)
  }
  \right)
  =
  {\textstyle
    \sqrt{\NC}  
    + 
    \frac{\delta\opNC}{2\sqrt{\NC}} 
    + 
    {\cal O}\left( \frac{1}{\sqrt{\NC}} \right)
  }
\label{eq:app_sqrtNC-FIN}\;,
\end{eqnarray}
where we have used the $\sqrt{\NC}$ scalability of $\delta\opNC$ operators to work out the leading-order terms. Since we are studying cases with a macroscopic BEC number, the two first contributions of $\sqrt{\op{N}_C}$ accurately describe the properties of the BEC fluctuations.

This relation can also be derived without the explicit knowledge of the scalability of $\delta\opNC$. To show this, we start from
\begin{eqnarray}
  \op{N}_C
  = {\cal N} - \opNN
\label{eq:app_NC1}\;,
\end{eqnarray}
based on definitions \eqref{eq:N_N}--\eqref{eq:N_connect}. An alternative form of Eq.~\eqref{eq:app_NC} becomes then
\begin{eqnarray}
  \sqrt{\op{N}_C} 
  = \sqrt{{\cal N} -\opNN}
  = 
 \sqrt{\cal N} \left({\textstyle 1 -\frac{\opNN}{\cal N}} \right)^{\frac{1}{2}}
\label{eq:app_sqrtNC1}\;.
\end{eqnarray}
Since $\opNN$ always generates a number smaller than or equal to the total atom number ${\cal N}$, fraction $\frac{\opNN}{\cal N}$ can be treated in the same way as a number that is smaller than or equal to one. Therefore, the square root always produces a converging Taylor expansion
\begin{eqnarray}
  \sqrt{\op{N}_C} 
  = 
  \sqrt{\cal N} \sum_{J=0}^\infty (-1)^J
  \left(
  {\textstyle 
    \begin{array}{c}
      \frac{1}{2}\\
      J
    \end{array}
    }
  \right)
  \left({\textstyle \frac{\opNN}{\cal N}} \right)^J
\label{eq:app_sqrtNC2}\;,
\end{eqnarray}
where we have used a general form of the binomial factorial
\begin{eqnarray}
  \left(
  {\textstyle 
    \begin{array}{c}
      n\\
      J
    \end{array}
    }
  \right)
  \equiv
  \frac{\Gamma(n+1)}{\Gamma(J+1)\,\Gamma(n-J+1)}
\label{eq:app_Gamma}\;.
\end{eqnarray}
expressed in terms of the gamma functions.\cite{Arfken:1985} Next, we work out the commutation properties of each $(\opNN)^J$ operator that appears in expansion \eqref{eq:app_sqrtNC2}.

It is straightforward to show that a commutator of $B_{\bf k}$ and $\opNN$ produces 
\begin{eqnarray}
  \comm{B_{\bf k}}{\opNN } 
  &=& 
  B_{\bf k}\,,\qquad
  \comm{\opNN} {B_{\bf k}}
  =
  -B_{\bf k} 
\label{eq:app_BcommNN}\;,
\end{eqnarray}
based on defintion \eqref{eq:N_N} and the usual boson commutation relations. When we apply result \eqref{eq:app_BcommNN} $J$ times, we find 
\begin{eqnarray}
  \comm{B_{\bf k}}{( \opNN )^J } 
  = 
  \left( (\opNN+1)^J - ( \opNN )^J \right) B_{\bf k}
  =
  B_{\bf k} \left( (\opNN)^J - ( \opNN -1)^J \right) 
\label{eq:app_NJ-comm-B}\;.
\end{eqnarray}
The corresponding commutator for the creation operator then becomes 
\begin{eqnarray}
  \comm{B^\dagger_{\bf k}}{( \opNN )^J } 
  = 
  B^\dagger_{\bf k}
  \left( ( \opNN )^J -(\opNN+1)^J \right) 
  =
  \left( ( \opNN -1)^J -(\opNN)^J \right) B^\dagger_{\bf k}
\label{eq:app_NJ-comm-B*}
\end{eqnarray}
that is obtained directly from Eq.~\eqref{eq:app_NJ-comm-B} by Hermitian conjugating it; we have also used the property that $\opNN$ is Hermitian. We next consider a generic operator $f\left( \opNN \right)$ that is expressed using a function $f(x)$. We also assume that $f(x)$ has a Taylor expansion that converges
\begin{eqnarray}
  f\left( \opNN \right)
  = 
 \sum_{J=0}^\infty a_J\,(\opNN)^J
\label{eq:app_f-Taylor}\;.
\end{eqnarray}
where $a_J$ are the specific Taylor-expansion coefficients. The explicit convergence criteria will be defined later once the  $f(x)$ function is chosen explicitly. With the help of commutators \eqref{eq:app_NJ-comm-B}-\eqref{eq:app_NJ-comm-B*}, we find
\begin{eqnarray}
  \comm{B_{\bf k}}{f(\op{N}_C)} 
  &=&
  \left( f(\opNC-1)-f(\opNC) \right) B_{\bf k}
  = 
  B_{\bf k} \left( f(\opNC)-f(\opNC+1) \right) 
\label{eq:app_COMM-F}\;,
\\
  \comm{B_{\bf k}^\dagger}{f(\op{N}_C)} 
  &=&
  B_{\bf k}^\dagger \left( f(\opNC)-f(\opNC-1) \right) 
  = 
 \left( f(\opNC+1)-f(\opNC) \right)  B_{\bf k}^\dagger
\label{eq:app_COMM-F*}
\end{eqnarray}
for any function $f(x)$ that has a convergent Taylor expansion.

As shown by Eq.~\eqref{eq:app_sqrtNC2}, $\sqrt{\opNC} = \sqrt{{\cal N} -\opNN}$ has a converging Taylor expansion that has the same form as Eq.~\eqref{eq:app_f-Taylor}. Therefore, we can directly apply result \eqref{eq:app_COMM-F} to determine the commutators between $B_{\bf k}$ and $\sqrt{\opNC}$:
\begin{eqnarray}
  \comm{B_{\bf k}}{\sqrt{\op{N}_C}} 
  &=& 
  \left(\sqrt{\opNC-1}- \sqrt{\opNC} \,\right) B_{\bf k}
  = -\frac{1}{\sqrt{\opNC-1}+ \sqrt{\opNC}} \, B_{\bf k}
\nonumber\\
  &=& 
  B_{\bf k} \left(\sqrt{\opNC}- \sqrt{\opNC+1} \,\right) 
  = - B_{\bf k}\,\frac{1}{\sqrt{\opNC}+ \sqrt{\opNC+1}} 
\label{eq:app_COMMsqrtNC}\;,
\end{eqnarray}
where the first step of both lines follows from identifications \eqref{eq:app_sqrtNC1}--\eqref{eq:app_sqrtNC2} while the last expression of both lines is obtained after we have multiplied both the numerator and denominator by $\sqrt{\opNC}+ \sqrt{\opNC\pm1}$. Analogous steps produce
\begin{eqnarray}
  \comm{B^\dagger_{\bf k}}{\sqrt{\op{N}_C}} 
  &=& 
  \frac{1}{\sqrt{\opNC}+ \sqrt{\opNC+1}} \, B^\dagger_{\bf k}
  =
  B^\dagger_{\bf k}\,\frac{1}{\sqrt{\opNC}+ \sqrt{\opNC-1}} 
\label{eq:app_COMMsqrtNC*}\;,
\end{eqnarray}
for the creation operator.

In this paper, we focus the analysis on cases where the BEC number is appreciable, i.e.~much larger than one. Therefore, the factor one within $\sqrt{\opNC \pm 1}$ becomes negligible, making $\sqrt{\opNC \pm 1}$  and $\sqrt{\opNC}$ essentially identical. Applying this limit to Eqs.~\eqref{eq:app_COMMsqrtNC}--\eqref{eq:app_COMMsqrtNC*}, we find a simplification
\begin{eqnarray}
  &&\comm{B_{\bf k}}{\sqrt{\op{N}_C}} 
  \xrightarrow{\NC \gg 1}
  -\frac{1}{2 \sqrt{\opNC}} \, B_{\bf k}  = -B_{\bf k}\, \frac{1}{2 \sqrt{\opNC}} 
\label{eq:COMMsqrtNCsimp2}\;,
\\
  &&\comm{B^\dagger_{\bf k}}{\sqrt{\op{N}_C}} 
  \xrightarrow{\NC \gg 1} 
  \frac{1}{2\sqrt{\opNC}} \, B^\dagger_{\bf k}
  =
  B^\dagger_{\bf k}\,\frac{1}{2\sqrt{\opNC}} 
\label{eq:COMMsqrtNCsimp2*}\;.
\end{eqnarray}
These forms suggest that both $B_{\bf k}$ and $B^\dagger_{\bf k}$ commute with the same  $\frac{1}{2\sqrt{\opNC}}$ contribution when $\NC$ becomes sufficiently large. Since $B_{\bf k}$ and $B^\dagger_{\bf k}$ span all bosonic operators, an operator that commutes with both of them must be a constant, not an operator. Therefore, the leading order contribution of commutators \eqref{eq:COMMsqrtNCsimp2*} must necessarily be
\begin{eqnarray}
  &&
  \comm{B_{\bf k}}{\sqrt{\op{N}_C}} 
  \xrightarrow{\NC \gg 1}
  -\frac{1}{2 \sqrt{\NC}} \, B_{\bf k}  
\label{eq:COMMsqrtNCsimpFIN}\;,
\\
  &&
  \comm{B^\dagger_{\bf k}}{\sqrt{\op{N}_C}} 
  \xrightarrow{\NC \gg 1}
  \frac{1}{2\sqrt{\NC}} \, B^\dagger_{\bf k} 
\label{eq:COMMsqrtNCsimpFIN*}\;.
\end{eqnarray}
It is clear that the lowest order contribution to ${\sqrt{\op{N}_C}}$ is a constant $\sqrt{\NC}$ while \eqref{eq:COMMsqrtNCsimpFIN}--\eqref{eq:COMMsqrtNCsimpFIN*} settles the leading order operator aspects of ${\sqrt{\op{N}_C}}$. Especially, commutation relations \eqref{eq:COMMsqrtNCsimpFIN}--\eqref{eq:COMMsqrtNCsimpFIN*} are satisfied if we replace ${\sqrt{\op{N}_C}}$ by $\sqrt{\NC} + \frac{\delta\opNC}{2\sqrt{\NC}}$. This is exactly the leading order contribution of relation \eqref{eq:app_sqrtNC-FIN}, which verifies that expansion \eqref{eq:app_sqrtNC-FIN} indeed describes ${\sqrt{\op{N}_C}}$ whenever the BEC number is appreciable. Interestingly, we do not additionally need to know how the BEC fluctuations behave in order to replace ${\sqrt{\op{N}_C}}$ by $\sqrt{\NC} + \frac{\delta\opNC}{2\sqrt{\NC}}$. Or conversely, the simultaneous validity of  results \eqref{eq:app_sqrtNC-FIN} and \eqref{eq:COMMsqrtNCsimpFIN}--\eqref{eq:COMMsqrtNCsimpFIN*} suggest that the BEC fluctuations must 
scale at most like $\sqrt{\NC}$. Equations \eqref{eq:app_COMMsqrtNC}--\eqref{eq:app_COMMsqrtNC*} can be utilized as a general starting point to include $\sqrt{\opNC}$ effects beyond the linearization \eqref{eq:sqrtBFIN}.


\bibliographystyle{elsarticle-num-names}
\bibliography{ref,Seminal,Reviews,EXP,TheoKin,TheoStat}

\end{document}